\PassOptionsToPackage{table}{xcolor}
\documentclass[conference]{IEEEtran}

\usepackage{url}
\usepackage{xurl}
\usepackage[hidelinks]{hyperref}
\usepackage{orcidlink}
\usepackage{graphicx}
\usepackage{xcolor}
\usepackage{amsmath}
\usepackage{array}
\usepackage{booktabs}
\usepackage{tabularx}
\usepackage{listings}
\usepackage{microtype}
\usepackage{flushend}
\usepackage{tikz}
\usepackage{amssymb}
\usetikzlibrary{calc,shapes.geometric,shapes.symbols,positioning,arrows.meta,shapes,backgrounds,fit,decorations.pathreplacing}

\definecolor{ink}{HTML}{111111}
\definecolor{panel}{HTML}{F8F8F8}
\definecolor{accent}{HTML}{A32020}
\definecolor{accentbg}{HTML}{FDF2F2}
\definecolor{headerbg}{HTML}{E5E8EB}
\definecolor{rowzebra}{HTML}{F6F8FA}
\definecolor{sectionbg}{HTML}{DCE1E6}

\newcommand{\iconCardholder}[1]{%
  \begin{scope}[shift={#1}]
    \draw[line width=0.95pt, color=ink, fill=white] (0,0.26) circle (0.16);
    \draw[line width=0.95pt, color=ink, fill=white]
      (-0.38,-0.38) -- (-0.38,-0.22)
      .. controls (-0.38,-0.03) and (-0.20,0.10) .. (0,0.10)
      .. controls (0.20,0.10) and (0.38,-0.03) .. (0.38,-0.22)
      -- (0.38,-0.38) -- cycle;
  \end{scope}
}
\newcommand{\iconServer}[1]{%
  \begin{scope}[shift={#1}]
    \foreach \y in {-0.4,-0.1,0.2} {
      \draw[line width=0.95pt, color=ink, fill=white, sharp corners] (-0.42,\y) rectangle (0.42,\y+0.24);
      \fill[color=ink] (-0.33,\y+0.07) rectangle (-0.25,\y+0.17);
      \draw[color=ink, line width=0.6pt] (-0.15,\y+0.12) -- (0.30,\y+0.12);
    }
  \end{scope}
}
\newcommand{\iconGateway}[1]{%
  \begin{scope}[shift={#1}]
    \draw[line width=0.85pt, color=ink, fill=black!4, sharp corners] (-0.36,-0.18) rectangle (0.48,0.32);
    \draw[line width=0.95pt, color=ink, fill=white, sharp corners] (-0.48,-0.32) rectangle (0.36,0.18);
    \node[font=\small\bfseries, text=ink] at (-0.06,-0.07) {$\rightleftarrows$};
  \end{scope}
}
\newcommand{\iconDatabase}[1]{%
  \begin{scope}[shift={#1}]
    \foreach \y in {-0.36,-0.08,0.20} {
      \draw[line width=0.95pt, color=ink, fill=white, sharp corners] (-0.36,\y) rectangle (0.36,\y+0.22);
      \draw[line width=0.55pt, color=ink] (-0.36,\y+0.11) -- (0.36,\y+0.11);
      \fill[color=ink] (-0.27,\y+0.04) rectangle (-0.21,\y+0.08);
      \fill[color=ink] (-0.27,\y+0.14) rectangle (-0.21,\y+0.18);
    }
  \end{scope}
}
\newcommand{\iconSecurity}[1]{%
  \begin{scope}[shift={#1}]
    \draw[line width=0.95pt, color=ink, fill=white]
      (0,0.42) -- (0.40,0.18) -- (0.40,-0.18) -- (0,-0.42) -- (-0.40,-0.18) -- (-0.40,0.18) -- cycle;
    \draw[line width=1.3pt, color=ink, line cap=square, line join=miter]
      (-0.16,-0.05) -- (-0.02,-0.20) -- (0.22,0.12);
  \end{scope}
}

\tikzset{
  msgsolid/.style={-{Stealth[length=2.6mm,width=1.8mm]}, line width=1.05pt, draw=ink},
  msgdash/.style ={-{Stealth[length=2.4mm,width=1.6mm]}, line width=0.9pt, draw=black!80,
                   dash pattern=on 3.5pt off 2.2pt},
  lifeline/.style={line width=0.75pt, draw=black!60, dash pattern=on 2.5pt off 2.5pt},
  actboxlbl/.style={font=\sffamily\bfseries\footnotesize, text=ink},
  actboxsub/.style={font=\sffamily\scriptsize, text=black!65},
  msglbl/.style={font=\footnotesize\bfseries, text=ink, align=center, fill=white, draw=black!70, line width=0.5pt, sharp corners, inner sep=2.5pt},
  msgsub/.style={font=\scriptsize\bfseries, text=black!85, align=center, fill=white, draw=black!45, line width=0.4pt, sharp corners, inner sep=2.0pt},
  stepbadge/.style={rectangle, sharp corners, fill=ink, text=white, font=\scriptsize\bfseries,
                     inner sep=0pt, minimum size=5.2mm},
}

\newcommand{\sqnum}[1]{%
  \tikz[baseline=(n.base)]{%
    \node[rectangle, sharp corners, fill=ink, text=white, draw=none,
      inner sep=1.2pt, minimum size=1.15em,
      font=\sffamily\bfseries\scriptsize] (n) {#1};%
  }%
}

\providecommand{\Description}[1]{}
\providecommand{\nolinkurl}[1]{\texttt{#1}}

\definecolor{codegreen}{rgb}{0,0.6,0}
\definecolor{codegray}{rgb}{0.5,0.5,0.5}
\definecolor{codepurple}{rgb}{0.58,0,0.82}
\definecolor{backcolour}{rgb}{0.95,0.95,0.92}

\lstdefinelanguage{ProVerif}{
  keywords={free, channel, type, fun, reduc, event, query, let, in, if, then, else, new, out, process, private},
  keywordstyle=\color{blue}\bfseries,
  ndkeywords={attacker, true, false, bitstring},
  ndkeywordstyle=\color{darkgray}\bfseries,
  identifierstyle=\color{black},
  sensitive=false,
  comment=[l]{(*},
  commentstyle=\color{codegreen}\ttfamily,
  stringstyle=\color{codepurple}\ttfamily,
  morestring=[b]",
  literate={->}{{$\to$}}1,
}

\lstdefinestyle{mystyle}{
    backgroundcolor=\color{backcolour},
    commentstyle=\color{codegreen},
    keywordstyle=\color{blue},
    numberstyle=\scriptsize\color{codegray},
    stringstyle=\color{codepurple},
    basicstyle=\ttfamily\small,
    breakatwhitespace=false,
    breaklines=true,
    captionpos=b,
    keepspaces=true,
    numbers=left,
    numbersep=5pt,
    showspaces=false,
    showstringspaces=false,
    showtabs=false,
    tabsize=2,
    frame=single,
    rulecolor=\color{black!45}
}

\renewcommand{\arraystretch}{1.14}
\makeatletter
\renewcommand{\paragraph}{\@startsection{paragraph}{4}{\z@}%
  {0.9ex plus 0.25ex minus 0.15ex}%
  {0.4em}%
  {\normalfont\normalsize\bfseries}}
\makeatother
\renewcommand{\dbltopfraction}{0.90}
\renewcommand{\dblfloatpagefraction}{0.85}

\IEEEoverridecommandlockouts
\renewcommand{\IEEEkeywordsname}{Keywords}

\begin{document}

\title{The 3DS Oracle: BIN Attack via the Network}

\author{
\IEEEauthorblockN{{\fontsize{10.5}{12}\selectfont Wasif Faisal\,\orcidlink{0000-0003-0696-5623}\textsuperscript{,}\IEEEauthorrefmark{1}\IEEEauthorrefmark{2}, Suprava Saha Dibya\,\orcidlink{0009-0007-9607-925X}\textsuperscript{,}\IEEEauthorrefmark{1}}}
\IEEEauthorblockA{\IEEEauthorrefmark{1}\textit{BRAC University}}
\thanks{\IEEEauthorrefmark{2}{\fontsize{8}{9.5}\selectfont Independent research initiated by Wasif Faisal in 2018. The title's ``oracle'' honors Professor Ross Anderson's ``elf sitting in a black box'' analogy for the random-oracle model and his coauthored 3-D Secure analysis~\cite{anderson2020securityengineering,murdoch2010vbv}.}}
}

\maketitle

\begin{abstract}
EMV 3-D Secure (3DS) requires a client-visible continuation boundary in browser-mediated flows; in some deployments, that boundary exposes a pre-authorization classification channel. This paper studies 3DS as a protocol, integration surface, and governance problem, combining standards analysis, architectural reconstruction, formal modeling, a documentation audit, disclosure history, and a bounded live evaluation. We show that client-visible integrations can reflect routing and authentication state before financial authorization, allowing 3DS to function as a profiling and screening channel for issuer controls and cardholder/account state rather than only as a defense. In the tested browser-channel Braintree/Cardinal path, the first client-visible result separated researcher-owned anchor cards from bounded adjacent same-range comparison PANs in six of seven distinct ACS/card-range cohorts; separate public evidence, including openly accessible attacker tooling, rotating-proxy support, and illicit paid 3DS card-testing services, shows that comparable signals are already operationalized in the wild. Our live study was deliberately bounded for ethical reasons, so the empirical ceiling reported here reflects researcher restraint rather than an attacker-side scalability limit, although we do not claim network-wide prevalence or candidate issuance. The deeper issue is systemic: decoupled pre-authorization authentication, weak requestor-boundary assumptions, identifier-linked routing invariants, and fragmented accountability let the exposure persist even when individual actors appear locally conformant. Meaningful remediation therefore requires consortium-level reform, including stricter minimization of client-visible pre-authorization responses, more uniform continuation behavior, and stronger client authenticity and session binding.
\end{abstract}
\vspace{0.5\baselineskip}

\begin{IEEEkeywords}
EMV 3-D Secure (3DS), payment card enumeration, pre-authorization response leakage, authentication BIN attack, carding-as-a-service (CaaS), residential SOCKS5 proxy egress, ethical disclosure, payment security, protocol security.
\end{IEEEkeywords}

\IEEEpeerreviewmaketitle

\section{Introduction}

\vspace{2pt}
\noindent
{\centering
  {\small\bfseries
  ``Does the online card payment landscape\\
unwittingly facilitate fraud?''\par}
  \smallskip
  {\footnotesize\upshape---Ali et al., IEEE S\&P 2017~\cite{ali2017distributed}\par}}
\vspace{3pt}

EMV 3-D Secure (3DS) sits at a pivotal point in online card payments: before clearing and before financial authorization, it determines how an attempted authentication is routed and classified. Because browser-mediated 3DS must expose a client-visible continuation boundary, some deployments turn that boundary into a pre-authorization classification channel. A caller can receive input-dependent 3DS outcomes that reveal issuer routing, authentication posture, and account recognition. That makes 3DS not only an authentication control, but also a profiling and screening channel. The problem is systemic rather than local: the signal emerges across the merchant, gateway, 3DS Server, Directory Server, and issuer ACS path, while responsibility is divided across actors that can each claim component-level conformity.

Nearly twenty-five years after the introduction of 3-D Secure, Ali et al.'s question remains as timely as ever. Their study showed that differences in merchant-side checks can help attackers infer missing card details during payment authorization~\cite{ali2017distributed}. Nearly a decade later, we examine an earlier point in the stack: whether an attacker can algorithmically construct Luhn-valid PAN candidates and use the first client-visible response of a tested 3DS integration as an account-recognition oracle, without requiring a stolen-card dump or data-breach corpus. Within that broader threat model, we also test the stronger adjacent-anchor case: whether one known-issued anchor card is sufficient to prioritize bounded same-range candidates. We evaluate that bounded adjacent-anchor hypothesis across seven researcher-owned anchor-card cohorts, each representing a distinct ACS/card-range cohort. The result is not a claim of universal exploitability or candidate issuance, but evidence of pre-authorization response-class separation before payment authorization ever begins.

This manuscript is intentionally broader than a narrow attack paper. It studies 3DS as a protocol-and-ecosystem arrangement through standards analysis, architectural reconstruction, formal modeling, frozen-documentation review, disclosure history, and a bounded live evaluation. In the tested browser-channel Braintree/Cardinal path, the first client-visible result associated with the initial AReq/ARes exchange separated researcher-owned anchor cards from bounded adjacent same-range comparison PANs in six of seven distinct ACS/card-range cohorts; the seventh yielded a uniform challenge-required class. Separate public evidence, including openly accessible attacker tooling, rotating-proxy support, and illicit paid 3DS card-testing services, shows that comparable signals are already operationalized in the wild, including distributed workflows that classify estimated credentials. Our live study, however, was deliberately bounded for ethical reasons; that bound should not be mistaken for attacker-side infeasibility, even though we do not claim network-wide prevalence or candidate issuance.

We call the first client-visible result examined here the \emph{lookup-stage result}: gateway JSON associated with the initial exchange, not a raw ARes message or a named EMVCo phase. Conceptually, it is the first point at which the system shows the caller how it is classifying the attempted authentication. It is also a branch point. Frictionless authentication may complete there, while \texttt{transStatus=C} denotes that a challenge is required, not completed. In the attacker-style off-context, no-browser path tested here, the caller stopped after the lookup-stage result and never sent the initial \texttt{CReq}. The later challenge/OOB interaction therefore never began in our tests, and no SMS, push, or OTP reached our test recipients; EMVCo does not define a cardholder SMS/push alert for the lookup-only or missing-initial-\texttt{CReq} path, though independent issuer notification systems were not instrumented. Ali et al. observed their signal during authorization while guessing secondary fields of a known PAN. Our observation arises earlier in the payment stack: at the initial lookup, before financial authorization and, on the challenge branch, before challenge completion. That timing matters operationally. Even without a completed challenge or downstream authorization, a frictionless or other success-labeled lookup result is already a high-value screening signal because it lets an attacker triage candidates before step-up friction, browser challenge execution, or issuer authorization ever occurs.

\paragraph{Account existence versus credential validity.} The distinction matters throughout this manuscript. In this paper, \emph{account existence} means that the issuer/ACS appears to recognize the PAN at lookup as corresponding to an account record; it does not establish that the card is current, unblocked, or approvable for payment. \emph{Credential validity} is the stronger operational claim that the presented credentials are usable in the relevant authentication and payment flow; 3DS authentication alone does not guarantee financial authorization. We therefore describe the observed lookup-stage response-class separation between researcher-owned anchor cards and adjacent comparison PANs as a pre-authorization account-existence/recognition signal that can filter or prioritize inputs, not as proof that every candidate is unissued or currently usable.

Req~104 assigns the ACS Transaction ID at the ACS, Req~105 directs the ACS to use the account number from the AReq to determine whether authentication is available or can be completed, and Req~411 requires the ACS and DS Transaction IDs to match when the DS creates an ARes on the ACS's behalf~\cite{emvco_core_2311_2023}. Thus, under correct field mapping and specification conformance, \texttt{acsTransID} $\neq$ \texttt{dsTransID} supports ACS-path attribution and strengthens the account-existence inference when read with the response class. The inequality alone is provenance evidence, not proof that an issuer ledger record was found; it is also not required for response-class filtering, as Cohort~E separates with equal IDs while Cohort~G is uniform despite distinct IDs (Table~\ref{tab:cohort_result}).

\begin{figure*}[t]
\centering
\resizebox{0.95\textwidth}{!}{%
\begin{tikzpicture}[>=Stealth]

\def\xA{0.0}     
\def\xB{4.1}     
\def\xC{8.5}     
\def\xD{12.8}    
\def\xE{17.0}    

\def\hw{0.14}    

\def\yIcon{11.60}
\def\yBoxTop{10.35}
\def\yBoxBot{8.65}
\def\yLifeTop{8.65}
\def\yLifeBot{-3.55}

\def\yS{7.75}
\def\yTa{6.90}
\def\yTb{6.15}
\def\yLk{5.20}
\def\yNote{4.40}
\def\yAReq{3.65}
\def\yDSCheck{2.90}
\def\yRoute{2.10}
\def\yBranch{1.25}
\def\yARes{0.40}
\def\yJson{-0.55}
\def\yCReq{-1.65}
\def\yTimeout{-2.65}
\def\yParadox{-4.05}
\def\yStop{-5.15}
\def\yLeg{-5.85}

\begin{scope}[on background layer]
  \draw[lifeline] (\xA,\yLifeTop) -- (\xA,\yLifeBot);
  \draw[lifeline] (\xB,\yLifeTop) -- (\xB,\yLifeBot);
  \draw[lifeline] (\xC,\yLifeTop) -- (\xC,\yLifeBot);
  \draw[lifeline] (\xD,\yLifeTop) -- (\xD,\yLifeBot);
  \draw[lifeline] (\xE,\yLifeTop) -- (\xE,\yLifeBot);

  \draw[fill=black!8, draw=black!75, line width=0.6pt, sharp corners]
    (\xA-\hw, \yS) rectangle (\xA+\hw, \yJson);
  \draw[fill=black!8, draw=black!75, line width=0.6pt, sharp corners]
    (\xB-\hw, \yS+0.28) rectangle (\xB+\hw, \yS-0.22);
  \draw[fill=black!8, draw=black!75, line width=0.6pt, sharp corners]
    (\xC-\hw, \yTa+0.28) rectangle (\xC+\hw, \yJson);
  \draw[fill=black!8, draw=black!75, line width=0.6pt, sharp corners]
    (\xD-\hw, \yAReq+0.28) rectangle (\xD+\hw, \yARes-0.22);
  \draw[fill=black!8, draw=black!75, line width=0.6pt, sharp corners]
    (\xE-\hw, \yRoute+0.28) rectangle (\xE+\hw, \yBranch-0.22);
  \draw[fill=accentbg, draw=accent, line width=0.6pt, dashed, sharp corners]
    (\xE-\hw, \yBranch-0.22) rectangle (\xE+\hw, \yTimeout);
\end{scope}

\iconCardholder{(\xA,\yIcon)}
\draw[draw=ink, fill=white, line width=0.95pt, sharp corners]
  (\xA-1.45,\yBoxBot) rectangle (\xA+1.45,\yBoxTop);
\node[actboxlbl, align=center, text width=2.8cm] at (\xA,\yBoxTop-0.50) {Off-Context Caller\\[2pt]\textcolor{accent}{\sffamily\bfseries\scriptsize Scripted HTTP Client}};
\draw[draw=black!35, line width=0.5pt] (\xA-1.25,\yBoxBot+0.52) -- (\xA+1.25,\yBoxBot+0.52);
\node[actboxsub, align=center, text width=2.8cm] at (\xA,\yBoxBot+0.26) {no browser / Web SDK};

\iconServer{(\xB,\yIcon)}
\draw[draw=ink, fill=white, line width=0.95pt, sharp corners]
  (\xB-1.35,\yBoxBot) rectangle (\xB+1.35,\yBoxTop);
\node[actboxlbl, align=center, text width=2.6cm] at (\xB,\yBoxTop-0.50) {Merchant Server\\[2pt]\textcolor{accent}{\sffamily\bfseries\scriptsize Public Storefront}};
\draw[draw=black!35, line width=0.5pt] (\xB-1.15,\yBoxBot+0.52) -- (\xB+1.15,\yBoxBot+0.52);
\node[actboxsub, align=center, text width=2.6cm] at (\xB,\yBoxBot+0.26) {checkout authorization};

\iconGateway{(\xC,\yIcon)}
\draw[draw=ink, fill=white, line width=0.95pt, sharp corners]
  (\xC-1.55,\yBoxBot) rectangle (\xC+1.55,\yBoxTop);
\node[actboxlbl, align=center, text width=3.0cm] at (\xC,\yBoxTop-0.50) {3DS Server / Gateway\\[2pt]\textcolor{accent}{\sffamily\bfseries\scriptsize Payment Gateway \& 3DSS}};
\draw[draw=black!35, line width=0.5pt] (\xC-1.35,\yBoxBot+0.52) -- (\xC+1.35,\yBoxBot+0.52);
\node[actboxsub, align=center, text width=3.0cm] at (\xC,\yBoxBot+0.26) {Braintree / CardinalCommerce};

\iconDatabase{(\xD,\yIcon)}
\draw[draw=ink, fill=white, line width=0.95pt, sharp corners]
  (\xD-1.85,\yBoxBot) rectangle (\xD+1.85,\yBoxTop);
\node[actboxlbl, align=center, text width=3.6cm] at (\xD,\yBoxTop-0.50) {Directory Server (DS)\\[2pt]\textcolor{accent}{\sffamily\bfseries\scriptsize Card Scheme Routing Tier}};
\draw[draw=black!35, line width=0.5pt] (\xD-1.65,\yBoxBot+0.52) -- (\xD+1.65,\yBoxBot+0.52);
\node[actboxsub, align=center, text width=3.6cm] at (\xD,\yBoxBot+0.26) {card scheme range routing};

\iconSecurity{(\xE,\yIcon)}
\draw[draw=ink, fill=white, line width=0.95pt, sharp corners]
  (\xE-1.45,\yBoxBot) rectangle (\xE+1.45,\yBoxTop);
\node[actboxlbl, align=center, text width=2.8cm] at (\xE,\yBoxTop-0.50) {Issuer ACS\\[2pt]\textcolor{accent}{\sffamily\bfseries\scriptsize Access Control Server}};
\draw[draw=black!35, line width=0.5pt] (\xE-1.25,\yBoxBot+0.52) -- (\xE+1.25,\yBoxBot+0.52);
\node[actboxsub, align=center, text width=2.8cm] at (\xE,\yBoxBot+0.26) {core ledger evaluation};

\draw[msgdash] (\xB-\hw,\yS) -- (\xA+\hw,\yS);
\node[msglbl] at ({(\xA+\xB)/2},\yS+0.28) {\sqnum{1} \texttt{client\_token} (public checkout bootstrap)};

\draw[msgsolid] (\xA+\hw,\yTa) -- (\xC-\hw,\yTa);
\node[msglbl] at ({(\xA+\xC)/2},\yTa+0.28) {\sqnum{2} Tokenization request (candidate PAN)};
\draw[msgsolid] (\xC-\hw,\yTb) -- (\xA+\hw,\yTb);
\node[msgsub] at ({(\xA+\xC)/2},\yTb-0.24) {response: \texttt{nonce} + \texttt{bin}};

\draw[msgsolid] (\xA+\hw,\yLk) -- (\xC-\hw,\yLk);
\node[msglbl] at ({(\xA+\xC)/2},\yLk+0.32) {\sqnum{3} \textbf{Direct HTTP lookup} (synthetic telemetry, no browser runtime)};
\node[msgsub] at ({(\xA+\xC)/2},\yLk-0.26) {scripted query submits candidate PAN credentials to pre-auth endpoint};

\draw[draw=ink, line width=0.7pt] (\xC+\hw,\yNote) -- (\xC+\hw+0.30,\yNote);
\node[draw=black!60, fill=black!4, line width=0.55pt, sharp corners, inner sep=2.5pt,
      font=\scriptsize\bfseries, text=ink, anchor=west]
  at (\xC+\hw+0.30,\yNote) {internal mapping untraced};

\draw[msgdash] (\xC+\hw,\yAReq) -- (\xD-\hw,\yAReq);
\node[msglbl] at ({(\xC+\xD)/2},\yAReq+0.28) {\sqnum{4} \textbf{3DS AReq} (scheme range routing)};

\node[draw=ink, fill=black!4, line width=0.55pt, sharp corners, inner sep=3.0pt,
      font=\scriptsize\bfseries, text=ink, align=center]
  at (\xD,\yDSCheck) {DS checks Card Range Table (Seq 3.133 [Req 97])};

\draw[msgdash] (\xD+\hw,\yRoute) -- (\xE-\hw,\yRoute);
\node[msglbl] at ({(\xD+\xE)/2},\yRoute+0.28) {Forward AReq to participating issuer [Req 105]};

\draw[msgdash] (\xE-\hw,\yBranch) -- (\xD+\hw,\yBranch);
\node[msglbl] at ({(\xD+\xE)/2},\yBranch+0.28) {ACS evaluates ledger; submits ARes (transStatus / reason code)};

\draw[msgdash] (\xD-\hw,\yARes) -- (\xC+\hw,\yARes);
\node[draw=accent, fill=accentbg, line width=0.9pt, sharp corners, inner sep=3.5pt,
      font=\scriptsize\bfseries, text=ink, align=center]
  at ({(\xC+\xD)/2},\yARes+0.42)
  {\sqnum{5} \textbf{Pass ARes to 3DSS}\,\textemdash\,\textbf{\color{accent}transStatus \& Reason}\\(e.g., Reason 08 / 13 / 98 / CAVV minted)};

\draw[msgsolid] (\xC-\hw,\yJson) -- (\xA+\hw,\yJson);
\node[msglbl] at ({(\xA+\xC)/2},\yJson+0.32) {\sqnum{6} \textbf{Pre-Auth Oracle Leak: Gateway reflects lookup JSON}};
\node[msgsub] at ({(\xA+\xC)/2},\yJson-0.26) {Client-visible response classes \& cryptographic proof; zero clearing holds};

\draw[line width=1.1pt, draw=accent, dashed, -{Stealth[length=2.5mm,width=1.7mm]}] (\xA+\hw,\yCReq) -- (\xE-\hw,\yCReq);
\node[draw=accent, fill=white, line width=0.9pt, sharp corners, inner sep=3.5pt,
      font=\scriptsize\bfseries, text=accent, align=center]
  at ({(\xA+\xE)/2},\yCReq)
  {\textbf{Challenge branch only (\texttt{transStatus=C}):} tested caller omits \texttt{CReq}};
\node[font=\tiny\bfseries\sffamily, fill=white, inner sep=1.5pt, text=ink]
  at ({(\xA+\xE)/2},\yCReq-0.38)
  {Normal browser flow continues with \texttt{CReq}/\texttt{CRes}; off-context caller stops at Step~6};

\node[draw=ink, fill=white, line width=0.8pt, sharp corners, inner sep=3.5pt,
      font=\tiny\sffamily, align=center]
  at (\xA+1.3, \yTimeout)
  {\textbf{Off-Context Scripted Caller}\\[1pt]
   Extracts response classes\\[1pt]
   \textbf{\color{ink}Zero Cardholder UI Rendered}\\[1pt]
   Clearing thresholds bypassed};

\node[draw=accent, fill=accentbg, line width=0.85pt, sharp corners, inner sep=3.5pt,
      font=\tiny\sffamily, align=center]
  at (\xE-1.4, \yTimeout)
  {\textbf{Silent ACS Session Timeout}\\[1pt]
   Pending \texttt{CReq} times out (Reason~14)\\[1pt]
   ACS-to-DS \texttt{RReq} sent silently\\[1pt]
   Cardholder blinded to probe};

\node[draw=accent, fill=accentbg, line width=0.95pt, sharp corners, inner sep=5pt,
      font=\scriptsize\sffamily, text=ink, align=center, text width=17.0cm]
  at ({(\xA+\xE)/2},\yParadox)
  {\textbf{Pre-Authorization Oracle Demarcation (Frictionless vs.\ Challenge Flow):}\\[2pt]
   Frictionless approvals (\texttt{Y}) yield immediate authentication cryptograms (CAVV) without cardholder friction. Challenge-required outcomes (\texttt{C}) branch to interactive step-up; by halting before \texttt{CReq}, the off-context caller extracts response classification while the pending session silently times out (Req~221/Reason~14), triggering zero cardholder SMS/push notifications~\cite{emvco_core_2311_2023}.};

\node[draw=ink, fill=white, line width=0.95pt, sharp corners,
      font=\scriptsize\bfseries, text=ink, inner sep=4.5pt]
  at ({(\xA+\xE)/2},\yStop)
  {\sqnum{!} \textbf{PRE-AUTHORIZATION ORACLE BOUNDARY:} Frictionless completion (\texttt{Y}) vs.\ Step-up challenge (\texttt{C}) \quad\textperiodcentered\quad Scripted caller extracts classification without clearing holds};
\draw[draw=black!40, line width=0.6pt] (\xA-1.2,\yLeg+0.42) -- (\xE+1.2,\yLeg+0.42);

\draw[msgsolid] (\xA-1.15,\yLeg) -- (\xA-0.35,\yLeg);
\node[font=\scriptsize\bfseries, text=ink, anchor=west] at (\xA-0.2,\yLeg) {observed off-context HTTP call};

\draw[msgdash] (\xB-0.8,\yLeg) -- (\xB,\yLeg);
\node[font=\scriptsize\bfseries, text=ink, anchor=west] at (\xB+0.15,\yLeg) {provider/network context (not traced)};

\draw[line width=1.1pt, draw=accent, dashed, -{Stealth}] (\xC+0.4,\yLeg) -- (\xC+1.2,\yLeg);
\node[font=\scriptsize\bfseries, text=accent, anchor=west] at (\xC+1.35,\yLeg) {aborted / omitted CReq path};

\fill[accent] (\xE-1.4,\yLeg-0.08) rectangle (\xE-1.28,\yLeg+0.08);
\node[font=\scriptsize\bfseries, text=accent!90!black, anchor=west] at (\xE-1.15,\yLeg) {pre-auth oracle response exposure};

\end{tikzpicture}
}
\caption{\textbf{Measured off-context HTTP path and the initial lookup-result boundary in the tested 3DS integration.} Solid arrows show the observed tokenization and direct lookup executed by a standalone off-context caller; dashed arrows show provider/network context, which was not internally traced. The initial client-visible result is integration output associated with the initial \texttt{AReq}/\texttt{ARes} exchange, not a raw ARes capture. It may be a frictionless completion, a request for a challenge (\texttt{transStatus=C}), or another status. A frictionless result completes without \texttt{CReq}/\texttt{CRes}; only the challenge-required branch proceeds to that exchange in a normal browser flow. In our tested off-context, no-browser path, the caller halted after the initial result and never sent the first \texttt{CReq}. That means the later challenge/OOB interaction never opened; no SMS, push, or OTP was observed on our test cards. For a missing initial \texttt{CReq}, Req~221 specifies an ACS-to-DS timeout \texttt{RReq}; the specification's push example belongs to the subsequent OOB challenge interaction, not the lookup-only path~\cite{emvco_core_2311_2023}. Independent issuer notification systems were not instrumented. The client-visible result alone does not establish account validity.}
\label{fig:intro_oracle_summary}\label{fig:offbrowser-boundary}
\end{figure*}
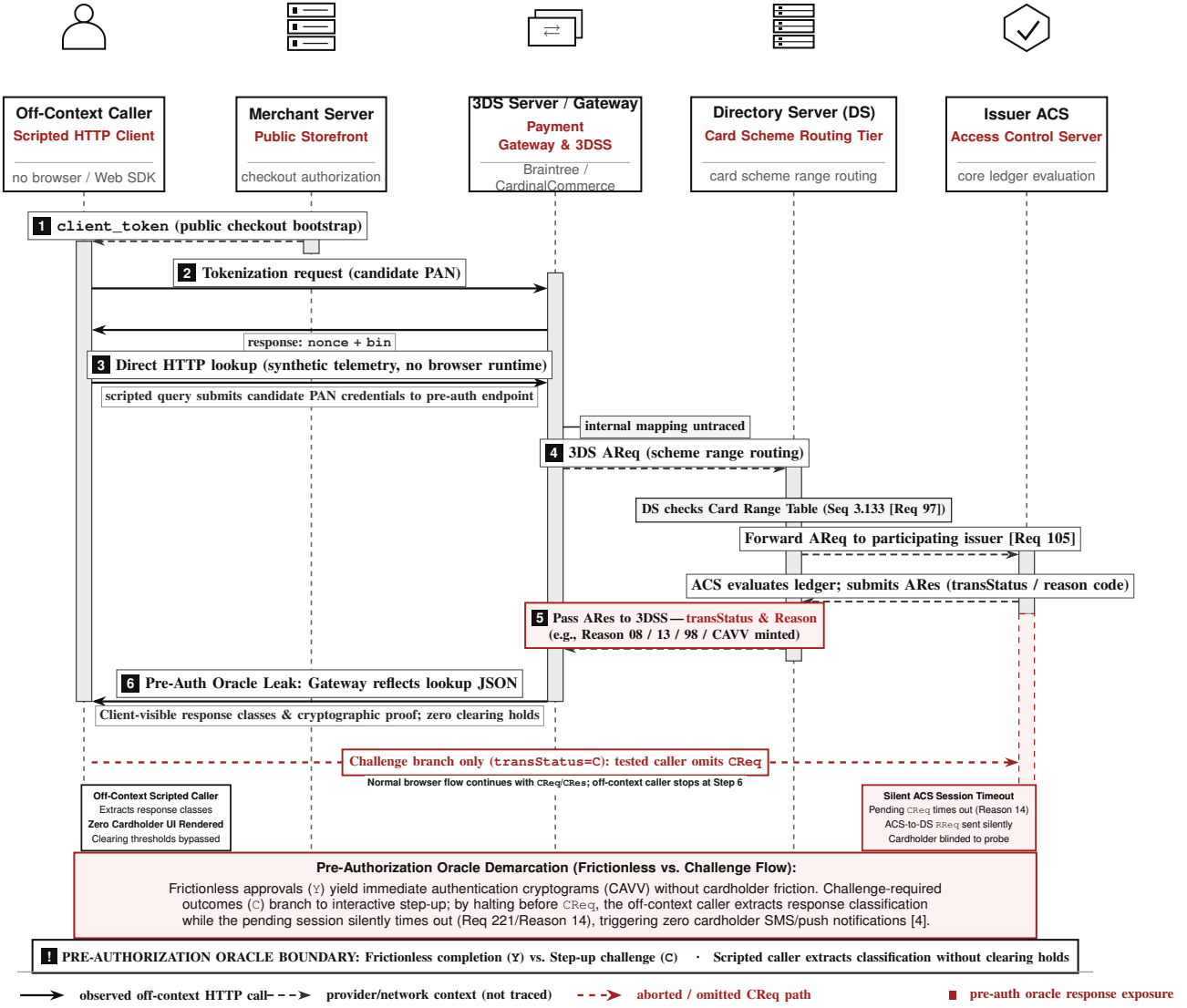

EMV 3-D Secure (3DS) governs cardholder authentication across infrastructure that supports card-based payments at global scale~\cite{emvco_agentic_2025}, resolving the challenge branch prior to financial authorization~\cite{emvco_core_2311_2023}. In this architecture, the 3DS Server routes an Authentication Request (AReq) via the scheme Directory Server (DS) to the issuer's ACS, which evaluates the account and returns an Authentication Response (ARes) back through the DS~\cite{emvco_core_2311_2023}. That DS-mediated routing path is the protocol boundary analyzed in this work and is illustrated in Figure~\ref{fig:intro_oracle_summary}.

The security significance is protocol-level, not scheme-specific: EMV 3DS defines a standardized AReq/ARes exchange routed through a scheme Directory Server, so the protocol boundary analyzed here exists wherever this flow is deployed. Any implementation that exposes input-dependent authentication outcomes at its client-facing integration can instantiate the same pre-authorization classification oracle. Our observations and analyses across major 3DS network implementations (Visa Secure, Mastercard Identity Check, and American Express) show that this is a concrete design risk, not a hypothetical property of one network; however, the evidence tiers differ and do not establish that every network exposes a useful distinction. The attack surface is shared by the protocol; actual leakage and exploitability depend on each implementation's routing, response normalization, and controls.

In published U.S. Patent Application US 2026/0134429 A1, Mastercard describes an authentication architecture operating by ``imposing an authentication phase distinct from an authorization phase''~\cite{mastercard2026abnormaltraffic}. Under this decoupled structure, the patent application describes a threat model where fraudsters rely on authentication responses to evaluate guessed credentials:
\begin{quote}
\centering
\bfseries
``\emph{Where the user is a fraudster, the approval, or other signals, from the first party, may be relied upon to verify the existence of a legitimate account. In this way, a fraudster may initiate a transaction with estimated credentials (e.g., account numbers, etc.), and rely on the signals from the first party, or to the first party, to indicate which estimated credentials are legitimate and linked to actual/legitimate accounts... whereby ones of the estimated or guessed credentials are verified as legitimate or valid, in which case, the credentials are sellable or usable for fraudulent purposes}''~\cite{mastercard2026abnormaltraffic}.
\end{quote}
\noindent To address such automated ``authentication BIN attacks,'' the application proposes a processing-network defense: ``feature data specific to the first party'' is scored at the Directory Server by a ``trained model'' that identifies ``abnormal network traffic,'' and participants are notified through risk scores, message extensions, and in-band reason codes such as the dedicated score 998 and reason code R-98~\cite{mastercard2026abnormaltraffic}. Two features of this disclosure are central to our study. First, the application does not dispute the existence of the oracle signal; it concedes that authentication responses can validate estimated credentials and instead attempts to detect the attack pattern around that signal. Second, its remedy is a trained-model defense, not a response-uniformity mechanism: the trained model relies on source-concentrated velocity features, so it gates \emph{when} an adversary can observe issuer-side responses, not \emph{whether} those responses distinguish accounts. The security consequence is therefore not merely credential validation: the same lookup-stage surface can expose issuer and product metadata, enrollment posture, and account-state signals, enabling pre-clearing profiling of issuing-bank controls and cardholder/account state before any payment attempt. Our independent, passive review of public threat telemetry identified paid 3DS card-testing services in carding channels and related public software~\cite{telegram_3ds_tooling_2026,independent_hosted_payment_abuse_audit_2026}. These services and operational code show that pre-authorization 3DS outcomes are already abused in paid carding-as-a-service (CaaS) workflows to classify estimated credentials and distribute requests across proxy egress at scale; CaaS denotes commercial offerings that bundle card data, tools, and support, not every individual tool~\cite{rapid7_caas_2026}. Our review documents these services and workflows but does not estimate market prevalence, transaction success rates, or operator revenue. These observations motivated this independent, public-interest research: to document the exposure and inform defensive responses without interacting with illicit actors or operating those services or tools. The subsequent disclosure process proved difficult: some reports received no substantive written response, while others were treated as intended behavior within individual component boundaries. This intended-by-design paradox---component-level conformity alongside unresolved cross-party risk---underscores the need for system-level analysis.

The title's network-path framing is motivated in part by a normative Directory Server (DS) stand-in rule, not by a claim that we observed internal routing. EMVCo Core Specification v2.3.1.1 [Req~411] states that when the DS creates an \texttt{ARes} on the ACS's behalf, the ACS Reference Number and ACS Transaction ID are set equal to their DS counterparts~\cite{emvco_core_2311_2023}. If these fields are surfaced in a client-facing integration response, their equality can be a clue consistent with a DS-created response. Our captures are integration JSON, not raw \texttt{ARes} messages or correlated internal traces; therefore, this is conditional provenance evidence, not direct observation of where a request was resolved.

If two identifiers are independently generated UUIDv4 values, the probability of accidental equality is $2^{-122}$ under the UUIDv4 random-bit model. The specification itself permits RFC~4122 identifiers of any compliant version, so that bound is an explicit assumption rather than a normative consequence of Req~104 or Req~411. It addresses coincidence only: it does not establish correct field mapping, independent generation, specification conformance, or the actual internal processing path. Accordingly, observed identifier equality supports a conditional DS-path inference when those assumptions hold; it does not by itself prove ACS non-involvement, candidate account status, or successful authentication.

\paragraph{Adversarial Implications: A Conditional Two-Stage Filtering Model.} The protocol analysis suggests two possible sources of response-class information, subject to the tested integration exposing the relevant fields. At the range level, identifier equality together with an early range-related reason code can support conditional attribution of a DS-generated response under [Req~411]; where the reason may indicate a non-participating or unallocated range, one representative query may be sufficient to prune that block in the analytical model ($Q_{\text{ACS}}=0$ for that DS-terminated request). Equality alone does not establish that every candidate is unallocated, and it says nothing about issuer-side logging or alerts. Within participating ranges, selected cohorts showed different lookup-stage response classes for researcher-owned anchor cards and adjacent comparison PANs. Such differences could help prioritize candidates and may also expose issuer-internal participating sub-ranges within a 6-digit BIN/IIN range that are more likely to correspond to issued accounts. Our observations do not establish the comparison PANs' issuance status or demonstrate recovery of additional live accounts beyond the known researcher-owned anchors. Because the live study deliberately capped the number of adjacent offsets tested per anchor, that bounded negative finding does not imply that a broader unauthorized range scan would fail to encounter other issued accounts. In particular, \texttt{transStatus=C} means that a challenge was requested, not that the account is issued, usable, or successfully authenticated.

We organize this study around four core research questions:
\begin{itemize}
\item \textbf{RQ1 (Protocol Demarcation):} How does the client-visible transaction ID equality rule ($\text{acsTransID} \equiv \text{dsTransID}$) under EMVCo [Req 411] support conditional attribution between Directory Server stand-in/termination and Issuer ACS evaluation?
\item \textbf{RQ2 (Off-Context HTTP Caller):} Can a standalone off-context HTTP caller replay valid merchant-scoped client authorization and submit synthetic device telemetry without executing genuine browser runtime logic, decoupling pre-authorization lookups from authentic user interaction?
\item \textbf{RQ3 (Reason Code Semantics \& Pruning):} How do scheme-specific transaction status reasons (e.g., codes 06 vs. 08 vs. 98) correlate with internal routing boundaries, and how can this inform search tree pruning?
\item \textbf{RQ4 (Scheme-Tier Defenses Under Distributed Egress):} To what extent do published scheme-level trained-model defenses, including the trained model of US 2026/0134429 A1, remain effective when the source-concentrated feature space they score is dispersed across distributed egress, and what does their in-band signaling reveal to an adaptive caller?
\end{itemize}

Our empirical investigation evaluates major international card networks (e.g., Visa, Mastercard) and diverse issuing bank routing behaviors, complemented by a deep analysis of a researcher-controlled merchant integration (PayPal Braintree with Visa-owned CardinalCommerce acting as the 3DS Server [3DSS]). We observe that issuing institutions outsourcing ACS processing to commercial multi-tenant cloud proxies (an observed deployment model across evaluated commercial card cohorts) frequently return challenge-required branches for candidate numbers within participating card ranges, while client-facing observation of Reason 98 illustrates in-band feedback consistent with published scheme-level defense proposals.

\paragraph{Implementation Differences and Mitigation Trade-offs.} Expanding beyond the merchant clearing-layer variations studied by Ali et al.~\cite{ali2017distributed}, our observations point to a trade-off at the pre-authorization authentication layer. Account-dependent checks may produce lookup-stage response-class separation; approving without an account-specific evaluation raises separate questions about the meaning and downstream use of authentication artifacts; and a uniform challenge policy may reduce response-class differences while adding friction and cart abandonment. Our client-facing observations do not establish a universal issuer architecture, account status, CAVV validity, or issuer-side SMS/push behavior. Range-level challenge proxying is therefore a conditional mitigation proposal, not a demonstrated universal fix; its effectiveness depends on consistent responses across the relevant inputs and has usability costs. The four research questions, principal findings, and evidence boundaries are summarized in Table~\ref{tab:intro_summary}.

\begin{table}[t]
\centering
\caption{Authentication BIN attack study: research questions, core findings, and evidence boundaries.}
\label{tab:intro_summary}
\footnotesize
\renewcommand{\arraystretch}{1.15}
\setlength{\tabcolsep}{4pt}
\rowcolors{2}{white}{rowzebra}
\begin{tabularx}{\columnwidth}{|>{\raggedright\arraybackslash}p{0.08\columnwidth}|>{\raggedright\arraybackslash}X|>{\raggedright\arraybackslash}X|}
\hline
\rowcolor{headerbg}
\textbf{Q} & \textbf{Core Result} & \textbf{Evidence Boundary} \\
\hline
\textbf{Q1} & Client-visible identity between $\text{acsTransID}$ and $\text{dsTransID}$ supports conditional attribution of a DS-created response under EMVCo [Req 411]. & Normative protocol invariant in EMVCo Core Spec v2.3.1.1; the $P < 2^{-122}$ bound applies only if the identifiers are independently generated UUIDv4 values. \\
\hline
\textbf{Q2} & A standalone off-context HTTP caller replayed merchant-scoped authorization, submitted synthetic telemetry, and received client-facing 3DS lookup responses. & Bounded live trace; browser-runtime execution was not required. Telemetry provenance and binding were not assessed. \\
\hline
\textbf{Q3} & Scheme-specific reason codes (06 vs. 08 vs. 98) expose routing-tier and account-recognition distinctions, enabling hierarchical search tree pruning ($\mathcal{O}(10^{9-k}) \to \mathcal{O}(1)$). & Evaluated across major payment schemes; verifies the dual-tier drop architecture ([Req 97] vs. [Req 105]). \\
\hline
\textbf{Q4} & The published scheme-level trained-model defense (Mastercard US 2026/0134429 A1) scores source-concentrated velocity features. In a controlled exploratory A/B comparison on the tested path, routing the Cardinal device-data path through distributed residential SOCKS5 egress removed the scheme-tier blocks observed under single-source probing on the identical PAN set, and the in-band R-98 signal itself served as an attacker calibration channel. & Controlled exploratory A/B comparison on identical PANs, alongside client-visible Reason 98 observations; bounded to the tested integration, with proxy-pool size limiting sustainable throughput. \\
\hline
\end{tabularx}
\end{table}

\par\smallskip
\textbf{Contributions and scope.} Because this manuscript is an extended protocol-and-ecosystem study rather than a narrow attack note, its value lies in how several evidence layers fit together. The main contributions are:
\begin{enumerate}
\item \textbf{Lookup-Stage Response-Class Analysis:} We examine the first client-visible integration result associated with the initial \texttt{AReq}/\texttt{ARes} exchange. It is a branch point, not a uniform pre-challenge phase: frictionless authentication may complete there, while \texttt{transStatus=C} requests a later challenge. In selected cohorts, response fields differed between researcher-owned anchor cards and bounded adjacent comparison PANs in the same participating range; this establishes observed response-class separation, not universal account-validity classification or direct observation of issuer-side backend decisions.
\item \textbf{Formalization of Transaction ID Equality \& Component Demarcation:} We formalize the EMVCo v2.3.1.1 [Req~411] rule ($\text{acsTransID} \equiv \text{dsTransID}$) alongside the RFC~4122 UUIDv4 collision bound ($P < 2^{-122}$) as an explicit assumption when those identifiers are independently generated UUIDv4 values. Given correct mapping of returned fields and specification conformance, identifier equality supports conditional attribution of a DS-generated response; it is not by itself proof of account or range status.
\item \textbf{Integration Boundary and Measurement Scope:} We report a tested off-context HTTP lookup path that did not execute the browser SDK and did not submit a financial authorization. The measurements concern client-visible integration output associated with the initial exchange, not raw network messages or a trace of internal DS/ACS processing. They therefore distinguish the tested lookup boundary from later challenge continuation and payment authorization without asserting that all issuer or scheme monitoring is bypassed.
\item \textbf{Conditional Range-Pruning Model:} We formalize the candidate space as a prefix tree and show analytically how a DS-side early-drop response, when paired with a reason code indicating a range boundary, could reduce search over that block from $\mathcal{O}(10^{9-k})$ to $\mathcal{O}(1)$ representative queries ($Q_{\text{ACS}}=0$ for a request terminated at the DS). Observed reason-code and identifier combinations are consistent with different emitting tiers in selected cases, but do not prove the issuance status of comparison PANs or establish a universal account classifier.
\item \textbf{Mitigation Trade-offs:} We discuss response minimization, consistent branch handling, and range-level uniform challenge proxying as possible mitigations. Uniform challenge responses may reduce differences in the lookup-stage class, but are conditional on consistent implementation and trade frictionless completion for additional user interaction. We do not claim that these measures eliminate all enumeration signals or establish issuer-notification behavior.
\end{enumerate}

\textbf{Ethics and Disclosure.} The primary study used researcher-owned cards as anchor controls and a bounded set of adjacent same-range comparison PANs; supporting live checks used controlled cards and comparison inputs. We limited the number of adjacent offsets per anchor, withheld full PANs, and stopped at the pre-authorization lookup boundary; we did not submit payment authorizations or settle transactions. We disclosed the issue through the vendor's official channel. Section~\ref{sec:ethics} details our ethical considerations.

\paragraph{Title terminology.} The fixed title, ``The 3DS Oracle: BIN Attack via the Network,'' names the card-network, Directory Server (DS)-mediated attack path, not a merchant-side 3DS Server as the attack endpoint; it does not claim a completed, network-wide attack. ``3DS'' names the EMV 3-D Secure authentication flow. ``Oracle'' denotes the client-visible result of a DS-mediated pre-authorization lookup: an off-context caller using a merchant-issued client token without executing the browser SDK can compare candidate PANs to infer account-recognition state and any exposed card or product metadata before authorization. The interface supports repeated lookup requests, and distributing them across callers could amplify the signal; our study did not test distributed operation or measure its scale. ``BIN attack'' names the authentication-BIN threat described in Mastercard Patent US~2026/0134429~A1 and the within-range candidate-comparison scenario studied here~\cite{mastercard2026abnormaltraffic}; it does not assert that every candidate was unissued, valid, or recovered for fraud. ``Via the network'' refers specifically to the AReq path from the first party's 3DS Server to the scheme's network DS. The patent places attack detection at that DS: it analyzes AReq feature data, flags suspected authentication BIN attacks, and forwards the flagged request toward the issuer ACS~\cite{mastercard2026abnormaltraffic}. EMV 3DS also permits DS-generated stand-in responses in specified cases~\cite{emvco_core_2311_2023}. Our caller initiated requests through Braintree/Cardinal and observed client-visible gateway JSON; it did not directly query or instrument the network DS or ACS. Thus, the DS is the network actor and protocol boundary at issue, while attribution of individual tested outcomes to DS or ACS processing remains conditional on returned identifiers and reason codes because no correlated internal messages were captured. The off-context caller also supplied synthetic client telemetry without executing a browser SDK. Calling these caller-supplied values \emph{spoofed telemetry} describes how the inputs were supplied; our evidence does not establish whether the ACS accepted, trusted, or used them. The title's ``oracle'' also honors Professor Ross Anderson's ``elf sitting in a black box'' analogy for the random-oracle model and his coauthored 3-D Secure analysis~\cite{anderson2020securityengineering,murdoch2010vbv}.

\paragraph{Evidence path.} The abstract's claims are developed in three stages: the 3DS route and protocol analysis (Sections~\ref{sec:demystified} and~\ref{sec:oracle_flaw}); the live-caller methodology and cohort results (the Methodology and Validation section and Section~\ref{sec:empirical_study}); and the architectural interpretation and proposed defenses (Section~\ref{sec:discussion}).
\section{Related Work}
\label{sec:related_work}
Our study connects card-enumeration research, prior critiques of 3DS authentication and risk signaling, and analyses of trust-boundary failures in multi-party payment integrations.

\paragraph{Authentication-BIN Probing, Distributed Guessing, and Layer Demarcation.} Official guidance on account testing, including Visa enumeration guidance~\cite{visa2021enumeration} and PCI SSC advisories~\cite{pci_ssc_enumeration_2020}, documents how automated scripts and botnets test estimated credentials against e-commerce merchants. Documented enumeration mitigations include CAPTCHA and bot controls, authorization-velocity monitoring, and network-level risk scoring such as Visa Account Attack Intelligence (VAAI)~\cite{visa2021enumeration,visa2026merchantenumeration,visa2024vaai}. Visa separately describes zero-amount Account Verification (AV) as an authorization flow for validating account information and recommends monitoring and blocking excessive failed AV attempts; the verification transaction itself is not a standalone enumeration defense~\cite{visa2026merchantenumeration}. These controls address merchant checkout, transaction authorization, tokenization, or network-processing surfaces and do not necessarily cover the pre-authorization 3DS lookup boundary examined here. In their seminal work on card payment fraud, Ali et al.~\cite{ali2017distributed} uncovered the \emph{distributed guessing attack}, showing how differences in merchant-side field validation let adversaries distribute sub-field guesses across independent merchants and evade per-merchant limits. Later work by Ali and van Moorsel~\cite{ali2019designed} analyzed browser-side telemetry and session replication vulnerabilities in early 3DS~2.0 deployments. However, prior enumeration work remained centered on merchant-side checkout diversity or browser-side spoofing, and treated 3DS primarily as a defensive control rather than as an attack surface in its own right.

\paragraph{Prior Critiques of 3-D Secure.} Murdoch and Anderson's analysis of early, password- and iframe-based 3DS documented weaknesses in user-facing authentication, enrollment, privacy, and incentive design~\cite{murdoch2010vbv}. Corella and Lewison's 2019 paper analyzed the 3DS~2 frictionless flow and its security, privacy, and usability trade-offs, and proposed cryptographic cardholder authentication as an alternative~\cite{corella2019frictionless}. In a separate 2019 technical report, they examined weaknesses in native-app challenge flows that could expose cardholder authentication data to a malicious merchant~\cite{corella2019flaws}. These works address cardholder authentication and app-level challenge integrity, not client-visible lookup-response classification or bounded adjacent-PAN discovery anchored on a known card. We examine that distinct boundary in a tested integration: whether initial lookup responses distinguish adjacent same-range PANs before challenge completion or financial authorization. At the protocol-design level, Anderson and Needham's Principle~3 warns against exposing signing or decryption operations as an opponent-controlled oracle~\cite{anderson1995robustness}. We invoke this as a design analogy, not as a claim that a cryptographic operation oracle is identical to the response oracle studied here.

We show that a distinct attack surface exists deeper in the payment stack: systemic implementation divergence at the 3DS network-routing and authentication layer, specifically between Directory Servers and Access Control Servers. Our attack differs fundamentally from Ali et al. across five technical dimensions:
(1)~\emph{Target Rail}: Ali et al. focus on merchant payment clearing; we exploit the pre-authorization 3DS routing rail (\texttt{AReq}/\texttt{ARes});
(2)~\emph{Adversary Starting State}: prior attacks assume a known 16-digit PAN and guess secondary fields (expiry/CVV); we treat one researcher-owned card as the bounded post-compromise anchor and probe adjacent same-range PAN inputs; candidate issuance was not independently established;
(3)~\emph{Oracle Mechanism}: prior work measures financial approval versus decline; we study authentication-tier reachability and response-class separation at the lookup boundary, using transaction-ID relations and ACS-associated reason codes (including Reasons 05, 08, and 13 in the observed traces) as conditional attribution evidence;
(4)~\emph{Operational Footprint}: clearing-layer guessing incurs gateway fees and can trigger cardholder alerts; our lookup-only test created no cardholder statement holds, never sent the initial \texttt{CReq}, and reached no challenge UI. Accordingly, no SMS, push, or OTP reached test recipients. EMVCo's specified initial-\texttt{CReq} timeout message is an ACS-to-DS \texttt{RReq}, not a cardholder alert~\cite{emvco_core_2311_2023}. These requests occur upstream of payment-clearing velocity monitors;
(5)~\emph{Role of 3DS}: Ali et al. treat 3DS as an obstacle; we show that some tested integrations can expose response-class differences at the initial lookup stage.

This decoupled pre-authorization attack surface is independently corroborated by Conroy et al. (Mastercard) in U.S. Patent Application US 2026/0134429 A1~\cite{mastercard2026abnormaltraffic}, which describes an adversary submitting ``estimated or guessed credentials'' and relying on authentication responses as signals of validity. The patent further proposes Directory Server detection using feature data, velocity signals, and a trained model, with risk communicated through a dedicated score 998 and attack-specific reason codes such as R-98 and R-99~\cite{mastercard2026abnormaltraffic}. Our work contextualizes that mechanism: we observe Reason 98 in live responses, and argue that any deeper oracle signal, where present, arises from architectural differences among issuer-side and scheme-side components rather than from the trained model alone.

\paragraph{3DS Risk Decisions, Protocol Invariants, and Uniform Proxies.} Prior literature examined how browser-derived telemetry affects 3DS risk scoring, including challenge avoidance through session replication and fingerprint spoofing~\cite{ali2018frauddetection,ali2019designed}. Formal analyses of EMV contact, contactless, and 3-D Secure protocols verified authentication guarantees and relay vulnerabilities~\cite{watanabe2025emv3ds,radu2022relay,basin2021emv}. However, existing work generally assumes that evaluating network-stage drop behavior requires internal traces. We show instead that EMVCo Core Specification v2.3.1.1 [Req 411] specifies transaction-ID duplication when the DS creates the \texttt{ARes} on the ACS's behalf; when the compared identifiers are independently generated UUIDv4 values, the resulting client-visible equality has collision probability bounded by $P < 2^{-122}$. As an interim mitigation for range-level pre-auth enumeration, we therefore analyze \emph{Uniform Range-Level Challenge Proxying}, while arguing that permanent closure requires consortium-level EMVCo reform.

\paragraph{Payment-Integration Trust Boundaries.} Widespread empirical studies document logic flaws and confused-deputy vulnerabilities in checkout architectures~\cite{wang2011shop,sun2014ecommerce,pellegrino2014blackbox,andow2019policylint,anwar2024wallet}. Our work examines the boundary where merchant-scoped client authorization meets the multi-party 3DS routing engine. We demonstrate that client-facing endpoints accept synthetic device telemetry from an off-context HTTP caller and release structured protocol artifacts before financial authorization, enabling adversaries to infer account-relevant issuer-side routing, enrollment, and state signals at the authentication boundary without needing payment authorization to complete.

Taken together, this prior work leaves a specific gap that our paper closes: it does not explain how pre-authorization 3DS routing can itself expose a scalable enumeration oracle at the network boundary.
\section{From Integration to Network Routing}
\label{sec:demystified}
The boundary studied here spans two connected but distinct layers: the client-facing gateway/3DS-service integration and the multi-party EMVCo protocol exchange. In the protocol, the 3DS Server selects a Directory Server (DS) using the card's Bank Identification Number (BIN). The DS checks account-range participation and ACS capability, either returning a DS-defined result or routing an Authentication Request (\texttt{AReq}) to the issuer's Access Control Server (ACS), which evaluates the request and returns an Authentication Response (\texttt{ARes}) through the DS~\cite{emvco_core_2311_2023}. While internal ACS decision logic remains opaque to external observers, empirical response outcomes clearly vary across issuing banks. Device collection, risk assessment, authentication, routing, and result presentation may occur in separate administrative components~\cite{braintree2024docs,emvco2022threeds,visa2021verification}. EMVCo specifies the 3DS Server--DS--ACS exchange; vendor guides and SDK documentation describe how a particular integration reaches that exchange and exposes results to a client~\cite{braintree_client_api,braintree_web_github}.

We use the standards and vendor documentation to reconstruct the tested integration and relate its lookup stage to the protocol route. This is an architectural mapping, not an internal network trace: it does not establish that the observed client endpoint contacts the DS directly, that each lookup maps 1:1 to an AReq, or that the DS generated every returned field. The definitions and workflow below distinguish client-facing artifacts from network messages and identify the release boundary analyzed later.

\subsection{Definitions}
\paragraph{Actors.}
At the integration layer, the \emph{Merchant Server} provisions client authorization and submits subsequent payment requests; while standard deployments assume the \emph{Client} is a legitimate cardholder operating a full browser runtime running vendor JavaScript SDKs (e.g., Cardinal Songbird.js), in our evaluation the \emph{Client} is a standalone off-context HTTP caller operating precisely as an automated attacker would by scripting direct API queries without a browser runtime; the \emph{Payment Gateway} authenticates Client API requests; and the \emph{Integrated 3DS Service} exposes device-fingerprinting and lookup interfaces. In this case study, Braintree and CardinalCommerce fill the gateway and integrated-service roles, respectively. These are integration roles, distinct from the EMVCo DS or issuer ACS. The DS routes and range-checks AReq messages; the ACS evaluates authentication and returns an ARes through the DS. Because external observers cannot access internal switchboards, we attribute measurements strictly to the client-facing integration~\cite{emvco_core_2311_2023,braintree2024docs}.

We reserve \emph{3DS Server} for the EMVCo actor that originates authentication requests and use \emph{Integrated 3DS Service} for the client-facing implementation role. Public client behavior does not identify whether these roles are co-deployed, which service operates the 3DS Server, or exactly how a lookup maps to an AReq.

\paragraph{Security-relevant artifacts.}
Seven artifacts are central to this workflow. We record each artifact's provenance because its source constrains what it can establish; the session identifiers and synthetic browser telemetry marked in Table~\ref{tab:artifact_provenance} were generated or supplied by the client, not issued as evidence of a genuine browser session.

\begin{table}[t]
\centering
\footnotesize
\caption{Security-relevant artifacts in the tested gateway/3DS-service integration, observed through scripted requests by a standalone off-context HTTP caller without browser SDK execution. These client-facing artifacts are distinct from network AReq/ARes messages. Artifacts marked $\dagger$ were client-generated session UUIDs or synthetic browser/device values accepted by the integration; cryptographic browser-session binding and downstream use were not assessed.}
\label{tab:artifact_provenance}
\renewcommand{\arraystretch}{1.2}
\setlength{\tabcolsep}{3pt}
\rowcolors{2}{white}{rowzebra}
\begin{tabularx}{\columnwidth}{|>{\raggedright\arraybackslash}p{0.22\columnwidth}|>{\raggedright\arraybackslash}p{0.26\columnwidth}|>{\raggedright\arraybackslash}X|}
\hline
\rowcolor{headerbg}
\textbf{Artifact} & \textbf{Direction} & \textbf{Key Properties} \\
\hline
\texttt{client\_}\-\texttt{token} (JWT) & Server $\to$ Client & Signed token for merchant-scoped Client API authorization; documented lifetime up to 24 hours; reuse observed in the tested HTTP workflow; origin/session binding not established~\cite{braintree_auth_overview,braintree_client_api_authorization} \\
\hline
Payment-method nonce & Braintree Gateway $\to$ Browser & Opaque reference for tokenized payment data; distinct from Client API authorization \\
\hline
Cardinal Auth JWT & Braintree Gateway $\to$ Browser & Server-signed JWT; bootstraps CardinalCommerce session \\
\hline
\texttt{Consumer}\-\texttt{SessionId}$^\dagger$ & Browser $\to$ CardinalCommerce & Client-generated UUID; \emph{not} a server-issued nonce \\
\hline
\texttt{Bin}\-\texttt{SessionId}$^\dagger$ & Browser $\to$ CardinalCommerce & Client-generated UUID; \emph{not} a server-issued nonce \\
\hline
Browser telemetry$^\dagger$ & Client $\to$ CardinalCommerce & Client-generated synthetic browser/device values were included in accepted telemetry-submission requests, followed by lookup responses; cryptographic browser-session binding and downstream use were not assessed \\
\hline
Lookup result & Braintree Gateway $\to$ Browser & Enrollment, ECI, liability-shift indicators, issuer and card metadata; non-empty \texttt{cavv} in one challenge-free observation \\
\hline
\end{tabularx}
\end{table}

The merchant server issues a \texttt{client\_token}; its \texttt{authorizationFingerprint} authorizes Client API requests for that merchant~\cite{braintree_auth_overview,braintree_client_api_authorization}.

In our bounded test, a standalone off-context HTTP caller replayed that authorization for checkout initialization and lookup without running a browser SDK. We use \emph{client-token replay} to mean cross-client reuse in the tested workflow---not token forgery, authorization for another merchant, or acceptance from an arbitrary browser origin. We did not test altered origin headers or binding to one-time use, a PAN, an amount, or a browser session. The documented token lifetime alone is not a vulnerability~\cite{braintree_auth_overview,braintree_client_api}.

The \emph{payment-method nonce} is an opaque, gateway-generated reference for tokenized payment data. The gateway also returns a Cardinal Auth JWT to initialize the 3DS-provider session. Both artifacts serve different purposes from the merchant-scoped Client API authorization discussed above.

The flow also uses client-generated session identifiers and client-submitted device telemetry. Crucially, in the browser channel (\texttt{02-BRW}), there is no cryptographic binding between client-supplied telemetry and the underlying browser runtime or hardware. Our standalone off-context HTTP caller supplied synthetic browser and device attributes directly within the telemetry-submission request, mirroring headless automated tooling. The integration accepted these synthetic payloads, and lookup responses followed. Notably, in our empirical validation of Cohort~D (and across additional calibration findings where specific BIN identifiers are strictly withheld to preserve confidentiality), this off-context caller obtained frictionless authentication (\texttt{status: "authenticate\_successful"}, \texttt{paresStatus: Y}) with a valid cryptographic CAVV and ECI~05, strongly indicating that the tested path lacks cryptographic verification of client telemetry provenance against a genuine browser runtime when queried by an automated off-context caller. Observed client-facing lookup responses also included enrollment status, issuer and card metadata, and liability-shift indicators.

\subsection{Unveiling the Tested Browser Flow}
\label{subsec:browser_flow}
At a high level, public Braintree integration material describes a checkout flow in which the merchant generates a client token, renders a page that collects customer payment information, performs 3DS verification, and then uses the returned nonce or authentication identifier on the server~\cite{braintree2024docs}. In the tested browser-facing path, the intended sequence comprises six distinct stages:

\begin{enumerate}
\setlength{\itemsep}{1.2pt}
\setlength{\parsep}{0pt}
\item \textbf{Client-token bootstrap.} The merchant server generates a \texttt{client\_token} and renders it into the checkout page. This artifact is intentionally delivered to the browser so that the SDK can communicate directly with Braintree client APIs.

\item \textbf{Card-data collection and tokenization.} The browser SDK collects the PAN and related card data and exchanges them for a payment-method nonce. In the browser integration model, the merchant does not directly process raw card data; instead, the browser receives a tokenized representation that can be used in later API calls.

\item \textbf{3DS verification start.} The browser invokes the 3DS component, typically through a call analogous to \texttt{verifyCard}, supplying the nonce, the BIN, and the transaction amount. Braintree Web SDK exposes an \texttt{onLookupComplete(data, next)} callback after the lookup response; calling \texttt{next()} continues the SDK flow, including a challenge when one is requested~\cite{braintree_web_github}.

\item \textbf{Device fingerprinting and telemetry collection.} During the 3DS stage, the browser executes Cardinal's device-fingerprinting logic in the tested path. Public 3DS documentation describes this stage as a browser-mediated method step used to collect device data for risk assessment before challenge execution~\cite{emvco2022threeds,visa2021verification,ali2019designed}. Crucially, under EMVCo Core Specification v2.3.1.1 Section 5.8.1 (Seq 5.95 [Req 255]), the \emph{3DS Method} is explicitly optional for an ACS. When omitted or bypassed, the 3DS Method Completion Indicator is set to \texttt{U} (\emph{Unavailable}), leaving the ACS without prior 3DS-Method interaction for that caller. EMVCo FAQ Q6~\cite{emvco_3ds_faq_2026} further states that verifying SDK authenticity under Req~7 is outside the scope of the Core Specification for the SDK integration model; that FAQ concerns the SDK path rather than our browser-based path, but it still illustrates that client-runtime authenticity checks are not fully specified by the core protocol itself. Separately, EMVCo FAQ Q85~\cite{emvco_3ds_faq_2026} records PCI SSC's position that PCI DSS v4.0.1 Req~6.4.3 (script inventory and integrity) does not apply to 3DS checkout scripts. Consequently, the tested path provides no protocol-level cryptographic attestation of genuine browser runtime execution, and browser telemetry becomes security-relevant state.

\item \textbf{Initial lookup result and branch selection.} Braintree Web SDK exposes the \texttt{onLookupComplete(data, next)} callback after the lookup response and before challenge continuation. This is a branch point, not necessarily a final authentication result: the result may complete frictionlessly, request a challenge, or report another status. The integration may also return a nonce for downstream server-side use~\cite{braintree_3ds_guide,braintree_web_3ds_docs}.
    \item \textbf{Frictionless completion, challenge continuation, or caller stop.} A frictionless result (for example, \texttt{transStatus=Y}) completes through the initial authentication exchange without \texttt{CReq}/\texttt{CRes}. A challenge-required result (\texttt{transStatus=C}) requests a later challenge; it is not proof that authentication succeeded. In the normal browser flow, the SDK can continue from \texttt{onLookupComplete} by calling \texttt{next()}, then handle the challenge exchange. Our off-context caller stopped after the initial lookup result and never sent the first \texttt{CReq}. The tested path did not request decoupled authentication (\texttt{threeDSRequestorDecReqInd} was not set to \texttt{Y} or \texttt{B}), and no \texttt{transStatus=D} result was observed. Only if that initial result requested a challenge would the ACS await \texttt{CReq}; if its timer expires, Req~221 specifies an ACS-to-DS \texttt{RReq}, while OOB push belongs to the later challenge interaction~\cite{emvco_core_2311_2023}. Because that branch never started in our no-browser test, no SMS, push, or OTP was observed. Challenge completion was outside our off-context evaluation.
\end{enumerate}

\subsection{Boundary Interpretation}
\label{subsec:boundary_interpretation}
Each stage above is operationally reasonable on its own. Merchants need a browser-usable token to initialize the SDK. The browser must participate in 3DS because device information and challenge UI are inherently client-facing. The server needs a post-verification nonce to submit a charge or store a payment method. The security difficulty emerges from the composition of these decisions.

Three properties are especially important for the rest of this paper. First, the initial authorization artifact is intentionally public because the browser must possess it. Second, the device-fingerprinting stage is also browser-mediated, which means security-relevant telemetry is collected outside the merchant server's direct control. Third, the initial lookup happens before monetary authorization and can therefore disclose structured state while the system is still deciding how to proceed. If these three elements are not strongly bound to one another, then the flow risks turning an internal decision point into an externally queryable surface.

Our analysis therefore focuses on a narrow but critical part of the overall 3DS pipeline: the pre-authorization boundary where a public client artifact authorizes client-side actions, telemetry is collected to support risk evaluation, and lookup output determines the next protocol branch. Furthermore, because internal ACS decision algorithms, risk scoring models (such as the gradient-boosted decision trees detailed in US Patent Application US 2026/0134429 A1), and velocity thresholds operate as an opaque black box to external callers, researchers cannot observe internal issuer scoring mechanics directly. However, as our measurements demonstrate, empirical authentication decisions clearly vary across issuing banks and ACS deployments.

\subsection{Differential Channel Architecture: Browser (02-BRW) vs.\ Native App (01-APP) vs.\ Split-SDK}
\label{subsec:channel_differences}
To avoid overgeneralizing web-based observations to the entire EMV 3DS standard, it is essential to delineate the cryptographic and architectural boundaries separating the three standardized client execution channels defined in EMVCo Core Specification v2.3.1.1 (Section~2.5 and Section~6.2):

\paragraph{1. The Server-to-Server Backbone (Mutual TLS).}
Communication across the 3DS network core---connecting the 3DS Server, payment scheme Directory Server (DS), and Access Control Server (ACS)---traverses private financial telecommunications backbones. Under EMVCo Section~6.1.2, every inter-server exchange (3DSS $\leftrightarrow$ DS and DS $\leftrightarrow$ ACS) is protected by mutual Transport Layer Security (mTLS) with X.509 digital certificates rooted in payment-system Certificate Authorities (CAs). The vulnerability uncovered in this paper is \emph{not} a cryptographic break of this backbone: an adversary cannot forge scheme CA certificates, inject messages directly into private scheme networks, or alter server-to-server exchanges in transit. Rather, the vulnerability is an \emph{integration-layer reflection leak}: back-end protocol decisions made across this authenticated core are reflected into an insufficiently sanitized, client-visible API before financial authorization occurs.

\paragraph{2. The Native Mobile App Channel (\texttt{01-APP}).}
In native mobile deployments, the 3DS Client is an embedded binary library (the 3DS SDK) running within the merchant's mobile application. The EMVCo specification establishes rigorous cryptographic bindings across this channel:
\begin{itemize}
\setlength{\itemsep}{1.2pt}
\setlength{\parsep}{0pt}
    \item \textbf{SDK Authenticity Verification (Seq~3.9 [Req 7]):} Under Section~6.2.1, the 3DS Server is normatively required to verify the authenticity of the 3DS SDK (validating the SDK Application ID and reference data) before processing the transaction. If verification fails, 3DS processing immediately ends.
    \item \textbf{Encrypted Device Information (Section~6.2.2.1):} The SDK collects native device telemetry and encrypts it directly to the Directory Server (\texttt{encData}) using JSON Web Encryption (JWE) with the DS public key. The DS decrypts this payload (Section~6.2.2.2) and forwards it to the ACS. The merchant server and intermediate gateways never possess the private key required to inspect or tamper with SDK telemetry.
    \item \textbf{Protected Challenge Channel (Seq~3.146):} When an interactive challenge is required, the ACS returns an \texttt{acsSignedContent} payload formatted as a JSON Web Signature (JWS) compact serialization, which the SDK cryptographically validates using the ACS public certificate before rendering the challenge interface. Subsequent challenge messages (\texttt{CReq}/\texttt{CRes}) travel over an SDK-to-ACS channel protected by end-to-end symmetric session keys derived during Diffie-Hellman key exchange.
\end{itemize}

\paragraph{3. The Split-SDK Channel Architecture.}
To accommodate modern merchant architectures where client-side SDK footprints must be minimized, EMVCo introduced the Split-SDK deployment pattern (Section~6.2.2.3):
\begin{itemize}
\setlength{\itemsep}{1.2pt}
\setlength{\parsep}{0pt}
    \item \textbf{Cryptographic Server Signature (Section~6.2.2.3):} A dedicated Split-SDK server collects device information and generates a time-stamped digital signature on transaction data (\texttt{sdkServerSignedContent}) using an X.509 certificate signed by a DS CA, employing PS256 or ES256 algorithms.
    \item \textbf{Directory Server Signature Verification (Section~6.2.2.4):} The Directory Server uses the DS CA public key to cryptographically validate the JWS signature. If validation fails, the DS terminates processing immediately.
\end{itemize}

\paragraph{4. The Browser Channel (\texttt{02-BRW}).}
In stark contrast to \texttt{01-APP} and Split-SDK, the Browser Channel lacks hardware-anchored execution containers and CA-signed client code:
\begin{itemize}
\setlength{\itemsep}{1.2pt}
\setlength{\parsep}{0pt}
    \item \textbf{Unattested Client Telemetry:} The 3DS Client is an off-the-shelf web browser. Telemetry parameters (EMVCo Table~A.1: \texttt{browserIP}, \texttt{browserUserAgent}, \texttt{browserAcceptHeader}, \texttt{browserLanguage}, \texttt{browserColorDepth}, \texttt{browserScreenHeight}, \texttt{browserScreenWidth}, \texttt{browserTZ}, \texttt{browserJavaEnabled}) are plain JSON properties collected via client-side JavaScript. They possess no digital signatures, cryptographic hashes, or hardware-backed attestation tokens.
    \item \textbf{Header-Derived Network Metadata:} Under Table~A.1, \texttt{browserIP} is defined merely as the client IP address observed in HTTP headers by the 3DS Requestor. It is not an attested network token and may reflect caller-controlled or intermediary-controlled network presentation rather than a trustworthy origin signal.
    \item \textbf{Optional 3DS Method (Seq~5.95 [Req 255]):} The 3DS Method is explicitly optional for an ACS. When omitted or blocked by browser privacy controls (e.g., third-party iframe blocking), the ACS evaluates the AReq with zero prior browser interaction or session cookies.
    \item \textbf{Gateway Lookup Reflection:} In the tested Braintree/CardinalCommerce path, the initial client-visible lookup result was returned before challenge continuation. This result is integration output associated with the initial \texttt{AReq}/\texttt{ARes} exchange, not a raw ARes capture. A frictionless result completes without \texttt{CReq}; only a \texttt{transStatus=C} result requests the later challenge exchange.
    \item \textbf{Pre-Challenge Alert Gap:} The caller stopped at the lookup result; for \texttt{transStatus=C}, it never transmitted the initial \texttt{CReq} and therefore never entered the ACS challenge/OOB interaction. In our tests, no cardholder-facing SMS, push, or OTP was observed. Under v2.3.1.1 Seq.~5.47 [Req~221], if the 30-second initial-\texttt{CReq} timer expires without an Error Message from the DS, the ACS must send the DS an \texttt{RReq} with Reason~14 and Challenge Cancellation Indicator~05~\cite{emvco_core_2311_2023}. FAQ Q93 addresses a different event: after a challenge has started, the v2.2 Req~343 \texttt{CRes}-to-Notification-URL behavior on the 10-minute timeout was removed in v2.3.1, and the Requestor should rely on the \texttt{RReq}~\cite{emvco_3ds_faq_2026}. The specification's banking-app push example belongs to the OOB challenge interaction; its communication method is outside 3DS scope~\cite{emvco_core_2311_2023}. Neither timeout path specifies a cardholder SMS/push alert. The patent separately contemplates timely network notifications to the issuer/acquirer ([0015], [0063]), which may be operational emails; investigation or remediation may follow minutes or hours later ([0064])~\cite{mastercard2026abnormaltraffic}. These are network-side notifications, not cardholder alerts. Thus, the tested lookup-only path has no specified 3DS cardholder alert, and we did not observe any SMS/push/OTP on our test cards; independent issuer notification systems were not instrumented.
\end{itemize}
This comparison bounds our empirical findings to the \textbf{Browser Channel (\texttt{02-BRW})}. The tested integration returned lookup-stage fields to an off-context client, but their value as account-related signals varies by response branch and issuer policy; this observation does not by itself establish a universal validity oracle or a result for native mobile SDK environments.

\subsection{The Dual-Tier Drop Architecture}
\label{subsec:dual_tier_drops}
A foundational insight of our study is that EMV 3-D Secure does not evaluate transactions as a monolithic black box. Instead, the protocol enforces a strict structural separation between two distinct evaluation tiers, visually contrasted in Figure~\ref{fig:ds_acs_id_divergence}:

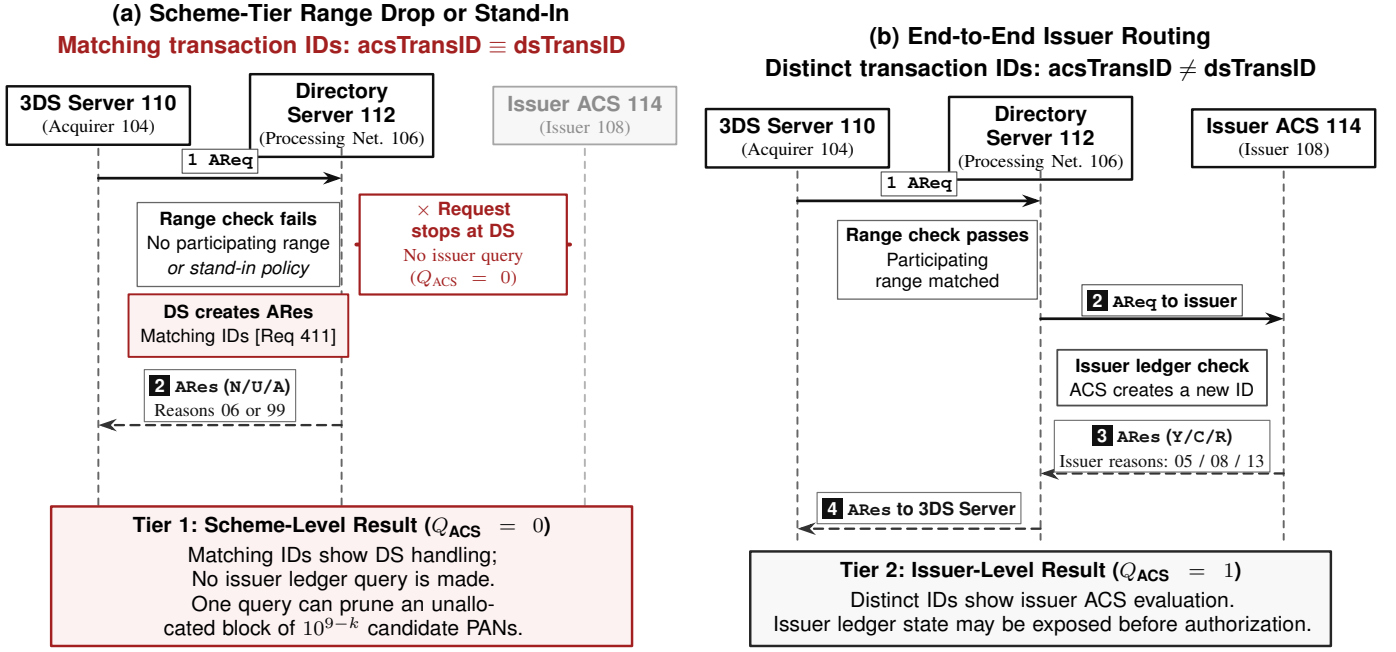
\begin{figure*}[t]
\centering
\begin{minipage}[t]{0.49\textwidth}
\centering
\resizebox{\linewidth}{!}{%
\begin{tikzpicture}[
    font=\small\sffamily,
    line cap=round,
    line join=round,
    >=Stealth,
    actor/.style={rectangle, sharp corners, draw=ink, fill=white,
      line width=0.95pt, minimum height=0.78cm, text width=2.25cm, align=center,
      font=\footnotesize\sffamily\bfseries},
    ghostactor/.style={rectangle, sharp corners, draw=black!35, fill=black!3,
      line width=0.7pt, minimum height=0.78cm, text width=2.25cm, align=center,
      font=\footnotesize\sffamily\bfseries, text=black!45},
    lifeline/.style={draw=black!60, line width=0.75pt, dash pattern=on 2.5pt off 2.5pt},
    ghostlifeline/.style={draw=black!30, line width=0.75pt, dash pattern=on 2.5pt off 2.5pt},
    msg/.style={-{Stealth[length=2.5mm,width=1.7mm]}, line width=1.05pt, draw=ink},
    resp/.style={-{Stealth[length=2.5mm,width=1.7mm]}, line width=0.9pt, draw=black!80, dashed,
                 dash pattern=on 3.5pt off 2.5pt},
    notebox/.style={rectangle, sharp corners, draw=black!60, fill=white,
      line width=0.5pt, align=center, font=\scriptsize\sffamily, inner sep=3.5pt},
    dsnotebox/.style={rectangle, sharp corners, draw=accent, fill=accentbg,
      line width=0.75pt, align=center, font=\scriptsize\sffamily, inner sep=4pt},
    barrier/.style={rectangle, sharp corners, draw=accent, fill=white,
      line width=0.85pt, align=center, font=\scriptsize\sffamily\bfseries, text=accent, inner sep=3pt},
    alertbox/.style={rectangle, sharp corners, draw=accent, fill=accentbg,
      line width=0.95pt, align=center, font=\footnotesize\sffamily, inner sep=4pt},
    steplabel/.style={font=\scriptsize\sffamily\bfseries, fill=white, draw=black!65, line width=0.45pt, sharp corners, inner sep=2.0pt},
    resplabel/.style={font=\scriptsize\sffamily\bfseries, fill=white, draw=black!55, line width=0.45pt, sharp corners, inner sep=2.0pt, text=ink}
  ]
  \def\xS{0.9}
  \def\xD{4.2}
  \def\xA{7.5}
  \def\yLifeBot{-5.15}

  \node[font=\small\sffamily, align=center, text width=8.2cm] at (4.2, 1.45) {
    \textbf{(a) Scheme-Tier Range Drop or Stand-In}\\[2pt]
    \textcolor{accent}{\textbf{Matching transaction IDs: $\mathbf{\text{acsTransID} \equiv \text{dsTransID}}$}}
  };

  \node[actor] at (\xS, 0.3) {3DS Server 110\\[-1pt]\normalfont\scriptsize (Acquirer 104)};
  \node[actor] at (\xD, 0.3) {Directory Server 112\\[-1pt]\normalfont\scriptsize (Processing Net. 106)};
  \node[ghostactor] at (\xA, 0.3) {Issuer ACS 114\\[-1pt]\normalfont\scriptsize (Issuer 108)};

  \draw[lifeline] (\xS, -0.15) -- (\xS, \yLifeBot);
  \draw[lifeline] (\xD, -0.15) -- (\xD, \yLifeBot);
  \draw[ghostlifeline] (\xA, -0.15) -- (\xA, \yLifeBot);

  \draw[msg] (\xS, -0.55) -- (\xD, -0.55);
  \node[steplabel] at ({(\xS+\xD)/2}, -0.32) {\texttt{1 AReq}};

  \node[notebox, text width=2.5cm] at (\xD-1.4, -1.45) {
    \textbf{Range check fails}\\[1pt]
    No participating range\\[1pt]
    \textit{or stand-in policy}
  };

  \draw[line width=1.5pt, accent] (\xD+0.2, -1.45) -- (\xA-0.2, -1.45);
  \node[barrier, text width=2.6cm] at ({(\xD+\xA)/2}, -1.45) {
    $\times$ Request stops at DS\\[1pt]
    \normalfont No issuer query ($Q_{\text{ACS}} = 0$)
  };

  \node[dsnotebox, text width=2.7cm] at (\xD-1.4, -2.55) {
    \textbf{DS creates ARes}\\[1pt]
    Matching IDs [Req 411]
  };

  \draw[resp] (\xD, -3.90) -- (\xS, -3.90);
  \node[resplabel, align=center] at ({(\xS+\xD)/2}, -3.50) {
    \sqnum{2} \texttt{ARes} (\texttt{N/U/A})\\[1pt]
    \normalfont\scriptsize Reasons 06 or 99
  };

  \node[alertbox, text width=7.6cm] at (4.2, -5.95) {
    \textbf{Tier 1: Scheme-Level Result ($Q_{\text{ACS}} = 0$)}\\[2pt]
    Matching IDs show DS handling;\\
    No issuer ledger query is made.\\
    One query can prune an unallocated block of $10^{9-k}$ candidate PANs.
  };

\end{tikzpicture}%
}
\end{minipage}%
\hfill%
\begin{minipage}[t]{0.49\textwidth}
\centering
\resizebox{\linewidth}{!}{%
\begin{tikzpicture}[
    font=\small\sffamily,
    line cap=round,
    line join=round,
    >=Stealth,
    actor/.style={rectangle, sharp corners, draw=ink, fill=white,
      line width=0.95pt, minimum height=0.78cm, text width=2.25cm, align=center,
      font=\footnotesize\sffamily\bfseries},
    lifeline/.style={draw=black!60, line width=0.75pt, dash pattern=on 2.5pt off 2.5pt},
    msg/.style={-{Stealth[length=2.5mm,width=1.7mm]}, line width=1.05pt, draw=ink},
    resp/.style={-{Stealth[length=2.5mm,width=1.7mm]}, line width=0.9pt, draw=black!80, dashed,
                 dash pattern=on 3.5pt off 2.5pt},
    notebox/.style={rectangle, sharp corners, draw=black!60, fill=white,
      line width=0.5pt, align=center, font=\scriptsize\sffamily, inner sep=3.0pt},
    acsnotebox/.style={rectangle, sharp corners, draw=black!70, fill=white,
      line width=0.7pt, align=center, font=\scriptsize\sffamily, inner sep=3.5pt},
    successbox/.style={rectangle, sharp corners, draw=black!85, fill=panel,
      line width=0.95pt, align=center, font=\footnotesize\sffamily, inner sep=4pt},
    steplabel/.style={font=\scriptsize\sffamily\bfseries, fill=white, draw=black!65, line width=0.45pt, sharp corners, inner sep=2.0pt},
    resplabel/.style={font=\scriptsize\sffamily\bfseries, fill=white, draw=black!55, line width=0.45pt, sharp corners, inner sep=2.0pt, text=ink}
  ]
  \def\xS{0.9}
  \def\xD{4.2}
  \def\xA{7.5}
  \def\yLifeBot{-5.15}

  \node[font=\small\sffamily, align=center, text width=8.2cm] at (4.2, 1.45) {
    \textbf{(b) End-to-End Issuer Routing}\\[2pt]
    \textbf{Distinct transaction IDs: $\mathbf{\text{acsTransID} \neq \text{dsTransID}}$}
  };

  \node[actor] at (\xS, 0.3) {3DS Server 110\\[-1pt]\normalfont\scriptsize (Acquirer 104)};
  \node[actor] at (\xD, 0.3) {Directory Server 112\\[-1pt]\normalfont\scriptsize (Processing Net. 106)};
  \node[actor] at (\xA, 0.3) {Issuer ACS 114\\[-1pt]\normalfont\scriptsize (Issuer 108)};

  \draw[lifeline] (\xS, -0.15) -- (\xS, \yLifeBot);
  \draw[lifeline] (\xD, -0.15) -- (\xD, \yLifeBot);
  \draw[lifeline] (\xA, -0.15) -- (\xA, \yLifeBot);

  \draw[msg] (\xS, -0.55) -- (\xD, -0.55);
  \node[steplabel] at ({(\xS+\xD)/2}, -0.32) {\texttt{1 AReq}};

  \node[notebox, text width=2.5cm] at (\xD-1.4, -1.35) {
    \textbf{Range check passes}\\[1pt]
    Participating range matched
  };

  \draw[msg] (\xD, -2.15) -- (\xA, -2.15);
  \node[steplabel] at ({(\xD+\xA)/2}, -1.90) {\sqnum{2} \texttt{AReq} to issuer};

  \node[acsnotebox, text width=2.6cm] at ({(\xD+\xA)/2}, -2.95) {
    \textbf{Issuer ledger check}\\[1pt]
    ACS creates a new ID
  };

  \draw[resp] (\xA, -4.25) -- (\xD, -4.25);
  \node[resplabel, align=center] at ({(\xD+\xA)/2}, -3.88) {
    \sqnum{3} \texttt{ARes} (\texttt{Y/C/R})\\[1pt]
    \normalfont\scriptsize Issuer reasons: 05 / 08 / 13
  };

  \draw[resp] (\xD, -5.00) -- (\xS, -5.00);
  \node[resplabel] at ({(\xS+\xD)/2}, -4.72) {\sqnum{4} \texttt{ARes} to 3DS Server};

  \node[successbox, text width=7.6cm] at (4.2, -5.95) {
    \textbf{Tier 2: Issuer-Level Result ($Q_{\text{ACS}} = 1$)}\\[2pt]
    Distinct IDs show issuer ACS evaluation.\\
    Issuer ledger state may be exposed before authorization.
  };

\end{tikzpicture}%
}
\end{minipage}
\caption{\label{fig:ds_acs_id_divergence}\textbf{Two outcomes across 3DS routing tiers.}
Panel~(a) shows a scheme-level range drop or stand-in: the Directory Server returns an ARes and mirrors the transaction IDs under EMVCo Req~411, yielding the analytical DS-created-response case ($Q_{\text{ACS}}=0$). Panel~(b) shows an issuer-routed request: the ACS generates a distinct ID and returns its result ($Q_{\text{ACS}}=1$). The paths map to Mastercard Patent US 2026/0134429 A1, Figs.~1 and~3; both occur before financial authorization.}
\end{figure*}

\begin{enumerate}
\setlength{\itemsep}{1.2pt}
\setlength{\parsep}{0pt}
\item \textbf{Tier 1: Scheme-Level Pre-Routing Range Verification (Directory Server).} Under EMVCo Core Specification v2.3.1.1 Seq~3.133 [Req 97] and Seq~3.134 [Req 20], the Directory Server inspects the Cardholder Account Number in the \texttt{AReq} against participating range allocation tables (\texttt{cardRangeData} [Req 460]) and verifies whether an Access Control Server is capable of authenticating that range under Seq~3.135 [Req 21]. When the account number falls outside an assigned range, when no ACS is capable, when protocol versions are incompatible (Seq~3.47 [Req 421]), or when scheme stand-in policy applies (Seq~3.225 [Req 412]), the Directory Server may synthesize the \texttt{ARes} on the ACS's behalf. Under Seq~3.155 [Req 411] and EMVCo FAQ Q34~\cite{emvco_3ds_faq_2026}, whenever the DS creates the \texttt{ARes} on the ACS's behalf, the DS is normatively required to set an ACS Transaction ID equal to the DS Transaction ID ($\text{acsTransID} \equiv \text{dsTransID}$) and an ACS Reference Number equal to the DS Reference Number. Thus, observed identifier equality is consistent with a DS-created response rather than an ACS-created response, assuming conformance and correct field mapping. If the identifiers are independently generated UUIDv4 values, accidental equality has probability $P < 2^{-122}$; equality alone does not establish why the DS created the response.
\item \textbf{Tier 2: ACS-Associated Evaluation.} If the PAN passes Tier~1, the Directory Server forwards the \texttt{AReq} to the issuer ACS. Under EMVCo Core Specification v2.3.1.1, the ACS executes Seq~3.142 [Req 104] to generate the ACS Transaction ID prior to later evaluation steps such as Seq~3.143 [Req 105]. The specification requires RFC~4122 canonical format but permits compliant identifiers of any supported version. Accordingly, distinct identifiers ($\text{acsTransID} \neq \text{dsTransID}$) are consistent with an ACS-created response and therefore with traversal beyond the DS-created-response case; they do not by themselves prove a ledger lookup, account status, or a particular backend decision.
\end{enumerate}

\paragraph{Authoritative Issuer-Originating Codes (Reasons 05, 08, and 13) and Scheme Drop Strategies.}
In EMVCo Table~A.1, transaction status reason codes exhibit distinct architectural origins:
\begin{itemize}
\setlength{\itemsep}{1.2pt}
\setlength{\parsep}{0pt}
\item \textbf{Observed ACS-Associated Codes (Reasons 05, 08, and 13):} In our traces, Reason Code $\mathbf{05}$ (\emph{Cardholder account expired}), Reason Code $\mathbf{08}$ (\emph{Card record not found}), and Reason Code $\mathbf{13}$ (\emph{Cardholder not enrolled in service}) appeared with distinct identifiers ($\text{acsTransID} \neq \text{dsTransID}$). That pattern is consistent with ACS-associated processing rather than a DS-created stand-in. Because the captures are client-facing integration JSON rather than correlated internal traces, these codes should be treated as conditional tier-attribution evidence, not as unconditional proof of issuer-side backend interrogation or definitive account-state outcomes.
\item \textbf{Scheme-Level Drop Codes (Reasons 06, 80, 87, 98, and 99):} In contrast to the issuer-originating codes above, scheme Directory Servers terminate processing at the routing boundary using a distinct set of reason codes. Under [Req~97]/[Req~20], when the account number falls outside an assigned participating range or no capable ACS exists, the DS synthesizes the \texttt{ARes} under [Req~411], duplicating transaction identifiers ($\text{acsTransID} \equiv \text{dsTransID}$, $Q_{\text{ACS}} = 0$). Visa signals non-participating range drops with Reason~87 (or Reason~06 for invalid-format boundaries); American Express SafeKey uses Directory Server Reason~80 for network-level range exclusions; Mastercard reserves Reasons~98 and~99 for suspected authentication BIN-attack termination. None of these scheme-tier codes require per-account ledger state, and all are normatively accompanied by identifier equality under [Req~411].
\end{itemize}

\subsection{Authentication--Authorization Separation and BIN Probing}
\label{subsec:auth_auth_separation}
Mastercard's patent disclosure (US 2026/0134429 A1) formalizes this structural decoupling, defining an architecture for ``imposing an authentication phase distinct from an authorization phase''~\cite{mastercard2026abnormaltraffic}. Under standard payment operations, the authentication phase (EMVCo Section~2.6, 3DSS $\rightarrow$ DS $\rightarrow$ ACS) is executed prior to compiling the financial clearing authorization request (Acquirer $\rightarrow$ Processing Network $\rightarrow$ Issuer; patent paragraphs [0024]--[0025]). As the patent explicitly observes in paragraph [0005]: \emph{``Where the user is a fraudster, the approval, or other signals, from the first party, may be relied upon to verify the existence of a legitimate account. In this way, a fraudster may initiate a transaction with estimated credentials... to indicate which estimated credentials are legitimate and linked to actual/legitimate accounts''}~\cite{mastercard2026abnormaltraffic}.

The tested integration returned a lookup-stage result before any financial authorization was submitted. That result could include fields derived from 3DS processing, but it is not a raw ARes capture and does not uniformly reveal whether an estimated credential is recognized. When identifiers and reason codes support conditional attribution, an observer may distinguish a DS-generated response from a response associated with ACS processing under the stated mapping and conformance assumptions. An ACS-associated response still does not, by itself, prove account activity. The lookup path incurred no financial authorization or clearing in our test; issuer-side logging, alerts, and monitoring were not measured. This separates the tested behavior from financial authorization without implying absence of all issuer visibility. PCI DSS and PCI 3DS requirements are discussed separately as control-scope context~\cite{pci_dss_v4_0_1,pci_3ds_core_v1}.
\section{Security Analysis}
\label{sec:oracle_flaw}
This section analyzes the lookup-stage result as a client-visible response boundary before financial authorization. Depending on the branch and the fields exposed by the integration, an observer may see a frictionless outcome, a challenge request, or another status, and may sometimes infer which protocol tier generated a response. We model the assumptions under which the [Req~411] transaction-ID invariant could support range-level pruning; we do not treat a lookup result as a universal account-status test.

\subsection{Adversary Model}
We use a remote web-attacker model for endpoint interaction, consistent with formal analyses of 3DS and adjacent payment protocols~\cite{watanabe2025emv3ds,basin2021emv,cortier2017emvmobile}. Crucially, the adversary operates as a standalone off-context HTTP caller---issuing direct programmatic HTTP requests without instantiating a browser runtime, rendering a graphical interface, or executing client JavaScript SDKs, exactly as automated card-testing attackers script headless queries at scale. The relevant capabilities are those exercised in the tested integration:

\paragraph{Capabilities.}
\begin{itemize}
\setlength{\itemsep}{1.2pt}
\setlength{\parsep}{0pt}
    \item \textbf{Checkout token acquisition.} Braintree documents that a merchant-generated \texttt{client\_token} is delivered to browser-side code to initialize the SDK~\cite{braintree_auth_overview,braintree_3ds_guide}. In our tested case, a visitor without an account-holder login obtained the token from the public card-entry checkout and reused it unchanged from a standalone off-context HTTP caller. Whether a visitor without an account-holder login can reach this checkout state is merchant-specific.
    \item \textbf{Scripted lookup requests.} Using valid authorization for the tested merchant, the caller issued direct HTTP requests without executing the browser SDK or runtime logic~\cite{braintree_client_api,braintree_client_api_authorization}, executing automated card-testing queries directly. The observed sequence does not establish arbitrary-origin acceptance or cross-merchant behavior.
    \item \textbf{Caller-chosen inputs.} The caller submitted Luhn-valid candidate PANs in selected BIN/IIN ranges and caller-constructed browser/device values. The telemetry-submission request was accepted and lookup responses followed; off-context HTTP testing could not establish browser-session cryptographic binding or downstream use. Luhn validity establishes checksum conformance, not that a PAN was issued~\cite{ali2017distributed,pci_ssc_enumeration_2020}.
\end{itemize}

Prior work tested distinct propositions at the 3DS~2.0 risk and transaction layers: Ali et al. varied browser-derived \texttt{machine\_data} in browser-mediated transactions and measured challenge and decline outcomes, while their later study replayed fingerprint data and ACS cookies in purchase flows and demonstrated challenge avoidance~\cite{ali2018frauddetection,ali2019designed}. In our bounded input-variation check, we varied caller-supplied synthetic telemetry but did not compare it with genuine browser telemetry or observe an issuer/ACS risk decision. The client-visible response class remained stable across the tested telemetry values; this does not establish browser-session cryptographic binding, downstream use, or that telemetry had no effect on hidden risk processing.

\paragraph{Constraints.}
\begin{itemize}
\setlength{\itemsep}{1.2pt}
\setlength{\parsep}{0pt}
    \item The attacker does not forge the merchant-issued credential, compromise merchant or gateway infrastructure, or access issuer, DS, or ACS internals.
    \item TLS and cryptographic primitives remain intact; the attacker has no account-holder login or session.
    \item The attacker does not control payment authorization or learn whether an adjacent comparison PAN is issued or active.
\end{itemize}

\paragraph{Goal.}
The threat-model objective is to assess whether client-visible 3DS responses could help distinguish candidate inputs and whether DS-side range responses could support pruning under the specified assumptions. The study does not independently verify that adjacent comparison PANs are unissued, does not demonstrate recovery of live account numbers, and does not measure issuer-side logs or alerts.

\subsection{Security Rules}
Given the flow in Section~\ref{sec:demystified}, we use four design objectives to reason about the lookup boundary. These are analytical recommendations, not requirements attributed to Braintree or EMV 3DS:
\begin{itemize}
\setlength{\itemsep}{1.2pt}
\setlength{\parsep}{0pt}
    \item \textbf{R1: Query-specific authorization.} Scope release of sensitive lookup results to the intended checkout or lookup operation, rather than relying only on merchant-scoped client authorization. This does not assume that the token is documented as browser-session-bound.
    \item \textbf{R2: Telemetry provenance.} Treat client-supplied device data as caller-controlled unless its origin is independently established; receipt or schema acceptance alone does not prove browser execution. This is a deployment-level goal, not a universal EMV 3DS requirement~\cite{vastel2018fpscanner,laperdrix2016beautybeast}.
    \item \textbf{R3: Pre-auth response minimization.} Before authorization, expose the minimum continuation state needed for the client to choose the challenge or challenge-free path, and only other values needed for the next authorized step. Where an authentication artifact is required downstream, expose it only through the intended transaction flow; do not return unrelated issuer or product metadata merely because the network or ACS needs other data.
    \item \textbf{R4: Abuse-resilient release boundary.} Apply rate, anomaly, and cost controls where lookup results are released; do not assume downstream authorization or network-side controls see every client-facing request.
\end{itemize}

These objectives address distinct properties: rate limits constrain query volume but do not provide response confidentiality or establish telemetry provenance. The revised ProVerif trace model follows the observed credential, session-initialization, client-generated-ID, synthetic-telemetry, and lookup sequence. Its queries check credential provenance and response-release reachability, while treating response contents as opaque. It does not derive the measured first-result separation or compare a fixed query against alternative hidden issuer states; those remain empirical observations. Neither analysis establishes candidate issuance, telemetry binding, or a gateway impossibility theorem.

The scenarios below separate observed request/result behavior from unmeasured scale and cost. They do not establish that a specific control was absent or that any hypothetical downstream consequence occurred.

\subsection{Hierarchical Search Tree Pruning via the Equality Invariant}
\label{subsec:tree_pruning}
To understand the adversarial power of the [Req 411] transaction ID equality invariant ($\text{acsTransID} \equiv \text{dsTransID}$), we formalize the search space of payment card numbers as a hierarchical prefix tree $\mathcal{T}_{\text{PAN}}$.

\paragraph{Hierarchical Account Structure} A standard 16-digit Primary Account Number $X \in \mathcal{S}_{\text{PAN}}$ is hierarchically partitioned into four distinct segments:
\begin{equation}
X = B \mathbin{\Vert} R \mathbin{\Vert} A \mathbin{\Vert} L
\end{equation}
where $B \in \{0..9\}^6$ is the Bank Identification Number (BIN) designating the issuing institution and scheme; $R \in \{0..9\}^k$ (typically $k \in \{2, 3\}$) represents the internal issuer sub-allocation block (defining contiguous allocations of $10^4$ or $10^5$ cards); $A \in \{0..9\}^{9-k}$ is the individual account identifier; and $L \in \{0..9\}$ is the Luhn checksum digit.

\paragraph{Algorithmic Complexity Reduction} In a brute-force card enumeration probe across a target 6-digit BIN $B$ and sub-range prefix $R \in \{0..9\}^k$, searching the unallocated block would naively require evaluating candidate accounts across the remaining $9-k$ account digits, incurring $\mathcal{O}(10^{9-k})$ queries (the final digit $L$ being fixed by the Luhn checksum; for an 8-digit BIN, this search space is $\mathcal{O}(10^{7-k})$). If these candidate queries reached the issuing bank, issuer-side anomaly detection or velocity controls would be positioned to observe the sweep.

However, the multi-tier EMV 3DS architecture adds a scheme-routing boundary. Under EMVCo Core Specification v2.3.1.1 Seq~3.133 [Req~97], the Directory Server checks $X$ against its participating card-range data (\texttt{cardRangeData} [Req~460]) before forwarding the AReq to an ACS. This check concerns scheme participation and routing; it does not reveal whether a specific PAN within a participating range is issued or active. If the submitted range is absent from the DS routing data, the DS may terminate processing at the network boundary, producing a DS-generated response such as:
\begin{equation}
\text{ARes} \gets \langle \text{`N'}, \text{Reason} \in \{06, 87\}, \text{acsID} \equiv \text{dsID} \rangle
\end{equation}

If a response's identifier equality and range-related reason code jointly support DS-side termination under [Req~411], the observer may infer that the request was not forwarded to the ACS for that response. This does not show that a PAN inside a participating range is unissued, nor does equality alone identify the reason for DS termination. For a block known to correspond to a DS routing range, the model's representative-query complexity for classifying that routing block is:
\begin{equation}
\mathcal{O}(10^{9-k}) \xrightarrow{\text{[Req 411] Rule}} \mathcal{O}(1)
\end{equation}
Because the Directory Server terminates the transaction at Tier~1 ([Req 97]), no message is dispatched to the issuer ACS:
\begin{equation}
Q_{\text{ACS}} = 0
\end{equation}
While this terminates processing before reaching the issuer's core ledger, we note that gateway-level and Directory Server-level logging may still record the transaction sequence.

\paragraph{Conditional Two-Stage Response Filtering Model}
The analytical model considers two possible sources of lookup-stage response differences: scheme-tier range handling and within-range authentication outcomes. A challenge-required result may reflect policy or account-specific processing, but its presence does not establish account activity; issuer policy, integration behavior, and unobserved processing can all affect the returned class. The model therefore describes potential response filtering, not a measured account-enumeration success rate:
\begin{enumerate}
    \item \textbf{Stage 1: Conditional range pruning:} If identifier equality and a range-related reason code jointly support DS-side termination under [Req~411], a representative response may permit pruning the corresponding block in the analytical model. Identifier divergence is consistent with a response associated with ACS processing under the same assumptions; it does not prove that a candidate is issued or active.
    \item \textbf{Stage 2: Within-range response classes:} In selected cohorts, researcher-owned anchor cards and adjacent comparison PANs produced different lookup-stage classes. Such observed separation could be useful for candidate prioritization if it persists under other inputs and deployments, and it may reveal issuer-internal participating sub-ranges within a 6-digit BIN/IIN range that are more likely to correspond to issued accounts. A \texttt{transStatus=C} result means only that a challenge was requested; it is neither a completed authentication nor a universal account-validity signal. The experiments did not establish the comparison PANs' issuance status or demonstrate recovery of additional live accounts beyond the known researcher-owned anchors. Because the tested adjacent offsets were deliberately capped for ethical reasons, this bounded result does not imply that a broader unauthorized scan of the participating range would fail to encounter other issued accounts.
\end{enumerate}

\paragraph{Formal Protocol Invariants} We formalize four universal protocol invariants that govern component attribution across 3DS execution tiers:
\begin{enumerate}
\item \textbf{The Universal Challenge Invariant:}
\begin{equation}
\begin{aligned}
\forall t: \quad & \text{transStatus}(t) = \text{`C'} \\
& \implies \text{acsTransID}(t) \neq \text{dsTransID}(t)
\end{aligned}
\end{equation}
Because payment scheme Directory Servers do not host cardholder interaction interfaces or challenge screens, challenge generation strictly requires an external Issuer ACS. Under EMVCo core specifications, this invariant holds normatively across evaluated scheme implementations.
\item \textbf{The Issuer Rejection Invariant:}
\begin{equation}
\begin{aligned}
\forall t: \quad & \text{transStatus}(t) = \text{`R'} \\
& \implies \text{acsTransID}(t) \neq \text{dsTransID}(t)
\end{aligned}
\end{equation}
Under EMVCo Table A.1, status \texttt{`R'} is issuer-originating in the protocol framing relevant here. On that basis, a returned \texttt{`R'} strongly indicates ACS-side handling rather than DS stand-in behavior.
\item \textbf{The Semantic Ledger Invariant:}
\begin{equation}
\begin{aligned}
\forall t: \quad & \text{transStatusReason}(t) \in \{05, 08, 13\} \\
& \implies \text{acsTransID}(t) \neq \text{dsTransID}(t) \\
& \quad\quad {}\land Q_{\text{ACS}}(t) = 1
\end{aligned}
\end{equation}
In EMVCo Table A.1, reason codes $\mathbf{05}$ (\emph{Cardholder account expired}), $\mathbf{08}$ (\emph{Card record not found}), and $\mathbf{13}$ (\emph{Cardholder not enrolled in service}) require authoritative state knowledge of individual cardholder records. Payment scheme Directory Servers maintain only prefix routing tables (\texttt{cardRangeData} under Seq 5.88 [Req 460]) and do not track cardholder expiration dates, account-record existence, or cardholder-level service enrollment registries. The observation of $\text{transStatusReason} \in \{05, 08, 13\}$ accompanied by distinct identifiers ($\text{acsTransID} \neq \text{dsTransID}$) therefore strongly supports conditional attribution to issuer-side ACS-associated processing ($Q_{\text{ACS}} = 1$), while stopping short of direct observation of a ledger query.
\item \textbf{The Frictionless Approval Invariant (`Y'):}
\begin{equation}
\begin{aligned}
\forall t: \quad & \text{transStatus}(t) = \text{`Y'} \\
& \implies \text{acsTransID}(t) \neq \text{dsTransID}(t) \\
& \quad\quad {}\land \text{CAVV}(t) \neq \emptyset \land \text{LiabilityShift}(t) = 1
\end{aligned}
\end{equation}
Listing~\ref{lst:oracle_response} shows the redacted frictionless response. For an issuer-approved participating account, the ACS returns \texttt{Y} with authentication cryptograms (CAVV/AEVV) and integration-reported liability indicators. Because Directory Servers cannot generate issuer cryptograms, this client-visible outcome strongly indicates ACS-side evaluation of the authentication result.
\end{enumerate}

\subsection{Abuse Path A: PAN-Dependent Lookup Results}
Repeatedly submitting candidate PANs within a target BIN/IIN range constitutes an authentication BIN attack (formalized by Conroy et al. in Mastercard Patent US 2026/0134429 A1 [0005], [0023]~\cite{mastercard2026abnormaltraffic}). In the tested architecture, an adversary operates across two decoupled tiers: at the merchant integration tier, the caller replays public-checkout merchant-authorized credentials (the client-side \texttt{client\_token}) without backend compromise; at the cardholder and payment tier, the caller operates as an unauthenticated probe submitting ``estimated or guessed credentials'' ([0005], [0013]) without cardholder login, session context, or clearing authorization. Specifically, the caller could:
\begin{enumerate}
    \item Obtain the checkout-delivered \texttt{client\_token} and associated \texttt{authorizationFingerprint}; Braintree documents client authorization validity of up to 24 hours, subject to earlier invalidation~\cite{braintree_auth_overview,braintree_client_api}.
    \item Construct Luhn-valid candidate PANs within a target participating BIN/IIN range and submit them as estimated credentials, without requiring a stolen-card corpus or data-breach feed. Where one known-issued anchor PAN is available, the attacker can further prioritize a bounded set of adjacent same-range candidates. Our live study operationalized this adjacent-anchor comparison with researcher-owned anchor cards and a small number of adjacent offsets per cohort.
    \item Use a standalone off-context HTTP caller to submit 3DS lookup requests with that valid authorization, executing scripted queries directly without browser JavaScript execution. We observed reuse of the merchant-issued \texttt{client\_token} in the tested path; the observation alone does not establish a violated browser-session binding.
    \item The client-facing endpoint returns structured issuer- and authentication-related fields, including enrollment status, ECI, liability-shift indicators, and issuer and product attributes. Under R3, these fields motivate a response-minimization review; we do not assert that the integration violated a documented vendor or 3DS requirement.
\end{enumerate}
The workflow does not require downstream authorization: it assumes a publicly known target BIN/IIN prefix, a valid merchant-scoped client authorization, and attacker-constructed Luhn-valid candidate PANs. In the stronger adjacent-anchor variant that we evaluate live, one known-issued anchor PAN further sharpens within-range prioritization. The reviewed carding-as-a-service (CaaS) artifacts show tools accepting PAN candidates and returning classifications based on the first pre-authorization 3DS result, operationalizing this input-to-signal workflow~\cite{telegram_3ds_tooling_2026,independent_hosted_payment_abuse_audit_2026}. These classifications support triage only; they do not establish candidate issuance or usability. Across evaluated issuer cohorts pairing researcher-owned known-issued anchor cards with adjacent candidate PANs, the anchor card's first-result class consistently separated from comparison candidates in legacy separating architectures, whereas modern cloud proxies (Cohort~G) exhibited a uniform challenge-required class for all inputs. The repeated sessions measure temporal response stability across fresh off-context HTTP queries. Candidate issuance was not independently established, so the results show bounded response-class separation, not a validated card-validity classifier or successful PAN enumeration. They are, however, consistent with an attacker learning issuer-internal participating sub-range structure beneath a 6-digit BIN/IIN range and using that structure to focus further probing on sub-ranges more likely to correspond to issued accounts. Luhn validity establishes checksum conformance only. The returned responses also included issuer/BIN-associated metadata, but we do not claim every field depended only on PAN or BIN. We did not capture an internal Directory Server exchange or identify which component generated the fields~\cite{emvco_core_2311_2023}.

The client-visible output can also include a 3DS Transaction Status Reason, standardized as \texttt{transStatusReason}. While checksum conformance alone does not establish commercial account issuance, the coupling of semantic reason codes with transaction identifier divergence provides robust architectural attribution. Specifically, in Cohort A (Card~A returning Reason~05, \emph{Expired card}) and Cohort C (Card~C returning Reason~13, \emph{Cardholder not enrolled in service}), the responses exhibited distinct transaction identifiers ($\text{acsTransID} \neq \text{dsTransID}$). Because Directory Servers do not store cardholder expiration dates or cardholder 3DS enrollment registries, the combination of semantic reason codes and identifier divergence strongly supports issuer-side ACS-associated processing ($Q_{\text{ACS}}=1$); comparison candidates returning Reason~08 with identifier divergence ($\text{acsTransID} \neq \text{dsTransID}$) are likewise consistent with issuer-side ACS handling after the request traversed the DS range check under [Req~104] and [Req~105], without converting that inference into direct observation of a ledger query. A separate exploratory review of response logs also recorded Reason~98 during repeated within-range Mastercard lookups, consistent with the network-level anomaly detection mechanism disclosed in Mastercard Patent Application US~2026/0134429~A1~\cite{mastercard2026abnormaltraffic}.

The observed input-associated response differences motivate review of what a client can learn before authorization; they do not establish the rate, coverage, or cost of systematic probing, which depend on deployment-specific controls and remain unmeasured~\cite{ali2017distributed,pci_ssc_enumeration_2020}.

\paragraph{Information Returned Per PAN Query and Initial Lookup Result}
The EMV 3DS authentication flow is separate from financial authorization~\cite{mastercard2026abnormaltraffic}. In this paper, the \emph{lookup-stage result} is the first client-visible response returned by the integration's lookup, associated with the initial \texttt{AReq}/\texttt{ARes} exchange. It is not a raw ARes captured from the network. This initial result marks a branch, not one uniform outcome: it may indicate frictionless authentication, request a challenge (\texttt{transStatus=C}), or report another status. A frictionless result completes without a challenge exchange; \texttt{transStatus=C} requests a challenge and is not a final authentication success. The off-context caller stops after receiving the result. Listing~\ref{lst:oracle_response} shows a frictionless observation (Card~D, Visa Consumer Debit; live authentication cryptograms, account nonces, and institution names redacted). The listing is an integration response, not a protocol capture. Rather than returning a boolean, the gateway response includes cryptographic artifacts, transaction identifiers, and issuer-related metadata.
\begin{lstlisting}[language={},basicstyle=\ttfamily\fontsize{7.0pt}{7.4pt}\selectfont,breaklines=true,columns=fullflexible,float=*,escapeinside={(*@}{@*)},caption={Redacted client-visible response from an off-context frictionless lookup (Card D, Visa Consumer Debit). Identifiers are synthetic; sensitive values are redacted.},label={lst:oracle_response}]
{
  "paymentMethod": {
    "type": "CreditCard",
    "nonce": "[REDACTED_PAYMENT_NONCE]",
    "description": "ending in [XX]",
    "consumed": false,
    "threeDSecureInfo": {
      "liabilityShifted": true,
      "liabilityShiftPossible": true,
      "status": "authenticate_successful",
      "enrolled": "Y",
      "cavv": "[REDACTED_28_CHAR_BASE64_CAVV]",
      "xid": null,
      // (*@\textcolor{accent}{\textbf{Distinct transaction IDs: acsTransID $\neq$ dsTransID}}@*)
      "acsTransactionId":
        (*@\textcolor{accent}{\textbf{"a1b2c3d4-e5f6-4a1b-8c2d-3e4f5a6b7c8d"}}@*),
      "dsTransactionId":
        (*@\textcolor{accent}{\textbf{"f8e7d6c5-b4a3-4f8e-9d2c-1b0a9f8e7d6c"}}@*),
      "eciFlag": "05",
      "acsUrl": null,
      "paresStatus": "Y",
      "threeDSecureAuthenticationId": "[REDACTED_AUTH_ID]",
      "threeDSecureServerTransactionId": "[REDACTED_3DSS_UUID]",
      "threeDSecureVersion": "2.2.0",
      "lookup": {
        "transStatus": "Y",
        "transStatusReason": null
      },
      "authentication": {
        "transStatus": "Y",
        "transStatusReason": null
      }
    },
    "details": {
      "bin": "[REDACTED_BIN]",
      "lastTwo": "[XX]",
      "lastFour": "[XXXX]",
      "cardType": "Visa",
      "cardholderName": null,
      "expirationYear": "[REDACTED_EXP_YEAR]",
      "expirationMonth": "[REDACTED_EXP_MONTH]"
    },
    "binData": {
      "prepaid": "No",
      "healthcare": "No",
      "debit": "Yes",
      "durbinRegulated": "No",
      "commercial": "Unknown",
      "payroll": "No",
      "issuingBank": "[REGIONAL FINANCIAL INSTITUTION]",
      "countryOfIssuance": "[REDACTED_COUNTRY]",
      "productId": "F",
      "prepaidReloadable": "Unknown",
      "business": "Unknown",
      "consumer": "Yes",
      "corporate": "Unknown",
      "purchase": "Unknown"
    }
  },
  "threeDSecureInfo": {
    "liabilityShifted": true,
    "liabilityShiftPossible": true
  },
  "lookup": {
    "acsUrl": null,
    "md": "[REDACTED_AUTH_ID]",
    "termUrl": "https://[GATEWAY_HOST]/merchants/[MID]/client_api/v1/payment_methods/[NONCE]/three_d_secure/authenticate?[SIGNED_JWT]",
    "pareq": null,
    "threeDSecureVersion": "2.2.0",
    "transactionId": "[REDACTED_GATEWAY_TXN_ID]",
    "acsRenderingType": null
  }
}
\end{lstlisting}

\paragraph{Anatomy of the Client-Visible Metadata Surface} As demonstrated in Listing~\ref{lst:oracle_response}, a single pre-authorization lookup query returns structured fields across four distinct operational domains:
\begin{enumerate}
    \item \textbf{Client-Facing Authentication Artifacts:} The response releases a 28-character Base64 value (\texttt{cavv}, anonymized in Listing~\ref{lst:oracle_response}) alongside an Electronic Commerce Indicator (\texttt{eciFlag} = \texttt{"05"}). As noted in Table~\ref{tab:client_visible_data_domains}, this field was observed in a challenge-free observation; its cryptographic validity and acceptance were not evaluated in financial clearing. Gateway-reported liability indicators (\texttt{liabilityShifted: true}) reflect integration-level flags rather than proof of legal loss transfer.
    \item \textbf{Multi-Domain Component Identifiers:} The response simultaneously exposes the 3DS Server transaction identifier (\texttt{threeDSecureServerTransactionId}), the Directory Server identifier (\texttt{dsTransactionId}), and the issuer Access Control Server identifier (\texttt{acsTransactionId}). As shown in Listing~\ref{lst:oracle_response}, the synthetic transaction UUIDs are distinct ($\text{acsTransactionId} \neq \text{dsTransactionId}$; sanitized with structure-preserving synthetic identifiers \texttt{a1b2c3d4...} vs.\ \texttt{f8e7d6c5...} to prevent operational log correlation), a pattern consistent with handling beyond the DS-created-response case and with ACS-associated processing, assuming correct field mapping and conformance. Under EMVCo [Req 411], identifier equality ($\text{acsTransactionId} \equiv \text{dsTransactionId}$) is specified when the Directory Server creates the \texttt{ARes} on the ACS's behalf; if the compared identifiers are independently generated UUIDv4 values, accidental equality has probability $P < 2^{-122}$. Thus, client-visible identifier comparison can support conditional routing-tier attribution without internal network vantage points.
    \item \textbf{Enriched Issuer and Product Metadata:} The gateway populates a structured \texttt{binData} object disclosing the issuing institution (anonymized), country of issuance (redacted), card tier (\texttt{productId}: \texttt{"F"}), and funding classification (\texttt{debit: "Yes"}, \texttt{prepaid: "No"}). As noted in Appendix~\ref{app:client_visible_data_domains}, the source of this profile (whether from scheme tables, gateway caches, or external BIN databases) is not verifiable from client-side responses alone.
    \item \textbf{Downstream Flow Credentials \& State Tokens:} The response furnishes gateway-internal authentication identifiers (\texttt{threeDSecureAuthenticationId}), payment-method nonces, session identifiers (\texttt{md}), and an authenticated completion endpoint (\texttt{termUrl}) embedding signed merchant JWT authorization tokens, disclosing internal gateway routing parameters to the client.
\end{enumerate}
The complete field-by-field protocol provenance mapping across EMVCo ARes specifications appears in Appendix~\ref{app:client_visible_data_domains}.

\subsection{Observed Path B: Synthetic Telemetry Submission and Lookup Progression}
In the tested path, an off-context HTTP client submitted synthetic browser/device telemetry without executing the browser runtime. The integration accepted the telemetry-submission request, and subsequent lookups returned responses. This establishes that browser-runtime execution was not required for this sequence; it does not establish cryptographic binding to a genuine browser session, validation as authentic, or downstream use. The observed sequence was:
\begin{enumerate}
    \item The caller obtains the \texttt{cardinal\allowbreak Authentication\allowbreak JWT} through the token-authenticated flow.
    \item The caller initializes a CardinalCommerce session via \texttt{POST /V1/Order/JWT/Init}, then locally generates \texttt{ConsumerSessionId} and \texttt{BinSessionId}; these are client-created UUIDs, not server-issued nonces.
    \item The caller submits synthetically generated browser/device data directly to \nolinkurl{/SaveBrowserData} rather than executing the genuine \texttt{Songbird.js} runtime collection.
    \item The caller issues lookup requests and receives responses.
\end{enumerate}
We varied synthetic telemetry in bounded input checks, but did not compare it with genuine browser telemetry or observe an issuer/ACS risk decision. Crucially, EMV 3DS does not guarantee that fresh browser-collected 3DS Method data will be available on every transaction: use of the 3DS Method depends on card-range configuration, ACS support, and recent prior execution under Section 5.8.1. An off-context caller that omits genuine method execution therefore deprives the ACS of browser-derived context for that transaction. The ACS is consequently left to evaluate the authentication request based on the parameters forwarded in the AReq message. In this experiment, external callers have no visibility into how each proprietary ACS calculates its internal decision, yet empirical outcomes clearly vary across ACS issuing banks. The observed response-class stability under tested values does not establish browser-session cryptographic binding, downstream use, or absence of influence on hidden risk processing.
If a deployment treats client-submitted telemetry as proof of genuine browser execution, this sequence could overstate the assurance telemetry provides. We did not measure downstream risk decisions or cardholder impact.

\subsection{Scenario C: Automated BIN-Probing Beyond the Bounded Test}
The standalone client could repeat lookups for the tested merchant using valid merchant-scoped authorization, so the observed request sequence was scriptable. This does not establish how many requests a production service would accept: we did not enumerate card ranges at scale, measure production rate limits, or test distributed or low-rate strategies. The Mastercard application proposes a network-side classifier using AReq-associated features, including velocity, while CardinalCommerce documents reason~98 as a suspected Mastercard authentication BIN attack label~\cite{mastercard2026abnormaltraffic,cardinal_transreason_2025}. These sources describe detection mechanisms, not their deployment, effectiveness, or coverage of this client-facing lookup. Our exploratory reason~98 observation does not establish that the detector evaluated the request, and we cannot determine whether the other lookups generated DS-visible AReqs. Any broader reconnaissance or evasion claim remains untested~\cite{pci_ssc_enumeration_2020}.

\subsection{Scenario D: Conditional Financial-Cost Pressure}
If a boundary accepts sustained probing, each request incurs provider-side work or a fee, and controls do not interrupt the activity, repeated requests could impose cost or load on a merchant. This is a conditional scenario, not an observed financial-denial-of-service event. The scenario is:
\begin{enumerate}
    \item An attacker repeats the tested lookup sequence using valid merchant-scoped authorization.
    \item If boundary-local controls permit the requests and each lookup consumes billable or scarce resources, the merchant may incur cost or operational load.
    \item Whether this condition can be sustained, what volume is accepted, and whether integration- or network-side controls interrupt it are untested.
\end{enumerate}
This scenario is consistent with broader concern over high-volume pre-transaction abuse~\cite{pci_ssc_enumeration_2020}. Actual exposure depends on pricing, rate limits, and provider-side controls; our study did not inspect billing records or measure production impact~\cite{braintree_fees_2025}. The separate sandbox experiment and the proposed network-side controls are discussed in the Evaluation and Discussion sections, respectively.

\subsection{Combined Risk, Cardholder Privacy, and Issuer Reissuance Costs}
The lookup does not itself execute or authorize a payment. In our off-context HTTP test, we stopped after the lookup result: we did not continue a challenge, enter or submit an OTP, or request payment authorization. 

\paragraph{Frictionless and Challenge Branches: Scope of the ACS Timeout Claim}
EMVCo distinguishes the frictionless flow (Section~2.5.1) from the challenge flow (Section~2.5.2). Both begin with an \texttt{AReq}/\texttt{ARes} exchange, so that exchange should not be called the frictionless phase in every case. A successful frictionless result completes without a challenge. An initial \texttt{transStatus=C} requests the challenge branch; it does not mean that the cardholder completed authentication. In a normal browser flow, the client continues with \texttt{CReq} to the ACS, receives \texttt{CRes}, and obtains the resulting authentication outcome. Braintree documents the \texttt{onLookupComplete(data, next)} callback before challenge continuation~\cite{braintree_3ds_guide,braintree_web_3ds_docs}. EMVCo places banking-app push in the OOB challenge interaction, not the initial lookup; FAQ Q91 discusses what a cardholder should do if that challenge push is not received~\cite{emvco_core_2311_2023,emvco_3ds_faq_2026}.

In our off-context tests, the attacker-like caller stopped after receiving the initial client-visible result and did not construct or post a \texttt{CReq}. The tested path did not request decoupled authentication (\texttt{threeDSRequestorDecReqInd} was not set to \texttt{Y} or \texttt{B}), and no \texttt{transStatus=D} result was observed. For an initial \texttt{transStatus=C}, Req~220 sets a 30-second timer for the first \texttt{CReq}; if it expires without an Error Message from the DS, Req~221 requires the ACS to send the DS an \texttt{RReq} with \texttt{transStatus=N}, Reason~14, and Challenge Cancellation Indicator~05~\cite{emvco_core_2311_2023}. This is protocol-side timeout reporting to the DS, not a cardholder SMS/push alert. FAQ Q93~\cite{emvco_3ds_faq_2026} addresses a different point: after a browser challenge has started and its 10-minute timeout occurs, the v2.2 Req~343 \texttt{CRes}-to-Notification-URL behavior was removed in v2.3.1; the Requestor should rely on the \texttt{RReq}. Because our caller never sent the initial \texttt{CReq}, that later browser-challenge notification is not the event under test. No SMS, push, or OTP reached the test recipients. EMVCo specifies no cardholder SMS/push alert for the lookup-only or missing-initial-\texttt{CReq} path; independently operated issuer notification systems were not instrumented.

\paragraph{Cardholder Confidentiality and Privacy Risk}
The off-context caller can receive lookup-stage authentication and account-related metadata without completing a challenge or initiating financial authorization. The response's meaning depends on the branch and issuer architecture. In selected cohorts, response classes differed between researcher-owned anchor cards and adjacent comparison PANs; this is response-class separation, not independent proof of candidate issuance or account status. Uniform challenge responses did not provide that separation. Because the off-context caller never initiated the later challenge/OOB branch with a \texttt{CReq}, no SMS, push, or OTP reached the test recipients. As demonstrated in Listing~\ref{lst:oracle_response}, observed fields included:
(1)~response classes that, in selected controlled cohorts, were consistent with account-lifecycle differences; candidate issuance was not independently verified;
(2)~cardholder 3DS enrollment posture (e.g., Reason~13 \emph{Not enrolled});
(3)~account lifecycle state (e.g., Reason~05 \emph{Cardholder account expired});
(4)~granular financial attributes, including card product tier (credit, debit, prepaid, commercial), currency, issuing financial institution, and geographical jurisdiction; and
(5)~in one frictionless observation, a non-empty \texttt{cavv} field and liability indicators; cryptographic validity and downstream use were not tested.
These fields create a privacy exposure because an external caller can receive issuer and product metadata without a cardholder-facing session. Response classes may support account-related inferences in selected integrations, but the client-visible output alone does not prove account existence or active status.

\paragraph{Ecosystem Risk: Dual Attacker Screening and Bank Reissuance Costs}
In carding threat models, adversaries may use pre-authorization response differences as a screening signal, but its value depends on the integration and issuer branch policy; a challenge-required result is not universally an account-validity signal. Public threat artifacts and gateway guidance describe automated tools that inspect 3DS responses~\cite{telegram_3ds_tooling_2026,independent_hosted_payment_abuse_audit_2026,stripe_card_testing_2026}. Mastercard Patent Application US~2026/0134429~A1 also describes how authentication responses may be used to evaluate estimated credentials~\cite{mastercard2026abnormaltraffic}. Some reviewed artifacts group frictionless, attempted, and challenge-required outcomes separately. Such grouping describes tool behavior, not protocol semantics: \texttt{transStatus=C} means that a challenge is requested and does not by itself confirm that the candidate corresponds to an issued or usable account.

Operationally, a response-class side channel could be used in two conditional ways:
(1)~\textbf{Prioritizing Candidates from Response Differences:} Where an integration returns different lookup-stage classes for researcher-owned anchor cards and adjacent same-range PANs, an adversary could use that observed separation to prioritize candidates after one-card compromise and to infer issuer-internal participating sub-ranges that are more likely to correspond to issued accounts. Our study did not establish candidate issuance, account balances, or a universal mapping from any single status to account activity.
(2)~\textbf{Screening Compromised Card Records:} A pre-authorization lookup may expose response differences without initiating a payment authorization, but this does not make it a universal or cost-free account check. Directory Server and ACS behavior varies by range and issuer; some results are integration-level signals, and a challenge-required result may leave a session pending if the caller does not send \texttt{CReq}. In our bounded no-browser tests, the caller never entered that later branch and no SMS/push/OTP was observed. EMVCo's specified response to a timed-out initial \texttt{CReq} is an ACS-to-DS \texttt{RReq}; the protocol does not define a cardholder alert at the lookup-only boundary~\cite{emvco_core_2311_2023}. Independent issuer notification systems were not instrumented. The study also did not establish that adjacent comparison PANs were unissued or test downstream fraud. Any operational screening value therefore remains conditional on the observed response class and applicable controls.
When account ranges are harvested by automated scrapers or flagged by scheme-level detection mechanisms (such as Directory Server Reason~98 notifications~\cite{mastercard2026abnormaltraffic}), issuing banks face severe financial and operational exposure. To preempt imminent Card-Not-Present (CNP) fraud and avoid clearing-stage chargebacks on accounts believed to be exposed or at elevated fraud risk, financial institutions are frequently forced to initiate \emph{emergency card reissuance}. Industry studies and banking trade group analyses estimate that reissuing a compromised payment card incurs direct costs between \$5 and \$15+ per card (encompassing physical plastic production, EMV chip personalization, and secure postal delivery)~\cite{pci_ssc_enumeration_2020,aba2015cardreissue,javelin2014cardreissue}. When automated sweeps enumerate wide sub-allocated BIN blocks, reissuance expenses rapidly scale into hundreds of thousands of dollars. Furthermore, this induces severe customer support disruption: in a documented real-world 3DS-rail BIN attack incident, G+D Netcetera reported a call-center queue of roughly 1,000 panicking cardholders overwhelming customer support~\cite{netcetera2025fraudalert}. We present these economic consequences and chaining pathways as risk hypotheses grounded in industry practitioner disclosures and scheme patent filings~\cite{mastercard2026abnormaltraffic,ali2017distributed,anwar2024wallet,yan2024stealingtrust}. Some researcher-owned anchors also fell into overlapping adjacent-card neighborhoods, which underscores a concentration risk: when issued accounts from one issuer are densely packed within the same participating sub-range, compromise of one anchor can place multiple nearby issued accounts inside a small probing window. Accordingly, issuers whose participating sub-ranges appear densely populated and whose lookup-stage response classes remain distinguishable should treat those ranges as a high-priority exposure for monitoring and incident planning; this study did not measure issuer-level account density or establish a universal risk threshold.

\subsection{Security Interpretation}
The measured concern is the release point: a separate client with valid merchant-scoped authorization received structured pre-authorization results, and first-result classes differed across inputs across separating issuer cohorts. This supports review of lookup authorization and response minimization; it does not establish PAN validity, a token-binding flaw, or a network-detector bypass. Integration-side release controls and DS-side detection address different stages. Determining whether they jointly cover a lookup requires correlated end-to-end traces showing which client requests produce AReqs and how each is handled. The security issue, therefore, is not merely that a lookup exists, but that the release boundary exposes structured pre-authorization signals whose semantics remain useful to an automated adversary even when downstream controls are present, including by revealing issuer-internal participating sub-range structure beneath a 6-digit BIN/IIN range.

\section{Methodology and Validation}
We use an evidence-tiered, integration-level methodology to keep the observation point and inference boundary explicit. Our empirical evaluation is hypothesis-driven and anchored in EMVCo 3DS transaction reason codes (\texttt{transStatusReason}, Core Specification Table~A.1). We test whether semantically distinct reason codes generated during Directory Server routing and Access Control Server evaluation are propagated to the client-facing pre-authorization interface in a way that creates an observable account-existence and ledger-separation oracle. We therefore separate five evidence tiers: documented protocol and client flow, one formal authorization-boundary model, one live Braintree/CardinalCommerce path, narrower Stripe and Adyen recurrence checks, and a public-documentation audit. As in prior EMV security work, these tiers remain analytically distinct~\cite{basin2021emv}. Specifications and public documents support claims about documented behavior; the model supports only assumption-bounded reachability; live requests support only the outputs visible at the tested integration boundary. No evidence tier establishes provider-wide behavior, prevalence, or network-detector effectiveness.

\subsection{Case Selection and Threat Model}
We selected one Braintree/CardinalCommerce integration deployed for a researcher-controlled merchant because its public checkout and documented client flow could be reconstructed and subjected to minimum-necessary live validation. This is purposive case selection, not a representative sample of gateways or merchants. The tested caller operates as a standalone off-context HTTP client---mirroring an automated attacker's scripted query engine without a browser runtime---and has no account-holder login but can obtain the merchant-issued \texttt{client\_token} through the public checkout and replay it unchanged. The threat model therefore assumes valid merchant-scoped client authorization and attacker-constructed Luhn-valid candidate PANs, not token forgery, merchant compromise, credential-free access, or a breach-sourced card corpus. Where one known-issued anchor PAN is available, the threat becomes more sample-efficient within a participating range. Empirically, we tested that adjacent-anchor condition across seven researcher-owned anchor-card cohorts rather than implying that all anchors were compromised. We examine whether this ordinary client-facing path exposes distinguishable pre-authorization results, not whether the caller can complete a payment.

The published application informs the threat model but is not a trained-model specification or ground truth for our observations. It describes a processing-network design in which recent first-party, acquirer, and BIN/range-associated feature data are scored to identify suspected authentication BIN attacks, with participants notified through risk scores, reason codes, and related protocol signals~\cite{mastercard2026abnormaltraffic}. But it does not provide a deployable model, evaluated dataset, threshold values, or measured detection performance. Because our client-facing requests were not correlated with internal AReq, Directory Server, or ACS traces, we cannot determine whether that described trained model evaluated them or attribute any observed reason code to it.

\subsection{Operational Checks and Measurements}
We translated R1--R4 into four observable checks. For each check, we distinguish observed request acceptance and client-visible responses from unobserved downstream processing or decision logic.

\begin{itemize}\setlength{\itemsep}{1pt}\setlength{\parskip}{0pt}
    \item \textbf{C1 (R1: query-specific authorization).} We obtained the merchant-issued authorization through the public checkout, replayed it unchanged from an off-context client, and recorded whether the documented request sequence reached lookup and returned a result. This tests cross-client use of valid authorization only; we did not vary \texttt{Origin} or \texttt{Referer}, and do not claim arbitrary-origin acceptance or a violated browser-session guarantee~\cite{braintree2024docs,braintree_client_api}.
    \item \textbf{C2 (R2: telemetry provenance).} Without executing the browser runtime, the off-context client submitted caller-generated telemetry. The telemetry-submission request was accepted and the sequence continued to lookup, but off-context testing could not establish cryptographic browser-session binding or downstream use in an issuer decision. We did not compare genuine and synthetic telemetry or test other flows~\cite{braintree_web_github,vastel2018fpscanner,laperdrix2020survey}.
    \item \textbf{C3 (R3: pre-authorization response minimization).} We tested whether standard scheme reason codes (e.g., Reason~08 \emph{No card record}, Reason~05 \emph{Cardholder account expired}, and Reason~13 \emph{Not enrolled}) were propagated to the client-facing pre-authorization interface in a way that separated researcher-owned known-issued anchor cards from bounded adjacent comparison PANs in the same participating range. Comparison PANs were constructed as adjacent, Luhn-valid offsets around each anchor card. Each input was submitted across multiple fresh 3DS sessions under the off-context client; for each session, we recorded the initial lookup-stage response class, challenge requirement, and returned fields, stopping before challenge continuation or payment authorization. Candidate issuance status was not independently established; the comparison therefore tests whether the 3DS rail behaves as a response-separation oracle for adjacent-card probing, not a production classifier with known accuracy.
    \item \textbf{C4 (R4: abuse-resilient release boundary).} We recorded completion or client-visible results during the bounded primary trials but did not seek a production blocking threshold or stress the live service. In a separate Braintree Sandbox run, the script sustained 0.75 requests per second without a \texttt{429 Too Many Requests} response. That result describes only the sandbox run, not production throughput, cost, or network-side controls~\cite{pci_ssc_enumeration_2020,cardinal_transreason_2025}.
\end{itemize}

The response class is descriptive: it denotes the first client-visible enrollment/authentication result, with challenge-required outcomes distinguished as shown in Table~\ref{tab:cohort_result}. We compare each candidate only with the control in its selected cohort; a cohort counts as separating only when the control's first-result class differs from comparison candidates. The cohort is the unit for this architectural comparison. Multiple sessions per input assess within-input temporal stability. We report qualitative cohort patterns; the selected cohorts demonstrate structural protocol-level response separation without estimating global cardholder prevalence or cross-network classifier accuracy.

\subsection{Cross-Family Checks and Documentation Audit}
Separate from the repeated Braintree study, researcher-written headless clients exercised researcher-controlled test integrations using Stripe and Adyen (under their respective PaymentIntent/SetupIntent and standalone 3DS workflows) using controlled cards and comparison inputs. We recorded only the first requestor-visible result, stopping before challenge continuation or payment authorization. The Stripe observations included a 3DS-related \texttt{requires\_action}/\texttt{next\_action} state; the Adyen path returned \texttt{binLookup} information and a \texttt{ChallengeShopper} result~\cite{stripe_3ds_auth_flow_2026,adyen_result_codes_2026,adyen_binlookup_api_2026}. These checks had fewer inputs and repetitions than the primary study; they were not a matched cohort experiment, and we do not infer exact-account recognition, issuance ground truth, synthetic-telemetry acceptance, or behavior in other configurations. Per-provider input and repetition counts are not reported here, so we treat these as qualitative recurrence checks. We do not pool their outcomes with the primary Braintree cohort observations or report a classifier-accuracy estimate for them. Control of the test cards is not represented as authorization from the providers, networks, or issuers.

We separately exercised official Stripe and Adyen test-mode branches to confirm documented output placement. Stripe test mode reached three requestor-visible intent classes, while Adyen test mode reached its standalone \texttt{binLookup}/\texttt{ChallengeShopper} path. Because these environments prescribe simulated outcomes, these checks establish branch reachability only, not production issuer decisions or account recognition~\cite{stripe_test_3ds_2026,adyen_test_3ds_2026}.

For the documentation audit, we froze the core review on September~26, 2026. Two researchers independently coded 24 materially distinct 3DS~2.x integration paths against official public technical documentation or source, resolving disagreements by consensus. Eligible paths exposed enough detail to identify transaction initiation, telemetry collection, branching, continuation, callbacks, or lookup-result handling; marketing-only materials, legacy 3DS~1 documents, and architectures without a public Requestor-facing interface were excluded. We coded protocol phase, first observer, client continuation control, operation--credential binding, collection linkage, and any explicit outcome-independent pre-admission requirement. A separately reported 16-path extension expands document coverage without changing the frozen core corpus, yielding 40 path descriptions. The unit is an integration path, not a confirmed deployment or vulnerability. Documentation cannot establish private controls, live reachability, or deployment prevalence.

\subsection{Formal Validation}
We use ProVerif to analyze one explicit client-facing lookup-policy abstraction under the Dolev--Yao adversary defined in Section~\ref{sec:oracle_flaw}. The model represents valid merchant-scoped authorization and a backend that returns a BIN-indexed profile, so its queries address reachability and event correspondence at that boundary only. It does not model the application's proposed classifier, 3DS Server behavior, AReq/ARes processing, Directory Server routing, ACS decisions, account-level PAN validity, telemetry provenance, rate limits, or distributed evasion. The formal result is therefore complementary to, not predictive of, the live experiment~\cite{basin2021emv}.

\subsection{Scope, Supplementary Evidence, and Ethics}
The primary live validation used a tightly scoped client-facing path and researcher-owned anchor cards. It did not involve payment authorization or settlement, bulk scanning, or sustained production-rate replay. Separate Stripe and Adyen checks were narrower and are reported independently; neither they nor the documentation audit enlarge the primary response-class cohort. We did not inspect merchant billing records, internal gateway logs, or 3DS Server, DS, or ACS telemetry.
\begin{itemize}
    \item \textbf{Exploratory response-log review.} The reason-code observations reported in Section~\ref{sec:empirical_study} are separate from the primary controlled cohort evaluations. Exact supplemental volume and row-level records are withheld; we report presence or absence only, not rates or comparative detection effectiveness. Without correlated network traces, these records cannot identify which component assigned a reason or whether a detector evaluated the requests.
    \item \textbf{Harm minimization.} We limited live requests to small adjacent-card comparisons and boundary checks, stopped the Stripe and Adyen checks before challenge continuation, avoided payment authorization, settlement, and large-scale enumeration, and initiated responsible disclosure. The primary experiment used only a bounded number of adjacent Luhn-valid offsets around researcher-owned anchor cards, and the study does not measure production cost, throughput, or operational impact~\cite{menlo2012}.
\end{itemize}
These design choices bound inference to the tested integration and the reported client-visible outputs. Taken together, the methodology is designed to answer one bounded question well: what a client-facing 3DS integration visibly releases before authorization, and what that release does or does not justify inferring.

\section{Evaluation}
\label{sec:empirical_study}

We evaluate three evidence layers separately: modeled reachability, live client-facing behavior, and supplementary provider/documentation context. In the model, a caller with valid merchant-scoped authorization can reach an abstract BIN-indexed profile under the explicitly modeled release policy. In the primary live case, a caller without an account-holder login reused checkout-issued authorization from a standalone off-context HTTP client (operating precisely as an automated attacker would do by issuing direct scripted calls without a browser runtime); across representative issuer cohorts, the anchor card's first-result class consistently separated from that of adjacent same-range, Luhn-valid comparison PANs (while modern cloud proxies exhibited uniform challenge masking, as in Cohort~G). These result classes were stable across multiple fresh off-context sessions per input. Following payment-security \emph{existence-proof} methodology~\cite{wang2011shop,basin2021emv,sun2014ecommerce}, this demonstrates bounded response separation across diverse issuer deployments in the tested integration; it does not establish that any candidate was issued or support estimates of classifier accuracy, prevalence, or production-scale throughput~\cite{cremers2017comprehensive,fett2016comprehensive}. The network-level DS/ACS flow provides protocol context only; we did not capture it, and the supplementary evidence is not pooled with the primary study.

\subsection{Formal Reachability at the Integration Boundary}
We model the client-facing lookup boundary using ProVerif, a symbolic protocol verifier~\cite{blanchet2001efficient}. The model (Appendix~\ref{app:proverif}) encodes one candidate policy: a lookup service releases metadata to a caller presenting valid client authorization, with no additional origin, session, or transaction check. Its reachability query asks whether the attacker can obtain \texttt{bin\_profile(sampled\_bin)}; that query is satisfied by the modeled process, in which the backend returns the profile after the gateway accepts the merchant-scoped token. This is assumption-bounded reachability, not an observation about the deployed service. The model has no source-identity, proxy-rotation, distributed-caller, telemetry-provenance, rate-limit, or detector component, so it cannot establish whether those mechanisms are feasible or evade a control. It does not simulate the 3DS Server, network DS, AReq/ARes, or ACS; it proves neither credential-free access nor anything about DS visibility or detection.

\subsection{Live Client-Facing Integration Validation}
We performed bounded live validation on a researcher-controlled merchant deployment across diverse issuer ACS cohorts and card brands. Researcher-owned known-issued anchor cards were systematically evaluated against adjacent same-range Luhn-valid comparison PANs across multiple fresh 3DS sessions under a standalone off-context HTTP caller. The merchant's public checkout exposed merchant-scoped client authorization without requiring an account-holder login; we did not obtain or use credentials for other merchants. All live requests, including telemetry and lookup, used a standalone off-context HTTP client (operating just like an automated attacker would do without executing browser runtime logic); no browser, browser-automation framework, or Web SDK was executed. Thus, browser execution was not required for this tested request sequence. We did not vary \texttt{Origin} or \texttt{Referer}, and therefore do not claim arbitrary-origin acceptance. We also did not capture or inspect the internal 3DS Server--DS--ACS exchange, network logs, or detector decisions. We made no payment authorization or settlement attempt and limited requests to the bounded cohort comparisons~\cite{menlo2012}.

\begin{table*}[t]
\centering
\footnotesize
\caption{Primary study: seven researcher-owned known-issued anchor cards across Visa and Mastercard, labeled A--G, each in a distinct ACS/card-range cohort. Each anchor card was compared with three adjacent, Luhn-valid same-range PANs. The 28 unique inputs each had five fresh sessions (140 observations); repetitions assess within-input stability.}
\label{tab:cohort_result}
\renewcommand{\arraystretch}{1.15}
\setlength{\tabcolsep}{4pt}
\rowcolors{2}{white}{rowzebra}
\begin{tabularx}{\textwidth}{|>{\centering\arraybackslash}p{0.05\textwidth}|>{\centering\arraybackslash}p{0.09\textwidth}|>{\raggedright\arraybackslash}p{0.34\textwidth}|>{\raggedright\arraybackslash}p{0.34\textwidth}|>{\centering\arraybackslash}X|}
\hline
\rowcolor{headerbg}
\textbf{Cohort} & \textbf{Network} & \textbf{Anchor-card result (Issued Control)} & \textbf{Comparison PANs (Adjacent Candidates)} & \textbf{Result} \\
\hline
A & Visa & \texttt{05}: expired card ($\text{acsTransID} \neq \text{dsTransID}$) & \texttt{08}: no account ($\text{acsTransID} \neq \text{dsTransID}$) & Separating \\
\hline
B & Mastercard & \path{challenge_required}, ACS URL ($\text{acsTransID} \neq \text{dsTransID}$) & \texttt{08}: no account ($\text{acsTransID} \neq \text{dsTransID}$) & Separating \\
\hline
C & Visa & \texttt{13}: not enrolled ($\text{acsTransID} \neq \text{dsTransID}$) & \texttt{08}: no account ($\text{acsTransID} \neq \text{dsTransID}$) & Separating \\
\hline
D & Visa & \path{authenticate_successful}, CAVV ($\text{acsTransID} \neq \text{dsTransID}$) & \texttt{08}: no account ($\text{acsTransID} \neq \text{dsTransID}$) & Separating \\
\hline
E & Mastercard & \path{authenticate_attempt_successful}, CAVV ($\mathbf{\text{acsTransID} \equiv \text{dsTransID}}$) & \texttt{08}: no account ($\text{acsTransID} \neq \text{dsTransID}$) & Separating \\
\hline
F & Visa & \path{challenge_required}, ACS URL ($\text{acsTransID} \neq \text{dsTransID}$) & \texttt{08}: no account ($\text{acsTransID} \neq \text{dsTransID}$) & Separating \\
\hline
G & Visa & \path{challenge_required}, ACS URL ($\text{acsTransID} \neq \text{dsTransID}$) & Challenge, ACS URL ($\text{acsTransID} \neq \text{dsTransID}$) & Uniform \\
\hline
\end{tabularx}
\end{table*}

\begin{table}[t]
\centering
\footnotesize
\caption{Aggregate results from the expanded seven-control integration validation. Each PAN input retained its response class across five fresh sessions. The repeated observations measure response stability, not independent PAN samples; candidates were not independently labeled for issuance. These are client-facing integration outputs, not directly captured DS or ACS responses.}
\label{tab:empirical_results}
\setlength{\aboverulesep}{0pt}
\setlength{\belowrulesep}{0pt}
\renewcommand{\arraystretch}{1.15}
\rowcolors{2}{white}{rowzebra}
\begin{tabularx}{\columnwidth}{@{}Xccc@{}}
\toprule
\rowcolor{headerbg}
\textbf{Observed outcome} & \textbf{Cohorts} & \textbf{PAN inputs} & \textbf{Sessions} \\
\midrule
Known-issued anchor card differed from all three adjacent same-range candidates & 6 & 24 & 120 \\
Anchor card and all three adjacent candidates returned a uniform challenge-required class & 1 & 4 & 20 \\
\bottomrule
\end{tabularx}
\end{table}

\paragraph{Dataset and Observations} Table~\ref{tab:cohort_result} maps the researcher-owned anchor cards to their first-result classes across seven distinct ACS/card-range cohorts spanning evaluated Visa and Mastercard cases; Table~\ref{tab:empirical_results} aggregates the evaluated cohorts. Each evaluated input returned a stable class across multiple fresh sessions under the standalone off-context HTTP client. Most cohorts separated the known-issued anchor card from all adjacent same-range candidates; a uniform challenge-required class was also observed across all inputs in Cohort~G. For Cohort~A (Reason~05, \emph{Expired card}) and Cohort~C (Reason~13, \emph{Cardholder not enrolled in service}), both anchor cards exhibited distinct transaction identifiers ($\text{acsTransID} \neq \text{dsTransID}$). These outcomes are consistent with ACS-associated processing rather than a DS-created stand-in, supporting conditional tier attribution under EMVCo [Req~104] and [Req~105]. In Cohort~E, identical identifiers ($\text{acsTransID} \equiv \text{dsTransID}$) are consistent with a Directory Server stand-in attempt pattern under [Req~411], where the scheme creates the authentication response on the issuer's behalf. For comparison candidates returning Reason~08, distinct transaction identifiers ($\text{acsTransID} \neq \text{dsTransID}$) are consistent with ACS-associated negative responses after the adjacent comparison PANs passed the Directory Server range check under [Req~97]. These fields support conditional tier attribution, not direct observation of a ledger query or definitive proof of no-account-record status. Candidate issuance was not independently established, so these results show bounded response-class separation for adjacent-card probing, not a card-validity classifier. They are also consistent with sparse issuer sub-allocation beneath a participating 6-digit BIN/IIN range, where only some next-two-digit sub-ranges appear more likely than others to correspond to issued accounts. Full PANs are withheld; only cohort-level results are reported.

\paragraph{Input-Variation Check} Across five input-variation runs, we varied transaction amount, caller-supplied synthetic browser/device telemetry, and submitted expiration/CVV2 values; the initial baseline used the researcher-owned control card's actual expiration and CVV2. The first-result response class remained unchanged across the tested settings. This is a bounded observation about the client-visible class, not the complete response or an internal risk decision: without correlated AReq/ARes or ACS records, we cannot determine whether a component ignored, normalized, or used any varied field in downstream processing.

\paragraph{Returned Fields and Fraud Weaponization} Client-visible responses included enrollment and authentication-related statuses, issuer and card-product metadata, ECI, and liability-shift indicators. Under frictionless authentication (Cohort~D) and stand-in attempts (Cohort~E), the pre-authorization lookup returned \texttt{authenticate\_successful} (\texttt{paresStatus: Y}, ECI~05) and \texttt{authenticate\_attempt\_successful} (\texttt{paresStatus: A}, ECI~01) with non-empty \texttt{cavv} cryptograms without interactive cardholder friction. For an attacker, these outcomes are especially valuable not because they prove downstream payment success, but because they arrive before challenge completion or financial authorization and therefore support early candidate triage. In the threat model formalized in Mastercard Patent US 2026/0134429 A1 ([0013], [0036]), obtaining an authentication success or proof before financial authorization is treated as evidence that an estimated credential corresponds to a legitimate account record, rendering the credential sellable or weaponizable in downstream card-not-present fraud without triggering clearing-stage decline thresholds. These are integration outputs: we did not trace their provenance, independently validate the CAVV, or test downstream authorization. The observation establishes that this integration returned success-labeled authentication fields without browser execution, not verified end-to-end authentication or downstream acceptance. Issuer identifiers and sensitive values are redacted; the fields' utility for downstream abuse is unmeasured~\cite{ali2017distributed,anwar2024wallet}.

\paragraph{Supplementary Cross-Network Response-Code Observation} This exploratory client-visible log review is separate from the primary controlled cohort evaluations. Reason~98 appeared during repeated within-range Mastercard lookups, but was not observed in the distinct Mastercard ACS/range cohorts examined under SOCKS5 ingress. In the reviewed Visa-prefix records, same-range responses consistently carried reason~08, rendered as ``No Account'' in our logs (``No card record'' in CardinalCommerce's mapping)~\cite{cardinal_transreason_values}. The raw Mastercard records label reason~98 ``reserved,'' while CardinalCommerce's changelog gives it a Mastercard-specific suspected-attack label (Table~\ref{tab:network_reason_observation})~\cite{cardinal_transreason_2025,cardinal_transreason_values}. The consistent Visa response class may offer a more straightforward candidate-filtering signal in these observations; it does not establish PAN validity or reliable enumeration. Exact volumes, per-request timestamps, and correlated 3DS Server/DS/ACS traces are unavailable, so we report only the observed response patterns. The records do not identify the source or trigger of either reason, establish comparative detection effectiveness, or show detector activity or the absence of other controls.

\paragraph{Validation Cost and Financial Harm Minimization}
An important ethical constraint in payment-system research is avoiding unnecessary third-party impact. Our live validation was strictly restricted to researcher-controlled anchor cards and bounded adjacent candidate PANs across the evaluated cohorts; it did not include payment authorization or settlement. We did not observe a merchant-side charge, but did not inspect billing records and therefore make no claim that upstream processing was cost-free. The study does not measure production-scale cost or a financial denial-of-service event~\cite{menlo2012}.

\begin{table}[b]
\centering
\footnotesize
\caption{Exploratory client-visible reason-code observation. Exact supplemental probe volume is withheld; entries are not detection-rate estimates.}
\label{tab:network_reason_observation}
\setlength{\aboverulesep}{0pt}
\setlength{\belowrulesep}{0pt}
\renewcommand{\arraystretch}{1.18}
\rowcolors{2}{white}{rowzebra}
\begin{tabularx}{\columnwidth}{@{}p{0.32\columnwidth} p{0.22\columnwidth} X@{}}
\toprule
\rowcolor{headerbg}
\textbf{Input Cohort} & \textbf{Reason~98} & \textbf{Observed Pattern} \\
\midrule
Mastercard, repeated same-range lookups & Observed & Exploratory logs include Reason 98; in-band feedback consistent with US 2026/0134429 A1. \\
Mastercard probes, distinct ACS cohorts & Not observed & In the examined distinct-ACS Mastercard cohorts under SOCKS5 ingress, Reason 98 was not observed; this exploratory non-observation does not establish absence of detection or other controls. \\
Visa, same-range probes & Not observed & All reviewed captures carried Reason 08; does not establish absence of Visa controls. \\
\bottomrule
\end{tabularx}
\end{table}

\subsection{Supplementary Cross-Family and Documentation Findings}
Narrow supplementary checks conducted on researcher-controlled test integrations using Stripe and Adyen returned non-uniform first-result classes in their tested configurations. Stripe exposed a 3DS-related \texttt{requires\_action}/\texttt{next\_action} state around a server-created intent before client continuation; Adyen returned \texttt{binLookup} output and \texttt{ChallengeShopper}, a challenge-required result. These are bounded observations at distinct integration boundaries, not evidence that either provider accepts arbitrary PAN queries, shares Braintree's authorization model, or exposes an exact-account classifier. No challenge was completed and no payment authorization was attempted~\cite{stripe_3ds_auth_flow_2026,adyen_result_codes_2026,adyen_binlookup_api_2026}.

In the 24-path core documentation corpus, 13 paths placed the first documented branch or progression observer at the merchant backend, six at a browser callback or state machine, four in a provider-hosted component, and one at an external-result input. Eighteen paths documented branch-specific output somewhere in the integration. This 18/24 count is not a count of client-visible or attacker-reachable oracles: the audit does not establish caller-selected candidate processing, admission bypass, or candidate separation for these paths. The separately reported 16-path extension raises descriptive coverage to 40 paths but is not merged into those core-corpus counts. These are documentation findings, not live tests, counts of vulnerable providers, or a prevalence estimate.

\paragraph{CaaS Context and 3DS Card-Testing Tools.} A passive public-source review, separate from our controlled tests, identified public advertisements for paid card-testing services and publicly hosted code with 3DS-related functionality. The paid offerings and associated 3DS-capable code show that pre-authorization outcomes are already used in CaaS card-testing workflows; not every repository reviewed is itself a paid offering~\cite{rapid7_caas_2026}. Three public Telegram captures display advertised Braintree-oriented batch classifications alongside Stripe-oriented outcomes (\texttt{requires\_action}, declined, invalid)~\cite{telegram_3ds_tooling_2026}. An independently led hosting audit recorded related payment-abuse code repositories~\cite{independent_hosted_payment_abuse_audit_2026}. Some artifacts show batch processing and rotating-proxy support, indicating that distributed proxy-assisted screening is already an operational abuse pattern rather than a merely hypothetical extension. We did not execute the tools or measure throughput, success rates, or their use against the evaluated cohorts. Official Stripe documentation warns of automated scripts harvesting 3DS issuer responses~\cite{stripe_card_testing_2026}, while G+D Netcetera practitioners reported separate live 3DS-rail velocity attacks exceeding ten transactions per second~\cite{netcetera2025fraudalert}. These sources document in-the-wild CaaS abuse, but they are not pooled with our primary measurements. Our caller stopped before browser challenge continuation, and issuer-side alert behavior was not instrumented.

\subsection{Mapping the Tested Integration}
We examined official gateway documentation and sample code to reconstruct the implementation layer and then used EMVCo materials to map its 3DS stage to the network protocol. This is a case-study reconstruction, not a claim that documentation or the client API exposes internal DS processing. The ProVerif model (Appendix~\ref{app:proverif}) captures one possible lookup policy: release data to a caller presenting valid client authorization. It does not establish that Braintree intended the token to be bound to a particular browser session or that token reuse itself is a root-cause flaw.

Our analysis of Braintree's official authorization and 3DS documentation, the public Web SDK interface, and maintainer guidance indicates that the tested boundary is consistent with a Braintree 3DS-enabled browser integration pattern: merchants provision browser-usable \texttt{client\_token}-based authorization for 3DS, the SDK launches a common \texttt{verifyCard} flow with nonce, BIN, and amount, CardinalCommerce derives issuer and device-data context from that BIN, and the browser reaches the documented \texttt{onLookupComplete(data, next)} callback before challenge continuation~\cite{braintree_auth_overview,braintree_3ds_guide,braintree_web_3ds_docs,braintree_issue_681}. The SDK also exposes a browsing-session identifier as \texttt{analyticsMetadata.sessionId} and includes analytics metadata with Client API requests. The reviewed public documentation and SDK source do not describe that identifier as an authorization-binding control or specify that the \texttt{client\_token} must be used only from the browser session that obtained it~\cite{braintree_web_client_module,braintree_auth_overview,braintree_client_api}. Therefore, the observed separate-process execution provides empirical evidence of multi-session replay of valid merchant-scoped \texttt{client\_token} authorization, rather than an explicit browser-session-binding bypass. Our direct observations remain limited to the tested live path.

\paragraph{Token Lifecycle and Reuse Semantics} We inspected the client authorization used in the tested integration. The captured token's \texttt{exp} claim and Braintree's public documentation both indicated a validity window of up to 24 hours, subject to earlier invalidation in specified cases~\cite{braintree_auth_overview,braintree_client_api}. This is a documented lifetime for client authorization, not evidence by itself of a vulnerability.
\begin{itemize}
\item \textbf{Client-side authorization.} The \texttt{client\_token} and associated authorization fingerprint authorize supported Client API operations, including 3DS, during their validity window. Whether a checkout requires an account-holder login is merchant-specific.
\item \textbf{Vault-level rights.} The payload grants \texttt{manage\_vault} rights and \texttt{Braintree:Vault} scope, which are sufficient to tokenize arbitrary card data.
\item \textbf{No verification by default.} The token encodes \texttt{verify\_card\_by\_default:~false}, meaning card verification is not enforced at the API boundary.
\item \textbf{Separate-client acceptance.} In our bounded validation, a standalone off-context HTTP client (operating just like an attacker would by issuing raw scripted API queries) using the tested integration's valid authorization submitted lookup requests. This does not establish whether additional session or origin checks apply in other flows, nor that browser binding was guaranteed.
\end{itemize}
In our bounded PoC validation, the client authorization was reused across the tested initialization, telemetry-submission, and 3DS lookup steps. We describe this as reuse of a client credential rather than token rebinding.

Because the captured authorization can remain valid for up to 24 hours, an attacker who obtains it from a public checkout could potentially reuse the same token for successive PAN-specific lookups during that window, subject to earlier invalidation and any merchant- or gateway-side controls. This creates a plausible scale-amplification risk without requiring a new token for each lookup. We reused the credential only in our bounded PAN-varying workflow; we deliberately did not test sustained or high-volume replay against the researcher-controlled live service. We therefore make no claim about achievable throughput, whether velocity controls would intervene, or unrestricted production-scale enumeration.

\subsection{Theoretical Abuse Modeling and Cost Asymmetry}
To discuss potential implications, we distinguish the observed request path from untested scale and cost scenarios~\cite{pci_ssc_enumeration_2020}. The standalone off-context HTTP workflow is scriptable (operating just as an automated attacker would do), but the live study does not establish how many requests production deployments accept, whether other merchants expose the same behavior, or what costs any deployment would incur.

An adversary with access to multiple network egress identities, potentially including residential-proxy services, or the ability to alter caller-supplied telemetry could complicate controls that rely only on a single source identifier or unverified client values. The primary bounded cohort study deliberately used no proxy services, did not vary source identities, and did not distribute requests or test telemetry manipulation against the live service. In a separate exploratory check, however, we re-probed---through distributed residential SOCKS5 egress on the Cardinal device-data path---a set of Mastercard PANs that had been hard-blocked under single-source probing with client-visible Reason~98 feedback; every previously blocked card received an issuer-level response (Section~\ref{subsec:deconstruct_patent}). This A/B result is bounded to the tested integration and a modest proxy pool, which limits sustainable throughput; it does not measure production-scale evasion. But the boundedness here is an ethical and experimental limit, not evidence that proxy-dispersed screening is operationally implausible. The defensive implication is to validate telemetry provenance and assess abuse controls using server-side and aggregate signals rather than trusting client-supplied values or relying on one source identifier alone.

\subsubsection{Automation Along the 3DS Path and Enumeration Defense Demarcation}
Large-scale BIN enumeration is a recognized threat to global payment systems~\cite{pci_ssc_enumeration_2020,visa2021enumeration}. Visa's enumeration guidance recommends CAPTCHA controls and velocity checks for authorization attempts and declines~\cite{visa2021enumeration}; its merchant guidance also describes bot controls, rate limits, tokenization-layer controls, and network-level risk scoring such as Visa Account Attack Intelligence (VAAI)~\cite{visa2026merchantenumeration,visa2024vaai}. The 2026 guide treats zero-amount Account Verification as an authorization flow for validating account information and recommends monitoring and blocking excessive failed AV attempts; it does not present zero-amount verification itself as a standalone anti-enumeration defense~\cite{visa2026merchantenumeration}. These measures address checkout, authorization, tokenization, and network-processing paths, but do not by themselves establish coverage of the client-visible pre-authorization 3DS lookup boundary studied here. Our empirical measurements show that, for the tested merchant, a standalone off-context HTTP client reusing merchant-issued authorization could repeat lookups without executing the browser SDK. Because browser-channel telemetry lacked cryptographic execution binding, synthetic telemetry submitted by this caller elicited frictionless authentication (\texttt{status: "authenticate\_successful"}, \texttt{paresStatus: Y}) with valid CAVVs across representative anonymized cohorts (with specific BINs strictly de-identified). Repetition was therefore possible under the tested conditions without browser execution. The RDX reason code is operational context, not evidence that this request path was detected or blocked~\cite{cardinal_transreason_2025}.

A G+D Netcetera practitioner interview describes a separate production BIN-attack incident involving a large German merchant. The speaker estimated traffic at more than 10 transactions per second and said card-testing traffic roughly doubled the bank's normal monitored volume; the interview also recounts a call-center queue of about 1,000 affected cardholders. The speakers characterize 3DS-based PAN testing as occurring across multiple ACSs. These are provider-reported observations, not independently audited measurements, and the interview does not establish that the incident used the Braintree/Cardinal client-facing path studied here or that our tested interface could sustain comparable throughput~\cite{netcetera2025fraudalert}.

A separate G+D Netcetera product article claims that attackers can test several thousand card numbers within minutes and obtain hundreds of valid numbers. The article does not describe a measurement method, sample, or underlying dataset for this estimate. We therefore treat it as a vendor claim about possible attack scale, not as a measured rate or prevalence estimate~\cite{netcetera2025binattackarticle}. Together, these sources provide industry context for production-scale 3DS card testing; they do not establish the scale or impact of our bounded measurements.

\subsubsection{Sandbox Rate and Production Cost Boundaries}
In a separate Braintree Sandbox experiment, the automated client sustained 0.75 requests per second under the tested setup, and the observed responses did not include \texttt{429 Too Many Requests}. This demonstrates repeatability only in that sandbox run. We did not measure production throughput, merchant charges, billing terms, production rate limits, or network-side detection. The experiment therefore does not quantify a financial-denial-of-service risk; cost pressure remains conditional on provider controls and contractual pricing.

\subsection{Remediation Roadmap}
Remediation should address both integration release and network routing boundaries while keeping their responsibilities distinct:
\begin{itemize}
    \item \textbf{Scope lookup authority.} Bind sensitive lookup access to merchant, operation, and transaction context; impose origin or session binding only where the integration defines it as a security requirement.
    \item \textbf{Require execution-bound telemetry provenance.} Accept telemetry only when coupled to verifiable browser execution context.
    \item \textbf{Minimize pre-auth response semantics.} Preserve the minimum continuation state needed for the client to select the challenge or challenge-free path, while limiting ancillary account-dependent statuses, reasons, and metadata before authorization.
    \item \textbf{A Conditional Interim Stopgap: Range-Level Uniform Challenge Proxying.} While amending EMVCo standards requires multi-year consortium consensus and client SDK sanitization is prone to gateway implementation drift, Access Control Servers can deploy an immediate tactical countermeasure across participating card ranges: \emph{Uniform Range-Level Challenge Proxying}. Under this policy, when an AReq is routed to an ACS for any candidate PAN within an active participating card range, the ACS strictly suppresses early negative ledger rejections ($\text{transStatus} \in \{\text{`N'}, \text{`R'}\}$ with reason codes 05, 08, or 13) during the frictionless AReq phase.
        Account ledger validation is deferred to the interactive CReq challenge phase inside the cardholder's secure browser sandbox, which automated scraping bots cannot solve across distributed proxy networks.

    \item \textbf{Harden boundary-local abuse controls.} Apply velocity, anomaly, and cost-aware controls directly on lookup-release interfaces.
    \item \textbf{Measure DS visibility.} In an authorized environment, correlate client-facing lookups with 3DS Server--DS AReq/ARes traces and evaluate network-side detector coverage rather than assuming that the integration request is or is not visible to the DS.
    \item \textbf{Align documentation with security invariants.} Integration guidance should make binding requirements explicit and testable to reduce cross-party implementation drift.
\end{itemize}

Together, these steps target the architectural cause rather than only suppressing observable abuse symptoms.

\section{Discussion}
\label{sec:discussion}
This section asks a narrower question: once pre-authorization 3DS signals escape to the requestor boundary, can processing-network detection, issuer-local countermeasures, or consortium reform actually close the oracle?

\subsection{Analyzing U.S. Patent Application US 2026/0134429 A1}
\label{subsec:deconstruct_patent}

\begin{figure*}[t]
\centering
\begin{minipage}[t]{0.485\textwidth}
\centering
\resizebox{\linewidth}{!}{%
\begin{tikzpicture}[
    font=\small\sffamily,
    line cap=round,
    line join=round,
    >=Stealth,
    actor/.style={rectangle, sharp corners, draw=ink, fill=white,
      line width=0.95pt, minimum height=0.78cm, text width=1.85cm, align=center,
      font=\scriptsize\sffamily\bfseries},
    lifeline/.style={draw=black!60, line width=0.75pt, dash pattern=on 2.5pt off 2.5pt},
    msg/.style={-{Stealth[length=2.5mm,width=1.7mm]}, line width=1.05pt, draw=ink},
    resp/.style={-{Stealth[length=2.5mm,width=1.7mm]}, line width=0.9pt, draw=black!80, dashed,
                 dash pattern=on 3.5pt off 2.5pt},
    steplabel/.style={font=\tiny\sffamily\bfseries, fill=white, draw=black!65, line width=0.4pt, sharp corners, inner sep=1.5pt},
    notebox/.style={rectangle, sharp corners, draw=black!60, fill=black!4,
      line width=0.5pt, align=center, font=\tiny\sffamily, inner sep=2.5pt},
    alertbox/.style={rectangle, sharp corners, draw=black!85, fill=panel,
      line width=0.95pt, align=center, font=\scriptsize\sffamily, inner sep=4.5pt}
  ]
  \def\xU{0.5}   
  \def\xM{2.4}   
  \def\xQ{4.3}   
  \def\xD{6.2}   
  \def\xA{8.1}   
  \def\yBot{-6.8}

  \node[font=\footnotesize\sffamily, align=center, text width=7.8cm] at (4.3, 1.35) {
    \textbf{(a) Mastercard Patent US 2026/0134429 A1 (FIG.~3)}\\[2pt]
    \textcolor{black!75}{\textbf{Claimed Method 300: Pause (318) and Email Alerts (320/322)}}
  };

  \node[actor] at (\xU, 0.3) {User 120\\[-1pt]\normalfont\tiny ``Fraudster''};
  \node[actor] at (\xM, 0.3) {First Party\\[-1pt]\normalfont\tiny 102};
  \node[actor] at (\xQ, 0.3) {Acquirer 104\\[-1pt]\normalfont\tiny 3DS Server 110};
  \node[actor] at (\xD, 0.3) {Processing Net. 106\\[-1pt]\normalfont\tiny Directory Server 112};
  \node[actor] at (\xA, 0.3) {Issuer 108\\[-1pt]\normalfont\tiny ACS 114};

  \draw[lifeline] (\xU, -0.15) -- (\xU, \yBot);
  \draw[lifeline] (\xM, -0.15) -- (\xM, \yBot);
  \draw[lifeline] (\xQ, -0.15) -- (\xQ, \yBot);
  \draw[lifeline] (\xD, -0.15) -- (\xD, \yBot);
  \draw[lifeline] (\xA, -0.15) -- (\xA, \yBot);

  \draw[msg] (\xU, -0.55) -- (\xM, -0.55);
  \node[steplabel] at ({(\xU+\xM)/2}, -0.35) {\sqnum{302} Txn req. (estimated PAN [0054])};

  \draw[msg] (\xM, -1.15) -- (\xQ, -1.15);
  \node[steplabel] at ({(\xM+\xQ)/2}, -0.92) {\sqnum{304} Submit authentication user};

  \draw[msg] (\xQ, -1.75) -- (\xD, -1.75);
  \node[steplabel] at ({(\xQ+\xD)/2}, -1.52) {\sqnum{306} Compile \& submit AReq};

  \node[notebox, text width=3.5cm] at ({(\xQ+\xD)/2}, -2.40) {
    \sqnum{308} \textbf{DS XGBoost Evaluation:}\\[1pt]
    Computes velocity over BIN/MAID [0033]
  };

  \draw[msg] (\xD, -3.05) -- (\xA, -3.05);
  \node[steplabel] at ({(\xD+\xA)/2}, -2.82) {\sqnum{310} Transmit AReq (flag / score 998)};

  \node[notebox, text width=2.6cm] at ({(\xD+\xA)/2}, -3.65) {
    \sqnum{312} \textbf{ACS Verification:}\\[1pt]
    Verifies attack \& ledger [0060]
  };

  \draw[resp] (\xA, -4.30) -- (\xD, -4.30);
  \node[steplabel] at ({(\xD+\xA)/2}, -4.07) {\sqnum{314} Compile \& submit ARes (R-98)};

  \draw[resp] (\xD, -4.90) -- (\xQ, -4.90);
  \node[steplabel] at ({(\xQ+\xD)/2}, -4.67) {\sqnum{316} Pass ARes (reason code)};

  \draw[line width=0.9pt, draw=ink] (\xQ, -5.45) -- (\xQ+0.4, -5.45) -- (\xQ+0.4, -5.75) -- (\xQ, -5.75);
  \node[steplabel, anchor=west] at (\xQ+0.45, -5.60) {\sqnum{318} Pause attack txns [0061]};

  \draw[resp] (\xD, -6.20) -- (\xA, -6.20);
  \node[steplabel] at ({(\xD+\xA)/2}, -5.98) {\sqnum{320} Email alert to Issuer 108 [0063]};
  \draw[resp] (\xD, -6.55) -- (\xQ, -6.55);
  \node[steplabel] at ({(\xQ+\xD)/2}, -6.33) {\sqnum{322} Email alert to Acquirer 104};

  \node[alertbox, text width=7.4cm] at (4.3, -7.80) {
    \textbf{Mastercard FIG.~3 Method 300 Claimed Defense Boundary:}\\[2pt]
    The patent assumes the attack is arrested at the 3DS Server (Step~318 pause)\\
    and via out-of-band operational email alerts (Steps~320/322).\\
    \textbf{Crucially, FIG.~3 omits any in-band return path to the caller!}
  };

\end{tikzpicture}%
}
\end{minipage}%
\hfill%
\begin{minipage}[t]{0.485\textwidth}
\centering
\resizebox{\linewidth}{!}{%
\begin{tikzpicture}[
    font=\small\sffamily,
    line cap=round,
    line join=round,
    >=Stealth,
    actor/.style={rectangle, sharp corners, draw=ink, fill=white,
      line width=0.95pt, minimum height=0.78cm, text width=1.55cm, align=center,
      font=\scriptsize\sffamily\bfseries},
    lifeline/.style={draw=black!60, line width=0.75pt, dash pattern=on 2.5pt off 2.5pt},
    msg/.style={-{Stealth[length=2.5mm,width=1.7mm]}, line width=1.05pt, draw=ink},
    resp/.style={-{Stealth[length=2.5mm,width=1.7mm]}, line width=0.9pt, draw=black!80, dashed,
                 dash pattern=on 3.5pt off 2.5pt},
    steplabel/.style={font=\tiny\sffamily\bfseries, fill=white, draw=black!65, line width=0.4pt, sharp corners, inner sep=1.5pt},
    notebox/.style={rectangle, sharp corners, draw=accent, fill=accentbg,
      line width=0.55pt, align=center, font=\tiny\sffamily, inner sep=2.5pt},
    alertbox/.style={rectangle, sharp corners, draw=accent, fill=accentbg,
      line width=0.95pt, align=center, font=\scriptsize\sffamily, inner sep=4.5pt}
  ]
  \def\xC{0.5}   
  \def\xM{2.4}   
  \def\xQ{4.3}   
  \def\xD{6.2}   
  \def\xA{8.1}   
  \def\yBot{-6.8}

  \node[font=\footnotesize\sffamily, align=center, text width=7.8cm] at (4.3, 1.35) {
    \textbf{(b) Measured Attack Reality (The In-Band Evasion Oracle)}\\[2pt]
    \textcolor{accent}{\textbf{In-Band Feedback (Step 6) and lookup-stage Cardholder Blindspot}}
  };

  \node[actor] at (\xC, 0.3) {Caller\\[-1pt]\normalfont\tiny ``Fraudster''};
  \node[actor] at (\xM, 0.3) {First Party\\[-1pt]\normalfont\tiny 102};
  \node[actor] at (\xQ, 0.3) {3DS Server\\[-1pt]\normalfont\tiny 110};
  \node[actor] at (\xD, 0.3) {DS 112\\[-1pt]\normalfont\tiny (Network 106)};
  \node[actor] at (\xA, 0.3) {ACS 114\\[-1pt]\normalfont\tiny (Issuer 108)};

  \draw[lifeline] (\xC, -0.15) -- (\xC, \yBot);
  \draw[lifeline] (\xM, -0.15) -- (\xM, \yBot);
  \draw[lifeline] (\xQ, -0.15) -- (\xQ, \yBot);
  \draw[lifeline] (\xD, -0.15) -- (\xD, \yBot);
  \draw[lifeline] (\xA, -0.15) -- (\xA, \yBot);

  \draw[resp] (\xM, -0.55) -- (\xC, -0.55);
  \node[steplabel] at ({(\xC+\xM)/2}, -0.35) {\sqnum{1} Reusable \texttt{client\_token}};

  \draw[msg] (\xC, -1.15) -- (\xQ, -1.15);
  \node[steplabel] at ({(\xC+\xQ)/2}, -0.92) {\sqnum{2} Direct HTTP tokenization (skips 102)};

  \draw[msg] (\xC, -1.75) -- (\xQ, -1.75);
  \node[steplabel] at ({(\xC+\xQ)/2}, -1.52) {\sqnum{3} Lookup + synthetic telemetry (no browser)};

  \node[notebox, text width=3.5cm] at ({(\xQ+\xD)/2}, -2.40) {
    \textbf{Step 308 Classifier Evasion:}\\[1pt]
    $v_{\text{browserIP}} \equiv 1$ (residential proxies)\\[1pt]
    $v_{\text{MAID}} \le 1/\text{hr}$ (token rotation)
  };

  \draw[msg] (\xQ, -3.05) -- (\xD, -3.05);
  \draw[msg] (\xD, -3.05) -- (\xA, -3.05);
  \node[steplabel] at ({(\xD+\xA)/2}, -2.82) {\sqnum{4} AReq routed to ACS 114};

  \draw[resp] (\xA, -3.70) -- (\xD, -3.70);
  \draw[resp] (\xD, -3.70) -- (\xQ, -3.70);
  \node[steplabel] at ({(\xD+\xA)/2}, -3.48) {\sqnum{5} ARes with Reason 98/99/CAVV};

  \draw[msg, draw=accent, line width=1.3pt] (\xQ, -4.35) -- (\xC, -4.35);
  \node[steplabel, draw=accent, text=accent] at ({(\xC+\xQ)/2}, -4.10) {\sqnum{6} \textbf{In-Band Leak: Gateway reflects ARes in client JSON!}};

  \node[notebox, text width=3.4cm] at ({(\xC+\xQ)/2}, -5.20) {
    \textbf{In-Band Tripwire Paradox:}\\[1pt]
    $\bullet$ Reason 98 leaks threshold $\tau$ (tunes evasion)\\[1pt]
    $\bullet$ Non-R-98 separates live vs. dead ledger accounts
  };

  \node[notebox, draw=ink, fill=white, text=ink, text width=3.5cm] at ({(\xQ+\xA)/2}, -5.20) {
    \textbf{Step-Up Evasion Path:}\\[1pt]
    Caller halts before \texttt{CReq}\\[1pt]
    ACS session times out silently\\[1pt]
    Req~221: ACS $\to$ DS \texttt{RReq}\\[1pt]
    Zero cardholder alerts triggered
  };
  \node[alertbox, text width=7.4cm] at (4.3, -7.80) {
    \textbf{The In-Band Evasion Reality:}\\[2pt]
    Gateway mirrors ARes outcome fields into client JSON (Step 6);\\
    off-context caller extracts response classification before clearing;\\
    halting before \texttt{CReq} isolates traffic from cardholder step-up UI,\\
    converting scheme defensive signals into an in-band calibration oracle.
  };

\end{tikzpicture}%
}
\end{minipage}
\caption{\label{fig:patent_method300_deconstruction}\textbf{Architectural deconstruction of Mastercard Patent US 2026/0134429 A1: (a) Proposed Method~300 versus (b) the measured client-facing path.}
Panel~(a) summarizes the patent's proposed DS classifier, attack flag, ACS reason code, and downstream acquirer pause. The patent's baseline flow passes ARes content from the 3DS Server to the first party before that party compiles an authorization request ([0024]--[0025]); the Step~318 pause follows ARes delivery to the 3DS Server ([0061]). The application contemplates timely network notifications to issuers/acquirers ([0015]); in Method~300, separate notifications are optional and may be operational emails ([0063]), while investigation or remediation may follow minutes or hours later ([0064])~\cite{mastercard2026abnormaltraffic}. Panel~(b) shows the tested integration returning gateway JSON associated with the initial AReq/ARes before authorization; the off-context caller stops there without sending \texttt{CReq}. This is not a raw ARes capture or a correlated internal trace. Reason~98 appeared in exploratory client-visible logs, consistent with the patent's in-band signal, but its generating component was not established. In a controlled exploratory A/B comparison on the identical PAN set, routing the Cardinal device-data path through distributed residential SOCKS5 proxies removed the scheme-tier blocks observed under single-source probing; multi-merchant rotation remains an additional plausible dispersion vector not measured here. Because the caller sent no initial \texttt{CReq}, it did not enter the subsequent challenge/OOB interaction. If the initial-CReq timer expires, Req~221 specifies an ACS-to-DS \texttt{RReq} with Reason~14; FAQ Q93 concerns a separate timeout after a challenge has started~\cite{emvco_core_2311_2023,emvco_3ds_faq_2026}. No SMS, push, or OTP reached test recipients. The EMV 3DS timeout messages are not cardholder alerts, and the patent's separate notifications target the issuer/acquirer, not the cardholder.}
\end{figure*}
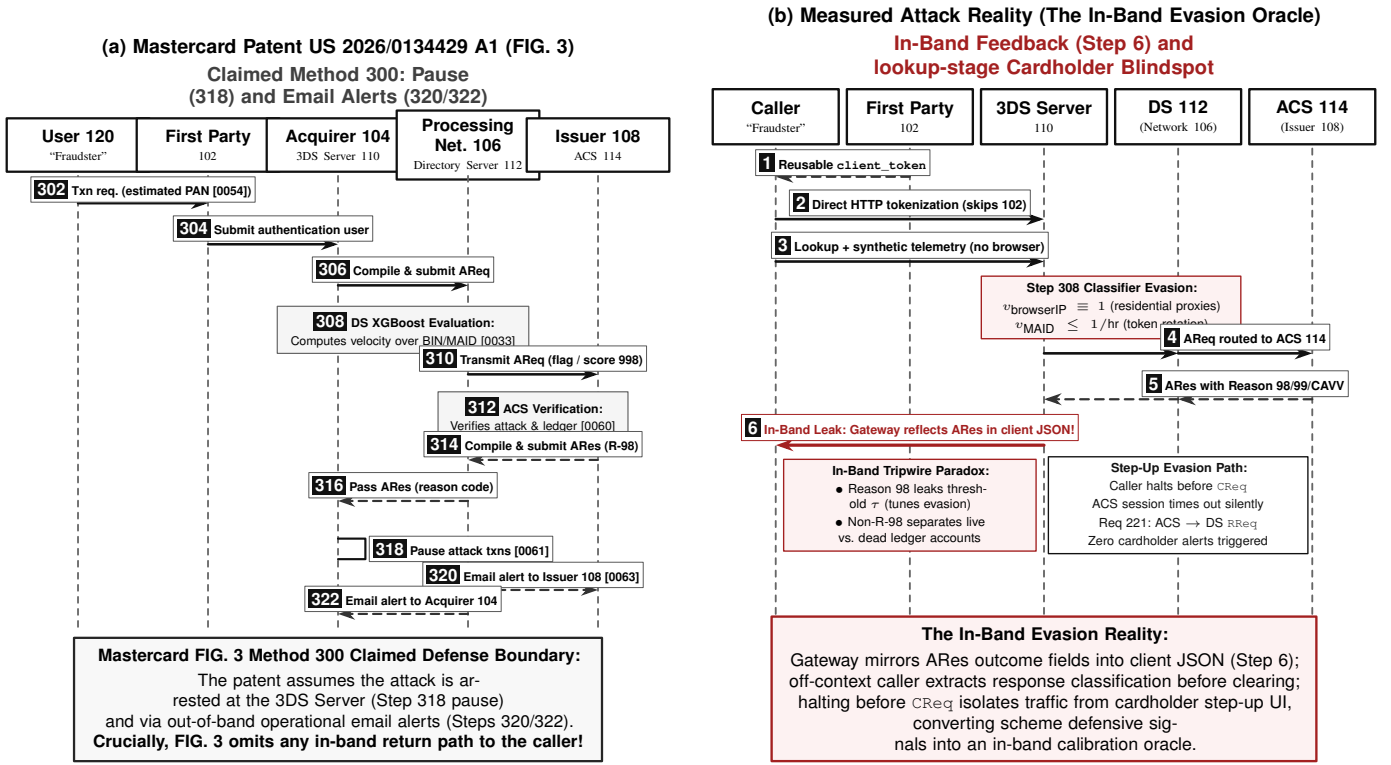

Figure~\ref{fig:patent_method300_deconstruction} exposes the central gap in US~2026/0134429~A1: the patent correctly recognizes the authentication BIN attack, but places mitigation at the processing-network layer while leaving the client-facing return path exposed~\cite{mastercard2026abnormaltraffic}. The disclosure formalizes the ``authentication BIN attack'' threat model, explaining that because 3DS operates by ``imposing an authentication phase distinct from an authorization phase'' ([0023]), fraudsters repeatedly submit transactions using ``estimated or guessed credentials'' and treat the resulting scheme signals as validation oracles ([0005], [0013], [0036]):
\begin{quote}
\centering
\bfseries
``\emph{...for a fraudster, who submitted an estimated or guessed credential, the authentication response, for example, is an indication that the credential is indeed legitimate or valid. The fraudster may leverage computer implementations (e.g., bots, automated programs, etc.) to initiate hundreds or thousands of authentications... whereby ones of the estimated or guessed credentials are verified as legitimate or valid, in which case, the credentials are sellable or usable for fraudulent purposes}''~\cite{mastercard2026abnormaltraffic}.
\end{quote}

The application discloses a processing-network model that scores feature data and velocity signals at the Directory Server, then propagates a dedicated risk score and reason code to downstream participants~\cite{mastercard2026abnormaltraffic}. Specifically, it describes an ``XGBoost model based on gradient boosting'' ([0034], Claim~3) with tuned hyperparameters (boosting rounds, learning rate, tree depth), log-transformed velocity distributions, and default value imputation for missing parameters. The model evaluates ``feature data specific to the first party'' and BIN snapshot profiles ([0015], [0030], [0032], Claims~1, 5), notifying issuers/acquirers via operational email alerts ([0038], [0063], Claim~6) or in-band via a 3DS \texttt{messageExtension} with a dedicated risk score 998 outside conventional score ranges and attack labels ([0040], Claim~7). In response, the Access Control Server returns a second reason code R-98 (\emph{Authentication BIN Attack}) ([0041], Claim~8), or the Directory Server injects an override response code R-99 ([0043]) or ``issues declines or other errors for credentials within the BIN range'' ([0044])~\cite{mastercard2026abnormaltraffic}.

\paragraph{Observation of Reason 98 and Threat Model Consistency.} In our exploratory review of live response logs, we observed instances of Reason 98 (\emph{Authentication BIN Attack}). This client-facing label is consistent with the threat model and signaling mechanisms described in published application US 2026/0134429 A1~\cite{mastercard2026abnormaltraffic}. However, as noted in Section~\ref{sec:empirical_study}, client-side observation of this label alone does not establish whether the response was generated by a Directory Server ML classifier, an ACS, or intermediate gateway mapping.

\paragraph{Feature Dynamics and Architectural Boundaries in Method 300.} The application describes compiling multi-dimensional feature data spanning first parties (that is, merchants), acquirers, and issuers, including velocity measures, Merchant-Acquirer IDs (MAIDs), client IP addresses, and snapshot profiles ([0027], [0029], [0030], [0058], Claims~1, 2)~\cite{mastercard2026abnormaltraffic}. While this feature-data model formalizes machine learning detection at the processing network, our empirical and protocol analysis reveals five fundamental architectural limits in Method~300's defense~\cite{mastercard2026abnormaltraffic}:

(1)~\textbf{Architectural Boundary Demarcation ([Req 97] vs. [Req 105]):} Mastercard's disclosure conditions mitigation on processing-network intervention, including attack-specific reason codes and declines within the BIN range ([0041], [0043], [0044])~\cite{mastercard2026abnormaltraffic}. However, once an AReq survives DS card-range pruning under Seq~3.133 [Req~97] (\texttt{cardRangeData} [Req~460]) and routes to the issuer ACS under Seq~3.143 [Req~105], response behavior can diverge across bank architectures. In our sampled cohorts, that divergence appears as response-class differences consistent with ACS-associated processing, not as direct proof of account existence or a universal issuer-side oracle.
(2)~\textbf{Velocity Feature Defeat under Distributed Egress:} The patent conditions detection on feature data specific to first parties and MAIDs ([0030], Claims~1, 2)~\cite{mastercard2026abnormaltraffic}, with the client IP address as a first-class device identifier ([0027]). In the browser channel (\texttt{02-BRW}), \texttt{browserIP} is an AReq field; in the tested Cardinal integration, the observed state flip is attributable to the change in TCP egress topology, consistent with TCP-layer derivation of the source identifier rather than application-layer forwarding headers. In a controlled A/B comparison, the identical set of Mastercard PANs that had been hard-blocked under single-source probing (with client-visible Reason~98 feedback) received issuer-level responses when the Cardinal device-data path was routed through distributed residential SOCKS5 egress: every previously blocked card obtained a real response, and no scheme-tier block recurred. The result is bounded to the tested integration and a modest proxy pool, which constrains sustainable throughput rather than the structural conclusion: the trained model keyed on source concentration gates \emph{when} an adversary observes issuer-side responses, not \emph{whether} those responses distinguish accounts.
(3)~\textbf{The \texttt{messageExtension} Conundrum and Error Code 203:} In paragraph [0040], the patent discloses signaling attacks to the ACS by injecting Risk Score 998 via a 3DS \texttt{messageExtension}~\cite{mastercard2026abnormaltraffic}. While EMVCo FAQ Q84~\cite{emvco_3ds_faq_2026} confirms DS authority to define proprietary extensions, EMVCo Section~5.9.2 mandates a \texttt{criticalityIndicator}. A critical extension risks rejection with Error Code~203 (\emph{Data element not recognized/supported}) by components lacking support, while a non-critical extension risks being ignored; this design tension helps explain the fallback override logic in paragraph [0043]~\cite{mastercard2026abnormaltraffic}, where the Directory Server overwrites the ACS reason code with override code R-99 in the ARes.
(4)~\textbf{The In-Band Tripwire Paradox (R-98 and R-99):} Surfacing explicit reason codes (R-98 and R-99) in client-visible responses ([0041], [0043])~\cite{mastercard2026abnormaltraffic} inadvertently introduces an in-band boundary oracle. Under PCI 3DS Data Matrix v1.3 Table~1, \texttt{transStatusReason} is unclassified and exempt from sensitive data protection~\cite{pci_3ds_data_matrix_v1_3}. In machine-learning terms, the network can provide the caller with label access that informs later request shaping. That observation supports a structural critique of in-band signaling without requiring a claim about any particular operational bypass threshold.
(5)~\textbf{The Gateway Return-Path Gap (Steps 318 vs.\ Step 6):} Mastercard's application itself identifies authentication responses as signals that can validate estimated credentials ([0005], [0013], [0036]) and describes the baseline flow passing ARes content to the first party ([0024]) before that party compiles an authorization request ([0025])~\cite{mastercard2026abnormaltraffic}. Method~300 detects and flags suspected attacks at the Directory Server, passes the ARes to the 3DS Server, and then describes the acquirer pausing associated transactions at Step~318 ([0061]). The application contemplates timely network-to-issuer/acquirer notifications ([0015]); the separate notifications in Method~300 are optional and may be operational emails ([0063]), while investigation or remediation may follow minutes or hours later ([0064])~\cite{mastercard2026abnormaltraffic}. These network-side mechanisms do not specify a cardholder SMS/push alert or guarantee that the first-party response is withheld or normalized before return. Our tested integration returned client-visible gateway JSON associated with the initial AReq/ARes before authorization, and the off-context caller stopped at that boundary without sending \texttt{CReq}. This establishes a pre-authorization response-class signal in the tested path, consistent with the patent's threat model; it does not show that the gateway returned a raw ARes or that every candidate's account validity was established.

\paragraph{In-The-Wild Operationalization of the Authentication BIN Attack Model.}
The threat model formalized in Mastercard's application---in which fraudsters deploy computer implementations to submit estimated credentials and rely on first-party signals ([0005], [0013], [0036])---is consistent with the broader CaaS threat model and the reviewed 3DS card-testing artifacts~\cite{rapid7_caas_2026,telegram_3ds_tooling_2026,independent_hosted_payment_abuse_audit_2026}. The artifacts show automated parsing of gateway-reflected pre-authorization outcomes, including frictionless, attempted, challenge-required, and negative classes. Attackers are unlikely to value these classes equally: frictionless or success-labeled attempted-authentication outcomes are operationally more attractive than challenge-required branches because they avoid step-up friction and can indicate weaker issuer-side risk discrimination at the lookup boundary. In one reviewed tool implementation, a first-result status of \texttt{C} or \texttt{R} is labeled ``Challenge required'' and returned as declined; this branch does not render a browser challenge or observe an OTP. The implementation is therefore coded to use the lookup-stage signal to classify estimated credentials before challenge completion or financial authorization. In our own measurements, a forged off-context caller obtained frictionless success-labeled responses in selected cohorts, which is consistent with weak issuer-side risk discrimination against off-context requests in those tested paths. This documents operational use of lookup-stage responses for account-recognition triage. It does not establish that every classified candidate corresponds to a live account or that downstream authorization would succeed, and we did not execute the reviewed tools or test them against our cohorts.

\subsection{The Multi-Party Infrastructure Dilemma: Consortium Governance and Asymmetric Burdens}
\label{subsec:multiparty_dilemma}

\paragraph{The Multi-Party Tragedy of the Commons: Ali et al. at the Network Infrastructure Layer.} In their seminal investigation of e-commerce payment fraud, Ali et al.~\cite{ali2017distributed} demonstrated that variations in merchant checkout form validation rules at the \emph{payment authorization and clearing layer} enabled distributed card guessing. In this paper, we demonstrate that a structurally identical yet fundamentally deeper vulnerability exists at the \emph{payment network routing and consortium governance layer}. The vulnerability is an infrastructure-level multi-party conflict of interest rooted in the historical origins and architectural structure of EMVCo.

Founded in 1994 by Europay, MasterCard, and Visa, EMVCo is network-dominated and owned by major card schemes. Specifications prioritize transaction velocity, frictionless conversion, and switching volume. Crucially, the protocol standard decouples authentication from financial authorization, explicitly declaring at Step~4 of the Frictionless Flow: \emph{``The 3DS Server communicates the result of the ARes message to the 3DS Requestor Environment... Note: 3-D Secure processing ends here. For Payment Authorisation, the subsequent steps apply''} (EMVCo Section~2.6). Across every architectural flow and security diagram (Figures~2.2, 2.6, 6.2), the standard adds the explicit footnote: \emph{``Dashed arrows and 3DS Requestor are not part of 3DS specification and are shown for clarity only''}, while mandating mutual TLS signed by the scheme DS CA for network routing (Section~6.1.2). In short, EMV 3DS is optimized for trusted inter-party routing, low friction, and rapid completion of the authentication phase. That design assumes the 3DS Requestor boundary is trustworthy; once that assumption fails, pre-authorization signals become externally queryable before clearing ever begins.

Because the network assumes that queries arriving over DS CA-authenticated channels originate from a trusted Requestor running genuine consumer checkouts, the protocol can expose internal routing and response state to downstream integrators. The specification itself makes a narrower point: EMVCo's diagrams repeatedly note that dashed arrows and the 3DS Requestor are shown for clarity and are not themselves part of the 3DS specification, while FAQ Q6 treats SDK-authenticity verification as outside the Core Specification for the SDK integration model and FAQ Q47 defines a frictionless transaction as complete for the ACS at the ARes stage~\cite{emvco_3ds_faq_2026}. We therefore describe the client-facing Requestor Environment as underspecified by the core protocol rather than formally governed by a Section~1.3 scope exclusion. Commercial 3DS Server vendors and payment gateways primarily optimize for checkout velocity and conversion; within that ecosystem, pre-authorization endpoints may feature permissive access thresholds. Card schemes similarly prioritize routing throughput while exposing identifier-linked distinctions that can support conditional prefix pruning under the assumptions discussed earlier. The burdens are asymmetric: issuing banks may face potential card-reissuance costs~\cite{aba2015cardreissue,javelin2014cardreissue}, support disruption, and downstream CNP fraud or chargeback exposure. Visa's rules provide for VIMP non-compliance assessments when an issuer's monthly Card-Absent Environment disputes or fraud meet or exceed program thresholds; the listed assessment reaches \$100,000 per month in months~10--12. In specified circumstances, Visa may prohibit an issuer from participating in the Visa Program if high dispute or fraud volumes cause undue economic hardship or harm to the goodwill of the Visa system~\cite{visa2026corerules}. Separately, a published VisaNet fee schedule lists monthly STIP assurance fees and a per-transaction Smarter STIP fee; these are STIP-related charges, not penalties for 3DS lookup probes~\cite{visanet_stip_fees_2022}. These are conditional potential costs, not automatic consequences of the tested lookup path. In the same attacker-like off-context pattern as a real scripted lookup, our caller never sent the initial \texttt{CReq}, so the subsequent challenge/OOB interaction discussed in the specification and FAQ never began; correspondingly, no SMS, push, or OTP reached our test cards, and the Core Specification does not define a cardholder alert on the lookup-only or missing-initial-\texttt{CReq} path~\cite{emvco_core_2311_2023,emvco_3ds_faq_2026}. Cardholders may also face privacy exposure and potential fraud-related disruption; our lookup-only tests did not measure realized losses.

\begin{table}[t]
\centering
\caption{Multi-Party Stakeholder Conflicts of Interest and Asymmetric Burdens across the 3-D Secure (3DS) Infrastructure.}
\label{tab:multiparty_stakeholders}
\footnotesize
\renewcommand{\arraystretch}{1.12}
\setlength{\tabcolsep}{2.2pt}
\rowcolors{2}{white}{rowzebra}
\begin{tabularx}{\columnwidth}{|>{\raggedright\arraybackslash}p{0.22\columnwidth}|>{\raggedright\arraybackslash}X|>{\raggedright\arraybackslash}p{0.29\columnwidth}|}
\hline
\rowcolor{headerbg}
\textbf{Domain / Party} & \textbf{Economic Priority \& Blindspot (How it Aids Attacker)} & \textbf{Net Asymmetric Burdens} \\
\hline
\textbf{Acquirer Domain} (Merchants, Gateways, 3DSS Providers) & Core business serves merchant checkout conversion and transaction routing. Operates upstream of clearing liability; in the observed configuration, the tested client-facing pre-auth endpoints accepted repeated candidate evaluations. & Minimal direct fraud/reissuance liability; routine per-lookup fees processed regardless of probe validity. \\
\hline
\textbf{Interoperability Domain} (Payment Schemes, DS) & Prioritizes routing velocity \& switching fees. DS stand-in [Req 411] can reveal a conditional prefix-pruning signal at the client-facing boundary; in-band R-98 leaks detection state. & Collects routine network fees on probe traffic; reactive scheme ML pushes detection onto endpoints. \\
\hline
\textbf{Issuer Domain} (Issuing Banks, ACS) & Prioritizes sub-500ms latency SLAs while curbing fraud. Fractured between mainframe checks (leaking Reasons 08, 01, 13) and cloud-hosted ACS challenge proxies; ACS silently times out (Reason 14), keeping probes invisible. & \textbf{Potential issuer burden}: reissuance costs~\cite{aba2015cardreissue}, support disruption~\cite{netcetera2025fraudalert}, and downstream CNP fraud/chargeback exposure; conditional VIMP assessments~\cite{visa2026corerules} and STIP usage fees~\cite{visanet_stip_fees_2022}. Visa rules permit prohibition from Visa Program participation in specified VIMP cases~\cite{visa2026corerules}. \\
\hline
\textbf{Cardholder} (Silent 4th Party) & Prioritizes privacy \& financial security. A caller stopping at the lookup result does not execute the browser challenge; no SMS/push reached test recipients, and EMVCo specifies no cardholder alert at this lookup-only boundary. & Covert profiling (account status, credit tier, bank); elevated fraud risk. \\
\hline
\textbf{Abuse Ecosystem} (Paid 3DS Testing / CaaS) & Paid services and operational code already use pre-authorization outcomes to screen or route candidates~\cite{telegram_3ds_tooling_2026,independent_hosted_payment_abuse_audit_2026}. & CaaS monetizes the signal; issuers and cardholders bear downstream exposure. Market scale and operator revenue are not quantified. \\
\hline
\end{tabularx}
\end{table}

\paragraph{Historical Precedents of Multi-Party Protocol Failures.}
This infrastructure-level failure mirrors foundational precedents in computer and communications security:
(1)~\textbf{Banking-Security Failures and Protocol Oracles:} Anderson's study of banking cryptosystem failures emphasizes implementation and management weaknesses rather than cryptanalytic breaks~\cite{anderson1993whycryptosystems}. Murdoch and Anderson's 3DS analysis separately examines authentication design and liability incentives~\cite{murdoch2010vbv}. Anderson and Needham's CRYPTO 1995 Principle~3 cautions against letting an opponent use a protocol participant as an oracle; we draw a design analogy, since their work concerns public-key protocol operations rather than PAN enumeration through 3DS responses~\cite{anderson1995robustness}. In the tested integration, callers without cardholder authentication can interrogate lookup-stage routing and response state using estimated or guessed credentials~\cite{mastercard2026abnormaltraffic}.
(2)~\textbf{SS7 and Diameter Cellular Signaling:} In mobile telecommunications, national carriers treated the core signaling network as a closed, trusted federation. When commercial roaming brokers, SMS aggregators, and third-party gateways gained access, attackers exploited location routing queries (e.g., Send Routing Info) to turn the global signaling grid into an unauthenticated surveillance and interception oracle without subscriber awareness.
(3)~\textbf{BGP (Border Gateway Protocol):} Internet routing prioritized autonomous domain flexibility and reachability over cryptographic origin validation, resulting in decades of prefix hijacking and traffic misdirection.

\paragraph{The PCI Regulatory Chasm: PCI DSS v4.0.1 vs.\ PCI 3DS Core and Data Matrix Blind Spots.}
This multi-party breakdown is directly enabled by an unaddressed regulatory chasm between the \emph{PCI Data Security Standard (PCI DSS v4.0.1)}~\cite{pci_dss_v4_0_1} and the \emph{PCI 3-D Secure Core Security Standard v1.0}~\cite{pci_3ds_core_v1}. Under PCI DSS v4.0.1 (effective March 2025), Requirement~6.2.4 formally identifies and condemns \emph{``Enumeration attacks---automated attacks that are actively exploited in payments and abuse identification, authentication, or authorization mechanisms,''} while Requirement~6.4.2 mandates automated technical solutions (e.g., WAFs) with \emph{``rate-limiting controls... to mitigate against brute-force attacks and enumeration attacks''} on public-facing web applications~\cite{pci_dss_v4_0_1,pci_ssc_enumeration_2020}. However, PCI DSS is entirely blind to 3DS (zero mentions across 2.2M characters). Under standard merchant assessment procedures (SAQ~A / SAQ~A-EP), merchants embedding client-side gateway tokenization iframes are certified as having outsourced cardholder data processing, placing gateway-hosted 3DS lookup endpoints outside the merchant's QSA audit scope. Conversely, the PCI 3DS Core Security Standard v1.0 governs the 3DS Environment (3DSS, DS, and ACS)~\cite{pci_3ds_core_v1,pci_3ds_assessor_guide_v1}, yet across its entire specification, the terms \emph{enumeration}, \emph{rate limit}, and \emph{brute force} appear zero times. PCI 3DS Core assumes client admission control is enforced outside the 3DE by the merchant. This blind spot is codified by the \emph{PCI 3DS Data Matrix v1.3} (September 2026)~\cite{pci_3ds_data_matrix_v1_3}, which defines protected ``3DS Sensitive Data Elements'' (Table~1) and explicitly establishes that unlisted elements are \emph{``not currently subject to the PCI 3DS Core Security Standard.''} Table~1 completely omits Primary Account Numbers (PANs), transaction identifiers (\texttt{acsTransID}, \texttt{dsTransID}), and outcome reason codes (\texttt{transStatusReason}), while version~1.3 formally removed \texttt{authenticationValue} (CAVV) from sensitive authentication data (Technical FAQ Q1~\cite{pci_3ds_core_faqs_2023}). This architecture produces a multi-party audit paradox: the merchant (SAQ~A-EP), gateway (clearing and 3DSS), scheme DS, and issuer ACS all hold active Attestations of Compliance (AoC), yet their integrated composition exposes a pre-authorization cardholder enumeration surface. This structural context is consistent with the approach disclosed in published patent application US~2026/0134429~A1~\cite{mastercard2026abnormaltraffic}: in the absence of explicit anti-enumeration mandates within 3DS core security standards, the patent describes an anomaly-detection approach at the scheme Directory Server (Method~300, dedicated Risk Score 998, Reason~98). However, because reason codes are excluded from the PCI 3DS Data Matrix Table~1~\cite{pci_3ds_data_matrix_v1_3}, commercial gateway integrations deserialize R-98 directly into client-visible JSON (Figure~\ref{fig:patent_method300_deconstruction}b, Step~6), converting the scheme's defensive signal into an unintended evasion calibration channel.

\paragraph{Systemic Implementation Divergence and the Issuer Trilemma.}
The observations and analytical scenarios suggest trade-offs in how issuers handle lookup-stage results, but they do not establish a universal issuer trilemma. Account-dependent decisions may produce different client-visible response classes; a response difference is not, by itself, proof of account existence. Approving without an account-specific evaluation raises separate questions about the meaning and downstream use of authentication artifacts, which we did not test. Returning a challenge-required result ($\texttt{transStatus=C}$) uniformly may reduce some response differences but replaces frictionless completion with a challenge. We did not complete challenges. Because the off-context caller omitted the initial \texttt{CReq}, the later OOB challenge interaction never began in our tests; no SMS/push/OTP was observed, and EMVCo defines no cardholder alert on the lookup-only path~\cite{emvco_core_2311_2023}. We do not infer OTP volume, account liveness, or telephony effects.

\paragraph{Deconstructing Naive Uniform Masking Proposals.} Industry proposals have suggested mitigating card enumeration by returning ``uniform responses'' or proprietary ``BIN Attack Scores''~\cite{netcetera2025binattackarticle,netcetera2025fraudalert}. However, naive uniform masking introduces severe operational and security hazards:
(1)~\textbf{Stand-In Hazards and Tokenization:} Returning uniform stand-in attempts ($\text{transStatus} = \text{`A'}$) causes Directory Servers to return authentication cryptograms (CAVVs) with ECI 01/06. Integrations that bind payment methods on pre-authorization lookup indicators alone would grant valid continuation tokens to unissued cards, polluting downstream ledgers.
(2)~\textbf{Protocol Invariants Bypass Generic Masking:} Even if status is held uniform at the Directory Server, EMVCo [Req 411] mandates $\text{acsTransID} \equiv \text{dsTransID}$ during stand-in. Under independent UUIDv4 generation ($P < 2^{-122}$), identifier equality distinguishes Directory Server stand-ins from authentic issuer ACS responses, bypassing uniform masking.

\paragraph{A Conditional Interim Stopgap: Range-Level Uniform Challenge Proxying.} As an immediate tactical measure prior to consortium specification revisions, Access Control Servers can deploy \emph{Uniform Range-Level Challenge Proxying} (the ``Blind Challenge URL'' pattern) across participating card ranges. Under this countermeasure:
(1)~\textbf{Suppression of Early Negative Ledger Rejections:} When an AReq is routed to an ACS for any candidate PAN within an active participating range, the ACS strictly suppresses early rejections ($\text{transStatus} \in \{\text{`N'}, \text{`R'}\}$ with failure reason codes, e.g., Reasons 08, 01, 13) during the frictionless AReq phase.
(2)~\textbf{Uniform Blind Challenge Outcome:} The ACS uniformly returns:
\begin{equation}
\begin{aligned}
\text{ARes} = \langle &\text{transStatus} = \text{`C'}, \text{acsURL} = \text{URL}_{\text{range-proxy}}, \\
&\text{acsTransID} = \text{UUIDv4}_{\text{fresh}} \rangle
\end{aligned}
\end{equation}
(3)~\textbf{Blinding the Pre-Auth Enumeration Oracle:} To a standalone off-context HTTP caller (operating precisely as an automated attacker would do by issuing direct scripted queries without a browser runtime), every tested input within the sampled range received the same first-result class: $\text{transStatus} = \text{`C'}$, an active challenge URL, and an independent UUID. Cohort~G showed a uniform first-result class across the tested inputs. Because candidate issuance and internal account states were not independently verified, this demonstrates sampled response uniformity rather than range-wide masking of every account state.
(4)~\textbf{Zero Cryptographic Leakage and Zero Liability Shift:} Unlike stand-in attempts ($\text{`A'}$), returning $\text{`C'}$ generates no CAVV/AEVV authentication cryptograms and grants no liability shift ($\text{liabilityShifted: false}$), preventing unauthorized payment tokenization.

\paragraph{The Zero-Variance Condition and the Breakout Risk.} Range-level challenge proxying is a conditional mitigation proposal, not an enduring or empirically validated resolution. For it to mask the lookup-stage class, responses would need to remain consistent across the relevant input range and fields, not merely share \texttt{transStatus=C}. A hybrid policy---for example, returning frictionless results ($\text{`Y'}$), attempts ($\text{`A'}$), or different reason codes for some inputs---could preserve observable distinctions. The present study does not claim a universal precision threshold or classifier accuracy for issued-account detection. But once a range exhibits stable non-uniform outcomes, even a low-frequency branch can furnish a clear prioritization signal for sub-ranges or candidates more likely to correspond to issued accounts; how strong that signal is in practice still depends on the observed response distribution, unknown candidate ground truth, and deployment controls.

\paragraph{The Structural Dilemma: Unified Common Branches vs.\ Frictionless UX.} A critical examination of EMVCo Core Specification v2.3.1.1 (Sections 2.5.1 and 2.6) reveals why this vulnerability exists: it is the direct architectural consequence of early protocol bifurcation. To satisfy commercial demands for sub-second e-commerce checkout velocity, EMVCo designed the frictionless flow to terminate processing early at the ARes boundary (\emph{``Note: 3-D Secure processing ends here''}), releasing the authentication result and cryptographic proof (CAVV) directly to client-facing APIs before any browser-mediated challenge occurs.

One proposed structural mitigation is to normalize the client-visible lookup result across the relevant range, for example by returning an opaque continuation artifact and deferring account-specific decisions until a protected continuation step. This is a design proposal, not an outcome established by our experiments.
However, under current specification constraints, this creates a severe operational trade-off: forcing all transactions through a common client container introduces network round trips (500--2,000\,ms), triggers third-party storage partitioning blocks in modern privacy-focused browsers (e.g., Apple Safari ITP and Google Chrome CHIPS), and disrupts native mobile SDK checkout flows.

\subsection{A Defensive Vulnerability Taxonomy for Pre-Authorization 3DS Integrations}
\label{subsec:vulnerability_taxonomy}
To synthesize our findings into an actionable framework for payment security researchers, gateway engineers, and standards bodies, we structure the observed attack surface into five distinct vulnerability classes:

\begin{enumerate}
    \item \textbf{3DS-AUD-001: Directory Server Created ARes Boundary Attribution.}
    \begin{itemize}
        \item \emph{Mechanism:} When a transaction is terminated or handled at the scheme routing tier (e.g., candidate PAN outside participating ranges under [Req~97]/[Req~20], no capable ACS under [Req~21], scheme stand-in under [Req~412], or version mismatch under [Req~421]), EMVCo Core Specification v2.3.1.1 Seq~3.155 [Req~411] and FAQ Q34~\cite{emvco_3ds_faq_2026} mandate setting $\text{acsTransID} \equiv \text{dsTransID}$.
        \item \emph{Architectural Root Cause:} Normative identifier duplication across protocol tiers. Notably, the official EMVCo 3DS Implementation Conformance Statement (ICS) Form Manual v1.12 (\S7.2.4)~\cite{emvco_ics_manual_2026} formally defines ``DS able to create the ARes message on the ACS's behalf'' as an active 2.3.1 protocol capability, yet the corresponding test specification (Test Requirements for DS v2.14~\cite{emvco_test_req_ds_2024}) does not specify test cases evaluating the client-side observability or privacy side effects of this identifier mirroring.
        \item \emph{Attribution Impact:} Under RFC~4122 independent entropy bounds ($P < 2^{-122}$), identifier equality supports conditional attribution of a DS-created response under [Req~411]. With a reason code that indicates a routing boundary, it may support pruning a corresponding DS routing block in the analytical model; equality alone does not establish that a PAN is unallocated or inactive.
        \item \emph{Defensive Remediation:} Protocol revisions must decouple DS stand-in identifiers from $\text{dsTransID}$, or gateways must strictly sanitize internal transaction IDs from client-facing payloads.
    \end{itemize}

    \item \textbf{3DS-AUD-002: Pre-Authorization Metadata Overexposure via Gateway Reflection.}
    \begin{itemize}
        \item \emph{Mechanism:} In the tested integration, the lookup-stage JSON included fields corresponding to protocol outcomes and integration metadata, including status, reason, enrollment, ECI, and liability indicators. The captures were client-facing gateway responses, not raw ARes messages, and do not establish direct deserialization of each field.
        \item \emph{Architectural Root Cause:} Integration design anti-pattern prioritizing front-end checkout flexibility over response minimization. This behavior mirrors the architecture of the compliance test harness: EMVCo Test Requirements for 3DS Server as SUT v2.16 (\S3.3.3)~\cite{emvco_test_req_3dss_2024} specifies that the 3DS Server forward the received ARes fields—including \texttt{acsTransID}, \texttt{dsTransID}, and \texttt{transStatus}—to the 3DS Requestor within the test-environment proprietary response (\texttt{pArs}), while EMVCo \S1.3 leaves production client-facing gateway filtering outside core specification scope.
        \item \emph{Attribution Impact:} Exposes response classes and product metadata to an off-context caller. In selected cohorts, response classes differed across tested inputs; the output alone does not establish account existence or activity.
        \item \emph{Defensive Remediation:} Consider returning an opaque continuation token and retaining nonessential protocol and profile fields server-side, while preserving the information needed to select and complete the appropriate authentication branch.
    \end{itemize}

    \item \textbf{3DS-AUD-003: Browser Channel Telemetry Provenance Gap.}
    \begin{itemize}
        \item \emph{Mechanism:} In the Browser Channel (\texttt{02-BRW}), risk-relevant device telemetry (Table~A.1) is collected via unauthenticated client JavaScript without digital signatures or hardware attestation. Furthermore, the 3DS Method is explicitly optional (Seq~5.95 [Req~255]).
        \item \emph{Architectural Root Cause:} Absence of hardware-anchored platform attestation in web standards, contrasting sharply with the Native App Channel (\texttt{01-APP}), where 3DS SDK authenticity is verified under PCI 3DS SDK Security Standard v1.1~\cite{pci_3ds_sdk_v1_1,pci_3ds_sdk_faqs_2022} (Seq~3.9 [Req~7]) and device info is encrypted to the DS via JWE (Section~6.2.2.1), and Split-SDK architectures, where transaction data is signed via JWS (Section~6.2.2.3; EMVCo Test Harness for Split SDK v1.15~\cite{emvco_test_harness_splitsdk_2025}).
        \item \emph{Attribution Impact:} Headless automated callers can supply arbitrary synthetic telemetry payloads that are readily accepted by gateway risk collectors.
        \item \emph{Defensive Remediation:} Integrate W3C Secure Payment Confirmation (SPC) and WebAuthn hardware-backed platform authenticators into browser payment flows to achieve cryptographic client attestation.
    \end{itemize}

    \item \textbf{3DS-AUD-004: Challenge-Abandonment and Notification Scope.}
    \begin{itemize}
        \item \emph{Mechanism:} Only an initial challenge-required result (\texttt{transStatus=C}) can leave an ACS session awaiting \texttt{CReq}. If the off-context caller stops at that point, it does not execute the browser challenge UI; the session may time out.
        \item \emph{Architectural Root Cause:} The initial lookup result branches into frictionless completion, challenge continuation, or another outcome. FAQ Q93~\cite{emvco_3ds_faq_2026} describes version-specific handling of the standard timeout \texttt{CRes} notification.
        \item \emph{Attribution Impact:} In the observed off-context path, the caller stopped before sending \texttt{CReq}, so no SMS, push, or OTP reached the test recipients. EMVCo's specified initial-\texttt{CReq} timeout report is an ACS-to-DS \texttt{RReq}, and its OOB push example belongs to the subsequent challenge interaction; neither defines a cardholder alert on the lookup-only path~\cite{emvco_core_2311_2023}. The patent contemplates network notifications to the issuer/acquirer, optionally by email, not cardholder alerts~\cite{mastercard2026abnormaltraffic}. Independent issuer notification systems were not instrumented.
        \item \emph{Defensive Remediation:} Correlate challenge-required lookup results with subsequent \texttt{CReq}/\texttt{CRes} events in authorized server-side logs; interpret abandonment in context rather than treating every omitted continuation as automated probing.
    \end{itemize}
    \item \textbf{3DS-AUD-005: Scheme Signaling Feedback Leakage.}
    \begin{itemize}
        \item \emph{Mechanism:} Attack-detection labels and reason codes generated by scheme-tier ML models (such as Reason~98 / R-98 under Mastercard Patent US 2026/0134429 A1 [0040]) are surfaced to client-facing return payloads.
        \item \emph{Architectural Root Cause:} In-band propagation of defensive security signals through the standard payment response payload rather than isolating them to out-of-band banking feeds.
        \item \emph{Attribution Impact:} The defensive tripwire becomes the adversary's tuning oracle: scripted callers observe exactly when probe rates exceed detection thresholds, enabling iterative evasion calibration.
        \item \emph{Defensive Remediation:} Scheme anti-abuse reason codes must be strictly classified as operational backend signals and filtered out of any client-facing response surfaces.
    \end{itemize}
\end{enumerate}

\subsection{Architectural Trade-offs: Scheme Heuristics, Tactical Stopgaps, and Consortium Reform}
\label{subsec:political_economy}
Published Application US 2026/0134429 A1 exposes a three-tier remediation trade-off: processing-network scoring of feature data and velocity signals, issuer-side conditional stopgaps, and eventual consortium reform~\cite{mastercard2026abnormaltraffic}:
(1)~\textbf{Tier 1: Post-Hoc Processing-Network Heuristics (Mastercard US 2026/0134429 A1):} Deploying Directory Server models that score feature data and velocity signals at the scheme tier~\cite{mastercard2026abnormaltraffic} flags anomalies via in-band reason codes (R-98) without consortium consensus; however, as shown in Section~\ref{subsec:deconstruct_patent}, this approach faces structural limits under distributed probing and other forms of traffic dispersion, while in-band signaling can inform the caller about detection state.
(2)~\textbf{Tier 2: Conditional ACS Mitigation (Range-Level Challenge Policy):} A uniform challenge-required result ($\texttt{transStatus=C}$) across a defined range is one possible way to reduce lookup-stage response-class differences without an immediate specification revision. It would not, by itself, establish account status or guarantee that all response fields are indistinguishable; effectiveness depends on consistent implementation and evaluation across branches. It also replaces frictionless completion with an interactive challenge for the affected range.
(3)~\textbf{Tier 3: Fundamental Consortium Protocol Reform (EMVCo Standards Overhaul):} An enduring architectural resolution cannot rely solely on tactical heuristics or ACS proxying. It requires consortium-level protocol updates across future EMVCo specifications. This intervention is exceptionally timely: EMVCo 3DS Approval Bulletin No.~31 (June 2026) has scheduled the sunset of protocol 2.2.0 testing for July 2027 (and ecosystem cutoff by July 2029), while Approval Bulletin No.~32 (July 2026) confirms that the draft of EMV 3DS Core Specification v2.4.0.0 is currently under active review with final publication scheduled for Q1 2027. Consortium protocol reform should target four core areas:
(1)~\textbf{Decoupling Stand-In Identifiers ([Req 411]):} Revising the specification so that Directory Server stand-in generation does not mirror the DS Transaction ID ($\text{acsTransID} \equiv \text{dsTransID}$), thereby removing one client-visible tier-attribution cue.
(2)~\textbf{Semantic Response Minimization:} Considering opaque continuation tokens and limiting client-visible reason codes and metadata where they are not needed to continue the selected authentication branch; such fields may expose response-class differences, but do not by themselves prove account existence or enrollment.
(3)~\textbf{Hardware-Backed Client Attestation (W3C Secure Payment Confirmation / WebAuthn):} Incorporating hardware-backed platform authenticators into the authentication branch, allowing verification to bind to verifiable browser execution context without releasing pre-authorization API signals to standalone off-context automated callers.
(4)~\textbf{Response-Normalized Ledger Clearing Integration:} Coupling 3DS authentication proofs to an active zero-dollar authorization hold ($\text{ISO 8583 0100}$, response $\text{`00'}$) before payment method binding or tokenization. Crucially, \emph{zero-dollar authorization is not an oracle-free panacea on its own}: it introduces secondary side channels including ISO 8583 decline code oracles, 300--800\,ms mainframe timing disparities, real-time banking push alerts, and assessment fees/VAMP thresholds. Zero-dollar clearing is therefore viable only when paired with (a)~\emph{Universal Authorization Response Normalization} (collapsing clearing declines into constant-time opaque rejections) and (b)~\emph{Pre-Flight Client Attestation} (W3C SPC/WebAuthn), preventing automated scripts from triggering clearing calls.

\subsection{Limitations and Scope Boundaries}
Our findings are strictly bounded to maintain scientific rigor and ethical proportionality:
(1)~\textbf{Zero Aggregate Disclosure:} To prevent operational intelligence exposure, we report structural bounds, mathematical collision probabilities ($P < 2^{-122}$), and normative invariants without disclosing confidential aggregate probe tallies.
(2)~\textbf{Initial Lookup Scope and Challenge Continuation:} Live probes targeted lookup endpoints and submitted no financial authorizations or charges. We observed client-facing integration results associated with the initial \texttt{AReq}/\texttt{ARes} exchange, not raw ARes messages. A frictionless result can complete at this stage; \texttt{transStatus=C} requests a later challenge and is not a final authentication result. Our attacker-like off-context caller, operating without the browser SDK, stopped before any \texttt{CReq}/\texttt{CRes}; the challenge/OOB branch therefore never started in our tests, and no SMS/push/OTP reached our cards. We did not instrument independent issuer notification systems, and we do not infer candidate issuance or final authentication outcomes. (3)~\textbf{ACS Decision Opacity and Bank Variance:} As an external study, we lack visibility into proprietary issuer ACS algorithms and correlated internal 3DS Server/DS/ACS traces. The observed reason/status classes and transaction identifiers vary across cohorts, but they do not by themselves establish a particular backend architecture or a universal mapping from \texttt{transStatus=C} to account activity. The 3DS Method is optional under EMVCo v2.3.1.1 Section~5.8.1 (Seq~5.95 [Req~255]); when omitted, the ACS may receive less prior browser telemetry, but its decision process remains unobserved in this study. (4)~\textbf{Browser Channel Scope vs.\ Native Mobile SDKs:} Our live empirical validation specifically targets the Browser Channel (\texttt{02-BRW}). As analyzed in Section~\ref{subsec:channel_differences}, native mobile app channels (\texttt{01-APP}) enforce SDK authenticity verification (Seq~3.9 [Req 7]), direct JWE device telemetry encryption to the Directory Server (Section~6.2.2.1), and ACS-signed challenge payloads (\texttt{acsSignedContent}), while Split-SDK architectures mandate JWS server signatures validated by the DS CA (Section~6.2.2.3). Consequently, our findings establish an integration-layer reflection vulnerability within browser-mediated checkouts, rather than an end-to-end cryptographic compromise of native mobile SDK protocols. (5)~\textbf{Cross-Network Generalization:} Our cross-scheme observations cover Visa, Mastercard, and American Express, not every card network or every combination of 3DS Server, Directory Server, and ACS. We therefore make a firm protocol-level claim about the shared AReq/ARes client-facing boundary and the conditions under which it can become a classification oracle; we do not claim that all six networks have been empirically tested or that every network currently exposes a useful signal. Establishing prevalence and exploitability for each remaining network requires implementation-specific testing. Response normalization or other controls may prevent the shared protocol boundary from producing an exploitable distinction.

Evidence across schemes is tiered, not balanced: the primary live validation is the Braintree/Cardinal path, Stripe and Adyen recurrence checks are narrower, and the wider scheme mapping also draws on exploratory observations, specifications, and public documentation. These evidence types are not interchangeable.

The discussion therefore points to a simple conclusion: where lookup-stage outputs vary with sensitive account-specific processing, response minimization and consistent branch handling are important architectural safeguards.

\section{Conclusion}
\label{sec:conclusion}
This paper examines the first client-visible lookup-stage result in a tested EMV 3DS integration. The result is a branch point, not a single pre-challenge signal: frictionless authentication may complete through the initial exchange, while \texttt{transStatus=C} requests a later challenge. In our off-context test, the caller stopped after the lookup result; it did not execute a challenge or submit financial authorization. Selected cohorts showed lookup-stage response-class separation between researcher-owned anchor cards and bounded adjacent comparison PANs, but these observations do not independently establish candidate issuance, account activity, or final authentication outcomes.

Our protocol analysis shows that Req~411's ACS/DS identifier equality can support conditional attribution when a DS creates an ARes on an ACS's behalf, assuming correct field mapping and specification conformance. Paired with a reason code that indicates a routing boundary, this invariant can support a model of coarse range pruning. The complexity reduction is analytical and applies to DS routing blocks under those assumptions; it is not evidence that a single lookup can classify arbitrary PANs or that the issuer has no visibility. In our tests, no SMS, push, or OTP reached our recipients because the off-context caller never sent the first \texttt{CReq}; the specified timeout report therefore remained the ACS-to-DS \texttt{RReq}, while OOB push belongs to a later challenge interaction. The protocol defines no cardholder alert at the tested lookup-only boundary~\cite{emvco_core_2311_2023}. We did not instrument independent issuer notification systems.

The structural weakness lies at the client-to-integrator boundary: EMV 3-D Secure underspecifies this interface and does not govern it with the same tightly defined security semantics as the DS/ACS core exchange, yet the tested integration returned identifier-linked pre-routing outcomes to an off-context caller before challenge completion or financial authorization. In contrast to Ali et al.'s authorization-stage guessing of expiry/CVV for a known PAN, the measured risk here is that an attacker can algorithmically construct Luhn-valid candidates and use the authentication rail as a pre-authorization oracle, with one-card compromise serving only as a stronger within-range prioritization aid rather than a universal prerequisite. The same signal can also expose issuer-internal participating sub-range structure beneath a 6-digit BIN/IIN range, allowing an attacker to focus subsequent probing on sub-ranges more likely to correspond to issued accounts. Reviewed public artifacts show real-world card-testing tools parsing these signals to screen or route candidates in CaaS workflows, including batch processing and rotating-proxy support~\cite{rapid7_caas_2026,telegram_3ds_tooling_2026,independent_hosted_payment_abuse_audit_2026}, demonstrating that the abuse is already operationalized in the wild even though its prevalence, success rates, and financial impact remain unquantified. The bounded live design should not be mistaken for attacker-side infeasibility; it is an ethical limit on the researchers, not evidence that attackers face the same ceiling. This architecture shifts asymmetric burdens onto issuing banks and cardholders, while routing intermediaries prioritize checkout conversion.

The findings motivate minimizing lookup-stage fields and handling frictionless and challenge branches consistently. Opaque continuation tokens or uniform challenge policies may reduce some client-visible distinctions, but they require evaluation across integrations and can add friction for legitimate users. The study does not establish a universal 3DS account-validity oracle or network-wide prevalence, but instead identifies a critical response-boundary exposure and a fundamental design concern for client-facing integrations. Thirty years after Anderson and Needham warned against opponent-controlled oracle use, our findings demonstrate that payment authentication protocols risk recreating this structural vulnerability at the ecosystem level. Addressing it requires treating the client-facing merchant/requestor boundary as a security-critical boundary within the global payment architecture.

\begin{samepage}
\vspace{2pt}
\noindent
{\centering
  {\small\bfseries
  ``Be careful when signing or decrypting data that you never let yourself be used as an oracle by your opponent.''\par}
  \smallskip
  {\footnotesize\upshape---Ross Anderson and Roger Needham, Principle 3~\cite{anderson1995robustness}\par}}
\vspace{2pt}

\end{samepage}

\section{Ethical Consideration}
\label{sec:ethics}
Because this paper studies a live payment-security boundary with dual-use implications, ethical restraint and disclosure discipline are part of the methodology. We conducted this research under established security-research ethics principles, including the Menlo Report~\cite{menlo2012}. Our operational boundaries were further informed by the DOJ's 2022 CFAA safe-harbor policy for good-faith security research~\cite{doj2022cfaa} and Harvard Law School's legal-risk guidance~\cite{harvard2024guide}.

\textbf{Intent and attribution.} This work evaluates a client-facing integration within the broader 3DS routing architecture; it does not allege that any gateway, payment network, merchant, or issuer intentionally created a vulnerability or acted improperly. We do not infer system-wide behavior or attribute internal processing to a gateway, 3DS Server, DS, or ACS without direct evidence. The term \emph{oracle} describes query--response behavior, not design intent. All analyses of published standards (EMVCo, PCI SSC), patent applications (Mastercard US~2026/0134429~A1), and vendor SDK documentation are conducted strictly for peer-reviewed academic inquiry and public security education under fair use and freedom of scientific expression. Our technical evaluation of protocol boundaries and architectural trade-offs implies no product disparagement, commercial malice, or breach of legal duty by any named entity, standards body, or financial institution.

\textbf{Stakeholder analysis.} We identified four stakeholder groups and defined explicit safeguards for each:
(1)~\textbf{Cardholders:} Probes used researcher-controlled accounts as anchors; live comparisons were limited to a small bounded set of adjacent Luhn-valid PAN offsets within the same participating ranges. No cardholders were recruited and no identity-linked records collected.
(2)~\textbf{Merchants:} Tests used the researcher-controlled merchant checkout to obtain valid merchant-scoped \texttt{client\_token} authorization, reused unchanged without account access or token forgery. We did not access merchant dashboards, settle payments, or test cross-merchant replay.
(3)~\textbf{Gateways \& Networks:} Traffic was restricted to bounded checks at the client-facing integration boundary; we did not probe DS or ACS infrastructure directly.
(4)~\textbf{Financial Institutions:} All reported artifacts omit raw PAN/BIN identifiers and avoid issuer-system access.

\textbf{Coordinated disclosure and dual-use analysis.} The disclosure chronology is in Appendix~\ref{app:disclosure_timeline} (Table~\ref{tab:app_disclosure_timeline}); raw PAN/BIN data remain withheld. We purchased no illicit CaaS offerings, operated no card-testing tools, and conducted no large-scale live experiments, aligning with dual-use research guidance~\cite{rescorla2005disclosure}.

\textbf{Statement on vendor response and availability.} On 14 April 2026, we submitted the initial gateway-integration report to PayPal through HackerOne (\texttt{\#3673424}, closed as ``Informative''), followed by network-tier disclosures to Visa HackerOne (\texttt{\#3978615}, closed as ``Informative'' under intended routing design reasoning), American Express HackerOne (\texttt{\#3979021}, under active program review), and Mastercard Bugcrowd (\texttt{8d647a18-e0b0-4c2e-8fd4-7840bbd82e49}, closed as out of scope for isolated product bounty triage), as well as filing a Private Vulnerability Report via GitHub Security Advisories (\texttt{GHSA-xjwr-jwm2-wvjr}) with the maintainers of \texttt{braintree/braintree-web}, which received no maintainer response. Across coordination attempts, participating entities maintained that individual components functioned according to their respective specifications, treating the cross-party integration side channel as an out-of-scope composition issue and prompting our escalation to CERT/CC. Following the 90-day disclosure window and CERT/CC's publication recommendation, we initiated direct protocol-level coordination across the payment architecture: submitting a formal specification inquiry directly to the standards consortium (EMVCo Query \texttt{\#20260827-17621D}), where the EMVCo 3DS Team noted submissions are processed under general non-confidential terms, and conducting direct technical outreach in September--October 2026 across major card schemes evaluated in our study (Visa Secure, Mastercard Identity Check/Standards, and American Express SafeKey). Visa and Mastercard contacts confirmed internal forwarding to their respective authentication, standards, and fraud-risk engineering teams, while American Express routing was requested through scheme security intake. Sensitive artifacts remain non-public pending responsible release; formal ProVerif models and sanitized logs (omitting full PANs/BINs) will be made available. We do not publish replay tooling against third-party merchants.

\vspace{4pt}
\section*{Generative AI Usage}
\vspace{3pt}
Generative AI tools assisted with formatting and editorial clarity; all technical claims, formal models, and final wording were verified by the authors.

\vspace{4pt}
\section*{Acknowledgments}
\vspace{3pt}
{\fontsize{9.5}{10.5}\selectfont
\setlength{\parskip}{1pt}
We thank S.~M.~Taiabul Haque and, in particular, Syed Rafiul Hussain for their extensive feedback on an earlier draft and assistance with writing. We also thank the CERT Coordination Center (CERT/CC), part of the Software Engineering Institute at Carnegie Mellon University, for its correspondence during the disclosure process and its recommendation that we proceed with public disclosure.

We dedicate this work to Mohammed Aamir Ali, whose research on online-payment security and 3-D Secure at Newcastle University inspired this study; to the memory of Professor Ross Anderson; to good-faith security researchers whose responsible work helps improve online-payment security; and to everyone who continues meaningful work despite setbacks. The title's use of ``oracle'' is a deliberate tribute to Professor Anderson's memorable ``elf sitting in a black box'' analogy for the random-oracle model, and to his coauthored analysis of 3-D Secure with Steven J. Murdoch.\par
}

\bibliographystyle{IEEEtran}
\bibliography{paper}

\appendices

\section{EMV 3DS Terminology Used in This Paper}
\label{app:emv3ds_terminology}
These definitions explain the terms as used in this study. EMVCo meanings apply to protocol fields; client-facing API values are described only as observed in the tested integration~\cite{emvco_core_2311_2023}.

\begin{itemize}
\setlength{\itemsep}{2pt}
\setlength{\parsep}{1pt}
\setlength{\topsep}{2pt}
\item \textbf{3DS authentication.} The authentication flow precedes financial authorization and may finish frictionlessly or continue into an interactive challenge. It does not itself perform payment authorization or clearing.
\item \textbf{Account existence and credential validity.} Account existence means that the issuer/ACS recognizes a PAN as associated with an account record at lookup; the account may nevertheless be expired, blocked, or otherwise unusable. Credential validity is the stronger claim that the presented credentials are current and usable in the relevant authentication and payment flow. A 3DS result can inform account-existence filtering but does not itself prove payment authorization or full credential validity.
\item \textbf{AReq/ARes and CReq/CRes.} AReq means Authentication Request: the initial message that starts a 3DS authentication. ARes means Authentication Response: the reply, which may indicate that more cardholder interaction is needed. In a challenge flow, CReq means Challenge Request and carries the request to begin or continue the challenge; CRes means Challenge Response and is the ACS's response in that exchange. These are protocol messages; the HTTP lookup and JSON result studied here are integration artifacts, not captured AReq/ARes messages.
\item \textbf{Initial lookup-stage result and branches.} This paper uses \emph{lookup-stage result} for the first client-visible integration response associated with the initial \texttt{AReq}/\texttt{ARes} exchange; it is not a raw ARes or a named EMVCo phase. The result is a branch point: a frictionless outcome completes without an interactive challenge; \texttt{transStatus=C} means the protocol status is ``challenge requested,'' not completed; other values may indicate a negative, unavailable, or attempted outcome. A caller stopping at this point has not completed any requested challenge. The status alone does not establish account existence or activity.
\item \textbf{Braintree lookup callback.} Braintree JavaScript v3 exposes \texttt{onLookupComplete(data, next)} after the lookup response; calling \texttt{next()} continues the SDK flow, including challenge handling when required. This is a callback/control-flow hook, not a \texttt{next\_action} response field or an EMVCo message/status. \texttt{verifyCard()} describes the broader verification flow~\cite{braintree_3ds_guide,braintree_web_3ds_docs}.
\item \textbf{3DS Server, DS, and ACS.} In the protocol path analyzed, the 3DS Server sends an AReq through the Directory Server (DS), which routes by participating card range or may return a permitted stand-in result; the issuer's Access Control Server (ACS) evaluates authentication. The tested internal route was not directly traced.
\item \textbf{Status and reason.} The protocol field \texttt{transStatus} reports the authentication outcome; its value \texttt{C} means a challenge has been requested. \texttt{transStatusReason} supplies a reason when one is present. Gateway fields such as \texttt{status}, \texttt{enrolled}, and \texttt{lookup} are not assumed equivalent without a documented mapping. In the reproduced traces, the integration label \texttt{challenge\_required} accompanied a mapped \texttt{transStatus=C}; this is an observed mapping, not a universal equivalence.
\item \textbf{Transaction identifiers.} \texttt{threeDSServerTransID}, \texttt{dsTransID}, and \texttt{acsTransID} identify the transaction at their protocol roles. Req~104 assigns the ACS Transaction ID at the ACS; when the DS creates an ARes on the ACS's behalf, Req~411 specifies ACS/DS identifier equality. Under correct mapping and specification conformance, inequality (\texttt{acsTransID} $\neq$ \texttt{dsTransID}) supports ACS-path attribution, while equality supports conditional DS stand-in attribution. For an attacker, that distinction helps interpret whether a client-visible signal is more consistent with issuer-side ACS evaluation or with an earlier scheme-level routing/stand-in result. These are provenance inferences; neither relation alone proves account existence or validity.
\item \textbf{PAN and BIN/IIN.} PAN means Primary Account Number; BIN/IIN and card-range references denote the prefix/range used in the tested routing comparison. Luhn validity establishes checksum conformance only, not issuance, account existence, or credential validity.
\item \textbf{Carding-as-a-service (CaaS).} Commercial offerings that bundle card data, card-testing tools, and support; not every related repository or tool is itself a paid service. Reviewed public artifacts document advertised services and 3DS-related tooling for classifying candidate lookups, but do not establish candidate validity, service reliability, prevalence, or success rates~\cite{rapid7_caas_2026,telegram_3ds_tooling_2026,independent_hosted_payment_abuse_audit_2026}.
\item \textbf{ECI and CAVV.} Electronic Commerce Indicator and Cardholder Authentication Verification Value fields observed in integration output. Their presence alone does not establish authorization, clearing, or cryptographic validity; downstream use was not tested.
\item \textbf{Off-context HTTP caller.} The scripted client used in the experiment, which submitted requests without running the genuine browser SDK. Client-supplied session identifiers and synthetic telemetry were accepted; browser-session binding and downstream use were not assessed.
\item \textbf{Traffic dispersion / residential SOCKS5 egress.} In this paper, \emph{traffic dispersion} means spreading requests across multiple network egress identities so source-concentration features are harder to trigger. Where the text refers to \emph{distributed residential SOCKS5 egress} or \emph{residential proxies}, it means routing the tested Cardinal device-data path through third-party residential proxy endpoints that exposed different apparent source IPs. This is an observed network-routing condition in the bounded exploratory A/B check, not a claim about general evasion success, throughput, or all deployment paths.
\end{itemize}

\section{Public Card-Checking Artifact}
\label{app:public_card_checking_artifact}
\begin{figure}[!h]
\centering
\includegraphics[width=\columnwidth]{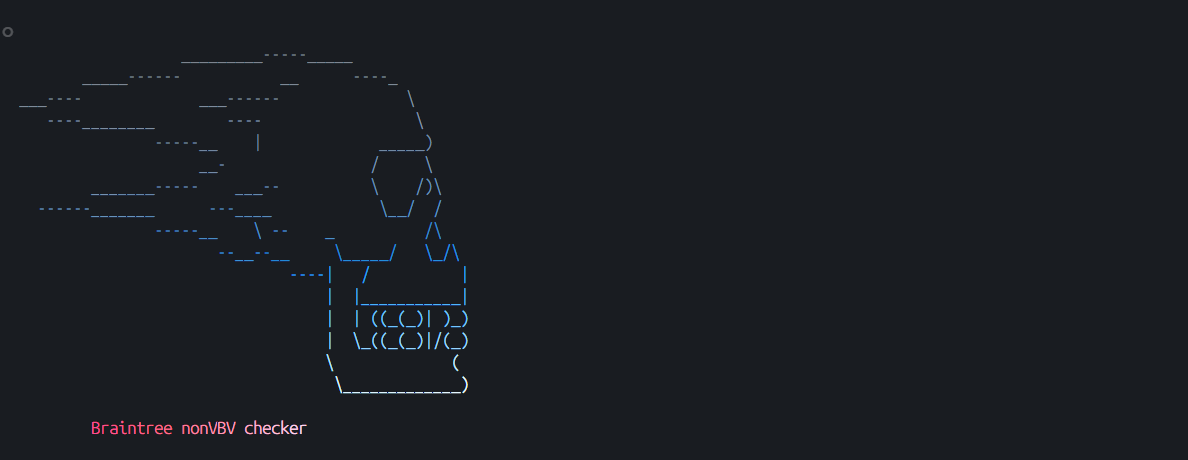}
\Description{Redacted terminal banner from the third-party BraintreeCHK card-checking tool; surrounding output is omitted.}
\caption{Redacted crop of a screenshot reproduced from archived public repository materials associated with the third-party BraintreeCHK GitHub repository~\cite{braintreechk_repo_2026}. Numerical counters and surrounding candidate and identifying data are omitted.}
\label{fig:braintreechk_status}
\end{figure}

Figure~\ref{fig:braintreechk_status} shows a tightly cropped terminal banner reproduced from publicly accessible repository materials preserved in an archived snapshot of the third-party BraintreeCHK GitHub repository under the account KianSantang777~\cite{braintreechk_repo_2026}. In that public presentation, the repository described BraintreeCHK as a card-checking tool and advertised Braintree VBV/3DS lookup functionality. These are claims made by the repository itself; we did not execute the tool or independently verify the advertised functionality, successful use, or operational reliability. Such functionality is consistent with card-testing workflows, but the artifact alone does not establish how the tool was used. The tool was not developed by the authors. Numerical counters and surrounding candidate and identifying data are omitted. This artifact is cited solely to document public presentation of attacker-advertised capabilities, not classification accuracy, candidate validity, prevalence, or real-world success.

\section{Client-Visible Data Domains}
\label{app:client_visible_data_domains}
Table~\ref{tab:client_visible_data_domains} provides the full mapping of observed client-visible fields to their documented protocol roles and the limits on the inferences they support. Protocol roles are not verified field provenance: aside from the ARes captured in the separate exploratory review, we captured no correlated internal messages or logs. The attacker-value column describes bounded information exposure, not a demonstrated attack outcome.

\begin{table*}[t]
\centering
\footnotesize
\caption{Full map of client-visible values, documented 3DS roles, and bounded attacker value exposed directly to a standalone off-context HTTP caller (operating precisely as an automated attacker would without browser SDK execution). Roles follow EMVCo; the reason~98 discussion also draws on the published application~\cite{emvco_core_2311_2023,mastercard2026abnormaltraffic}.}
\label{tab:client_visible_data_domains}
\setlength{\aboverulesep}{0pt}
\setlength{\belowrulesep}{0pt}
\renewcommand{\arraystretch}{1.0}
\setlength{\tabcolsep}{3pt}
\rowcolors{2}{white}{rowzebra}
\begin{tabularx}{\textwidth}{@{}p{0.24\textwidth}p{0.34\textwidth}X@{}}
\toprule
\rowcolor{headerbg}
\textbf{Client-visible value(s)} & \textbf{Protocol map (ACS / DS / integration)} & \textbf{Attacker value (bounded)} \\
\midrule
Authentication outcome: \texttt{transStatus}, \texttt{status}, \texttt{enrolled}, \texttt{authenticate\_\allowbreak successful}, \texttt{paresStatus}; version: \texttt{threeDSecureVersion} & The ACS evaluates authentication and may return an ARes outcome; the DS checks range support and routes requests. Some outcomes may be DS-defined. Several API fields are integration labels, not same-named ARes elements. & First-result classes differed between a known-issued anchor card and adjacent candidates across separating cohorts (while uniform challenge proxies masked outcomes). Candidate issuance was not independently established, so this is response separation, not a card-validity classifier. Version/enrollment labels alone do not establish account existence. \\
Reason: \texttt{transStatusReason}, especially reasons~98 (DS-defined) and~08 (ACS-defined) & An ARes reason may come from the ACS or be DS-defined; the published design describes both ACS R-98 and DS-side scoring. & Reason~98 appeared during repeated within-range Mastercard lookups, but not in the distinct ACS-associated Mastercard cohorts examined under SOCKS5 ingress. Reviewed Visa same-range captures consistently carried reason~08, a potentially clearer coarse response-filtering signal in these observations. Neither pattern establishes PAN validity, component attribution, or detector activity. \\
ECI/liability: \texttt{eciFlag}, \texttt{liabilityShifted}, \texttt{liabilityShiftPossible} & ECI may be provided by the ACS or DS. Liability-shift booleans are payment-system/integration indicators, not ARes elements. & Exposes authentication/liability posture before authorization; does not establish authorization approval, actual liability transfer, or predictive value. \\
Authentication value: \texttt{cavv} & The payment-system-specific authentication value may be generated or carried by the ACS or DS; \texttt{cavv} is the observed integration label. & One challenge-free control returned a non-empty value. It was not validated or tested in authorization, so downstream utility is unknown. \\
Transaction identifiers: \texttt{acsTransactionId}, \texttt{dsTransactionId}, \texttt{threeDSecureServer\allowbreak TransactionId}, \texttt{threeDSecure\allowbreak Authentication\allowbreak Id} & The ACS, DS, and 3DS Server identifiers correspond to protocol roles; \texttt{threeDSecure\allowbreak Authentication\allowbreak Id} is an integration field. & May support correlation with another data source; identifiers are not outcomes, credentials, or proof of detector activity. \\
BIN/product profile: \texttt{details.bin}, \texttt{cardType}, \texttt{binData.issuingBank}, \texttt{countryOfIssuance}, \texttt{productId}, \texttt{debit}, \texttt{prepaid}, \texttt{consumer} & Issuer/product metadata, not ARes fields. The product-profile supplier was not identified and is distinct from standardized DS Card Range Data. & Can group or prioritize probes by issuer/product attributes; metadata does not show that a PAN was issued or is active, and may be available from external BIN sources. \\
Challenge/return fields: \texttt{acsUrl}, \texttt{termUrl}, \texttt{md}, \texttt{pareq} & \texttt{acsUrl} identifies the ACS challenge endpoint; other fields are integration/compatibility values and do not show that a challenge completed. & Reveals challenge-flow metadata. The observed \texttt{termUrl} contained a redacted merchant-scoped authorization JWT; cross-merchant use was not tested. \\
\bottomrule
\end{tabularx}
\end{table*}

\begingroup
\linespread{0.95}\selectfont
\setlength{\parskip}{0.5pt plus 0.5pt minus 0.25pt}
\flushbottom
\section{Research and Coordinated Disclosure Timeline}
\label{app:disclosure_timeline}
This appendix records selected milestones from the multi-year investigation and disclosure chronology from its conceptual work beginning in 2018 through coordination milestones concluding October 15, 2026. Between 2020 and 2025, research focused on longitudinal specification audits across EMVCo v2.1--v2.3 releases and passive observation of publicly advertised card-testing services and public 3DS-related code; the CaaS literature provided broader threat context. The research maintained strict non-interaction with illicit actors or services. Outcomes are reported as received and distinguish technical dispositions from procedural declines and acknowledgements.

\paragraph{Intended-by-Design Paradox and Ecosystem Triage.} The recurring pattern across coordination attempts was the friction of addressing composed cross-boundary risk across independent stakeholders: the PayPal HackerOne report (\texttt{\#3673424}) received an ``Informative'' triage disposition, while Visa's HackerOne program similarly closed the lookup-stage 3DS report (\texttt{\#3978615}) as ``Informative'' under the reasoning that an attacker is presumed to already possess compromised card numbers and that client-facing lookup endpoints function as designed to route authentication challenges without a tangible breach of the CIA triad. In our response to triage, we clarified that the broader threat model does not require a compromised-card dump or breach-sourced corpus: an attacker can algorithmically generate Luhn-valid candidate PANs and use lookup-stage responses to classify them, while our live experiment specifically paired researcher-owned anchor cards with a bounded set of adjacent, same-range, Luhn-valid PANs to test the stronger adjacent-anchor variant. These candidates were not drawn from a compromised-card dataset, and their issuance status was not independently verified. We argued that observable lookup-stage response discrepancies can distinguish invalid inputs from responses associated with real, un-enrolled, or challenge-routable accounts before authorization, constituting a direct confidentiality violation under the CIA triad; a parallel scheme-level vulnerability report to American Express HackerOne (\texttt{\#3979021}) was accepted into active program triage; a subsequent GitHub Security Advisory (\texttt{GHSA-xjwr-jwm2-wvjr}) on \texttt{braintree/braintree-web} remained unacknowledged in maintainer triage without response, VulnCheck applied intended-behavior reasoning, and CERT/CC declined coordination while recommending direct public disclosure. The parallel CISA case was recorded as an automated acknowledgement. In parallel, our early preliminary research evaluated an isolated gateway integration path (client-token rebinding with $N=5$ live instruments). Methodological reflection and academic peer discussion highlighted that treating the phenomenon as a single-vendor implementation bug overlooked broader scheme-level defense interactions (such as Reason~98) and failed to assess protocol-wide prevalence. This directly motivated our expansion from isolated gateway triage to the comprehensive, multi-scheme EMV 3DS architectural analysis presented in this monograph. Despite the presentation of reproducible empirical artifacts and evidence of active in-the-wild carding exploitation, individual participants maintained that their respective components conformed to specifications, illustrating an intended-by-design paradox: behavior is treated as intended within an isolated component while cross-party anti-oracle properties remain unassessed and unmitigated. This paradox is rooted in the operational scope of the EMVCo Type Approval process, which evaluates syntactic interoperability rather than adversarial resistance or side-channel leakage: under Test Requirements for all SUT v2.14 (\S3.2.1), test account ranges explicitly fail the Luhn Mod-10 algorithm to avoid live account collisions, restricting laboratory evaluation to static mock profiles; Test Requirements for DS v2.14 (\S3.1.6) instructs Directory Servers to temporarily disable message limits during certification to avoid test aborts; and EMVCo Administrative Process v16 (\S1.1) and Testing FAQ Q10 explicitly define a Letter of Approval (LOA) as demonstrating syntactic interoperability on the date of testing while disclaiming all warranties of security. Consequently, a component may be certified as fully conformant with EMVCo specifications without being evaluated for cross-party account enumeration or pre-authorization oracle behavior. We distinguish this pattern from an explicit universal finding; we do not claim every recipient made that determination or that dispositions validate or refute our technical results.

\flushcolsend
\begin{table*}[t]
\centering
\footnotesize
\caption{Selected research and disclosure milestones through October 15, 2026, for the studied 3DS boundary issue. Outcomes are recorded as received; case-management dispositions are not independent determinations of the technical claim.}
\label{tab:app_disclosure_timeline}
\setlength{\aboverulesep}{0pt}
\setlength{\belowrulesep}{0pt}
\renewcommand{\arraystretch}{1.0}
\setlength{\tabcolsep}{3pt}
\rowcolors{2}{white}{rowzebra}
\begin{tabularx}{\textwidth}{@{}>{\raggedright\arraybackslash}p{0.13\textwidth}>{\raggedright\arraybackslash}p{0.20\textwidth}>{\raggedright\arraybackslash}X>{\raggedright\arraybackslash}p{0.24\textwidth}@{}}
\toprule
\rowcolor{headerbg}
\textbf{Date} & \textbf{Channel} & \textbf{Milestone} & \textbf{Written outcome} \\
\midrule
Since 2018 & Independent research initiated by Wasif Faisal & Conceptual literature review and threat modeling, followed by preliminary empirical testbed work. & Foundational security measurement framework established; protocol-level vulnerability research and defensive modeling ongoing. \\
2020--2025 & Longitudinal Specification \& Threat Audit & Passive audit of published EMVCo specifications (v2.1--v2.3), vendor integration documentation, and public adversarial threat intelligence identified by the authors; strictly observational with no financial purchases, actor interactions, or illicit infrastructure access. & Threat landscape contextualized; protocol-level compositional hypotheses formulated across multi-scheme specifications. \\
14--15 Apr. 2026 & PayPal HackerOne (\texttt{\#3673424}) & Initial gateway-integration report submitted on 14 April, covering the Braintree/Cardinal browser integration; vendor triage closed it after initial review. & Closed as ``Informative'' (standard integration behavior within component scope). \\
16 Apr. 2026 & MITRE CVE Program & Initial CVE assignment request submitted following vendor closure; formal intake acknowledged. & Assigned tracking reference \texttt{2026835} (not an assigned CVE). \\
1 May--31 Aug. 2026 & MITRE CVE Program & Full technical manuscript, proposed CWEs, in-the-wild abuse evidence, and protocol-level scope clarifications submitted under reference \texttt{2026835}. & Under MITRE intake; closed without CVE assignment at conclusion of review window (31 Aug. 2026). \\
16--18 May 2026 & CERT/CC (\texttt{VRF\#26-05-CJVGG}) & Broader coordination filing submitted with empirical and formal material; initial determination issued. & Declined; recommended setting target disclosure date and proceeding publicly if unaddressed. \\
1--9 Jun. 2026 & VulnCheck CNA & Alternate CNA review opened and evaluated in parallel with CERT/CC track. & Declined under intended-behavior reasoning; formally deferred to CERT/CC on 11 Jun. 2026. \\
9 Jun. 2026 & GitHub Security (\texttt{braintree-web}) & Private Vulnerability Report \texttt{GHSA-xjwr-jwm2-wvjr} filed with maintainers detailing telemetry provenance gaps and in-the-wild abuse. & Filed; author credited as finder on 11 Jun.; maintainers provided no written response during reporting window. \\
19 Jun. 2026 & Research Scope Expansion & Methodological evaluation of early single-vendor exploratory testbed prompted expansion from isolated gateway testing to multi-scheme architectural analysis. & Scope expanded to encompass full Visa, Mastercard, and American Express protocol-level evaluation. \\
13 Jul. 2026 & Disclosure window & Standard 90-day window from the original vendor filing elapsed. & 90-day target disclosure window elapsed following vendor report. \\
6--10 Aug. 2026 & CERT/CC (\texttt{VRF\#26-08-XWHPS}); CISA (\texttt{CCASE0209885}) & Narrowed filing submitted with coordination bundle; CERT/CC issued final determination; parallel CISA case opened. & CERT/CC reiterated public disclosure recommendation; CISA issued automated acknowledgement. \\
27 Aug.--3 Sep. 2026 & EMVCo (\texttt{\#20260827-17621D}) & Formal specification inquiry and technical memorandum submitted regarding pre-authorization PAN oracle; EMVCo 3DS Team responded that inquiries are processed under general non-confidential Terms of Use. & Assigned tracking reference; processed under general terms. \\
29 Aug. 2026 & Visa HackerOne (\texttt{\#3978615}) & Scheme-level vulnerability report filed covering lookup-stage 3DS response oracle and Cardinal integration path (\texttt{usa.visa.com}). & Closed as ``Informative'' (triage reasoned candidate card numbers are assumed known and endpoint routes challenges as designed without CIA triad breach). \\
30 Aug. 2026 & American Express HackerOne (\texttt{\#3979021}) & Scheme-level vulnerability report filed covering EMV 3DS lookup-stage PAN/account-state oracle requiring American Express review. & Under active program review / pending triage assessment. \\
29 Aug.--9 Sep. 2026 & Mastercard Bugcrowd (\texttt{8d647a18}) & Vendor submission filed detailing lookup-stage PAN disclosure and Mastercard calibration findings; closed by platform triage. & Closed as Not Applicable (out of scope for isolated bounty program). \\
1--2 Sep. 2026 & Payment Scheme Contacts (Visa Secure; Mastercard) & Formal protocol memoranda submitted to central security teams (referencing Bugcrowd \texttt{8d647a18} to request routing to Identity Check/3DS standards owners). & Filed; internal scheme routing initiated. \\
8 Sep. 2026 & Payment Scheme Contacts (Visa; Mastercard; Amex) & Revised technical memoranda (v1.3) submitted to Visa and Mastercard; protocol research memorandum submitted to American Express SafeKey. & Visa confirmed internal forwarding; Mastercard pending routing; Amex routing requested to SafeKey. \\
18--19 Sep. 2026 & Payment Scheme Contacts (Mastercard) & Protocol-level routing memo (v1.5) submitted to senior security engineering for Identity Check/standards teams. & Mastercard contact confirmed receipt and initiated internal team routing. \\
27 Sep. 2026 & Payment Scheme Contacts (Visa \& Mastercard) & Updated technical decision brief (v3.1) submitted outlining academic publication goals, threat telemetry, and structural remediation models. & Both networks confirmed receipt; internal routing ongoing. \\
7 Oct. 2026 & Payment Scheme Contacts (Mastercard) & Follow-up on protocol security routing and Bugcrowd scope limitations for cross-party architectural issues. & Mastercard contact confirmed memorandum forwarded internally to specialized teams. \\
15 Oct. 2026 & Payment Scheme Contacts (Visa \& Mastercard) & Comparative technical addenda submitted detailing Reason~98 observations, CardinalCommerce RDX specifications, and published patent application US~2026/0134429~A1. & Under ongoing technical review across Visa and Mastercard. \\
\bottomrule
\end{tabularx}
\end{table*}
\endgroup
\clearpage
\begingroup
\raggedbottom
\renewcommand{\dbltopfraction}{0.99}
\renewcommand{\textfraction}{0.01}
\renewcommand{\dblfloatpagefraction}{0.70}
\begin{figure*}[!t]
\centering
\section{Formal Verification Model}
\label{app:proverif}
\noindent
\textbf{Listing 2: ProVerif Trace Model of the Observed Client-Facing Flow.}
\vspace{1pt}
\begin{minipage}[t]{0.49\textwidth}
\vspace{0pt}
\begin{lstlisting}[language=ProVerif, numbers=none, basicstyle=\ttfamily\fontsize{6.2pt}{7.4pt}\selectfont, breaklines=true, columns=fullflexible, frame=single, xleftmargin=1pt, xrightmargin=1pt, aboveskip=1.5pt, belowskip=1.5pt, title={(a) Types, Events, and Queries}]
free checkout, cardinal_init, save_browser_data, lookup_api: channel.
free cardinal_init_reply, save_browser_reply, lookup_reply: channel [private].
free observed_query, observed_reply: channel [private].
type client_token. type cardinal_jwt. type merchant_id. type nonce.
type pan. type bin. type session_id. type telemetry. type lookup_response.
free tested_bin: bin.
free researcher_control_pan: pan.
free candidate_body_1, candidate_body_2, candidate_body_3: bitstring.
free synthetic_telemetry: telemetry.
free init_accepted, telemetry_accepted: bitstring.
fun candidate_pan(bin, bitstring): pan.
fun card_bin(pan): bin.
reduc forall b: bin, s: bitstring; card_bin(candidate_pan(b, s)) = b.
reduc card_bin(researcher_control_pan) = tested_bin.
fun issue_client_token(merchant_id, nonce, bitstring): client_token.
fun recover_client_merchant(client_token, bitstring): merchant_id.
reduc forall m: merchant_id, n: nonce, k: bitstring;
  recover_client_merchant(issue_client_token(m, n, k), k) = m.
fun issue_cardinal_jwt(merchant_id, nonce, bitstring): cardinal_jwt.
fun recover_cardinal_merchant(cardinal_jwt, bitstring): merchant_id.
reduc forall m: merchant_id, n: nonce, k: bitstring;
  recover_cardinal_merchant(issue_cardinal_jwt(m, n, k), k) = m.
event clientTokenIssued(merchant_id, client_token).
event cardinalJwtIssued(merchant_id, cardinal_jwt).
event cardinalSessionInitialized(merchant_id, cardinal_jwt).
event syntheticTelemetryAccepted(merchant_id, cardinal_jwt, session_id, session_id, telemetry).
event clientTokenAccepted(merchant_id, client_token).
event lookupRequestAccepted(merchant_id, client_token, pan, bin).
event lookupResultReleased(merchant_id, client_token, pan).
query m: merchant_id, j: cardinal_jwt;
  event(cardinalSessionInitialized(m, j)) ==> event(cardinalJwtIssued(m, j)).
query m: merchant_id, j: cardinal_jwt, cs: session_id, bs: session_id, telem: telemetry;
  event(syntheticTelemetryAccepted(m, j, cs, bs, telem)) ==> event(cardinalJwtIssued(m, j)).
query m: merchant_id, j: cardinal_jwt, cs: session_id, bs: session_id;
  event(syntheticTelemetryAccepted(m, j, cs, bs, synthetic_telemetry)).
query m: merchant_id, t: client_token;
  event(clientTokenAccepted(m, t)) ==> event(clientTokenIssued(m, t)).
query m: merchant_id, t: client_token, p: pan;
  event(lookupResultReleased(m, t, p)) ==> event(clientTokenAccepted(m, t)).
query m: merchant_id, t: client_token, p: pan;
  event(lookupResultReleased(m, t, p)).
\end{lstlisting}
\end{minipage}\hfill%
\begin{minipage}[t]{0.49\textwidth}
\vspace{0pt}
\begin{lstlisting}[language=ProVerif, numbers=none, basicstyle=\ttfamily\fontsize{6.2pt}{7.4pt}\selectfont, breaklines=true, columns=fullflexible, frame=single, xleftmargin=1pt, xrightmargin=1pt, aboveskip=1.5pt, belowskip=1.5pt, title={(b) Processes and Observed Request Order}]
let processMerchant(gateway_key: bitstring, cardinal_key: bitstring, m: merchant_id) =
  new n: nonce; new jn: nonce;
  let t = issue_client_token(m, n, gateway_key) in
  let j = issue_cardinal_jwt(m, jn, cardinal_key) in
  event clientTokenIssued(m, t); event cardinalJwtIssued(m, j);
  out(checkout, (t, j)).
let processCardinalService(k: bitstring, m: merchant_id) =
  !(in(cardinal_init, (j: cardinal_jwt));
    let jwtMerchant = recover_cardinal_merchant(j, k) in
    if jwtMerchant = m then
      (event cardinalSessionInitialized(m, j);
       out(cardinal_init_reply, init_accepted)))
  |
  !(in(save_browser_data, (j: cardinal_jwt, cs: session_id,
                           bs: session_id, telem: telemetry));
    let jwtMerchant = recover_cardinal_merchant(j, k) in
    if jwtMerchant = m then
      (event syntheticTelemetryAccepted(m, j, cs, bs, telem);
       out(save_browser_reply, telemetry_accepted))).
let processGateway(k: bitstring, m: merchant_id) =
  !(in(lookup_api, (t: client_token, p: pan));
    let tokenMerchant = recover_client_merchant(t, k) in
    if tokenMerchant = m then
      let b = card_bin(p) in
      event clientTokenAccepted(m, t);
      event lookupRequestAccepted(m, t, p, b);
      out(observed_query, (m, t, p, b));
      in(observed_reply, (m2: merchant_id, t2: client_token,
                          p2: pan, b2: bin, r: lookup_response));
      if m2 = m then if t2 = t then if p2 = p then if b2 = b then
        (event lookupResultReleased(m, t, p);
         out(lookup_reply, (p, r)))).
let processObservedResponse =
  !(in(observed_query, (m: merchant_id, t: client_token,
                        p: pan, b: bin));
    new r: lookup_response;
    out(observed_reply, (m, t, p, b, r))).
let processOffContextCaller =
  in(checkout, (t: client_token, j: cardinal_jwt));
  out(cardinal_init, j);
  in(cardinal_init_reply, (init_status: bitstring));
  new consumer_session_id: session_id; new bin_session_id: session_id;
  out(save_browser_data, (j, consumer_session_id, bin_session_id,
                          synthetic_telemetry));
  in(save_browser_reply, (telemetry_status: bitstring));
  out(lookup_api, (t, researcher_control_pan));
  in(lookup_reply, (control: pan, r0: lookup_response));
  out(lookup_api, (t, candidate_pan(tested_bin, candidate_body_1)));
  in(lookup_reply, (candidate1: pan, r1: lookup_response));
  out(lookup_api, (t, candidate_pan(tested_bin, candidate_body_2)));
  in(lookup_reply, (candidate2: pan, r2: lookup_response));
  out(lookup_api, (t, candidate_pan(tested_bin, candidate_body_3)));
  in(lookup_reply, (candidate3: pan, r3: lookup_response)).
process
  new gateway_key: bitstring; new cardinal_key: bitstring;
  new target_merchant: merchant_id;
  (!processMerchant(gateway_key, cardinal_key, target_merchant) |
   !processCardinalService(cardinal_key, target_merchant) |
   !processGateway(gateway_key, target_merchant) |
   !processObservedResponse |
   processOffContextCaller)
\end{lstlisting}
\end{minipage}
\end{figure*}
\clearpage
\renewcommand{\dbltopfraction}{0.90}
\renewcommand{\textfraction}{0.2}
\renewcommand{\dblfloatpagefraction}{0.85}
This ProVerif listing is a trace abstraction of the tested Braintree--Cardinal client-facing workflow, not a model of EMVCo AReq/ARes processing. It represents the two distinct credentials used in the observed flow: merchant-scoped Client API authorization (abstracting the client token and its authorization fingerprint) and a Cardinal Auth JWT. The caller obtains these through the tested checkout flow, uses the JWT to initialize a Cardinal session, creates fresh \texttt{ConsumerSessionId} and \texttt{BinSessionId} values locally, submits synthetic telemetry, and reuses the merchant-scoped authorization for one known-issued anchor card and three adjacent same-range candidate lookups. The candidate constructor records the experimental input design, \texttt{candidate\_pan(test\_bin, suffix)}; it keeps the BIN fixed to represent the within-range adjacent-card comparison, not a separate BIN lookup. It does not encode digits, compute a Luhn checksum, or assert that a candidate is issued. The old query \texttt{attacker(make\_pan(...))} was removed because a public constructor applied to public inputs makes that query trivially true. For compactness, the listing represents one anchor card and three adjacent candidates rather than all seven controls and five fresh sessions per input; those sample counts and observed classes remain experimental results. A private response source returns a fresh opaque envelope so the model can check token provenance and response-release reachability. It does not derive the observed first-result separation or Cohort~G's uniform challenge class; those remain empirical observations. The model does not establish telemetry provenance, binding between browser execution, session IDs, telemetry, or lookup, downstream telemetry use, token expiry or revocation, rate limits, detector behavior, financial authorization, CAVV validity, or internal 3DS Server--DS--ACS processing. It therefore cannot prove account validity, network routing, or general enumeration capability.

\endgroup

\section{Specification-Grounded Protocol Mapping: Normative Hypotheses vs. Multi-Scheme Reality}
\label{app:spec_grounded_mapping}
Figure~\ref{fig:concrete_empirical_artifacts} presents two redacted client-facing response patterns analyzed in this section. Their identifier and reason-code combinations support only conditional tier attribution, not direct observation of internal processing.

\begin{figure*}[t]
\centering
\begin{minipage}[t]{0.485\textwidth}
\begin{lstlisting}[language={},basicstyle=\ttfamily\fontsize{7.0pt}{8.2pt}\selectfont,breaklines=true,columns=fullflexible,frame=single,xleftmargin=1.5pt,xrightmargin=1.5pt,escapeinside={(*@}{@*)},title={(a) DS-Attributed Range-Drop Pattern (Inferred under Req 411)}]
{
  "paymentMethod": {
    "type": "CreditCard",
    "nonce": "[REDACTED_NONCE]",
    "description": "ending in [XXXX]",
    "threeDSecureInfo": {
      "status": "authenticate_frictionless_failed",
      "enrolled": "Y",
      // (*@\textit{\textbf{DS pattern inferred ($Q_{\text{ACS}}=0$)}}@*)
      "acsTransactionId":
        (*@\underline{\textbf{"e89ae1a9-de75-474a-9f9f-c1f623fa3622"}}@*),
      "dsTransactionId":
        (*@\underline{\textbf{"e89ae1a9-de75-474a-9f9f-c1f623fa3622"}}@*),
      "eciFlag": "07",
      "threeDSecureVersion": "2.2.0",
      "lookup": {
        "transStatus": "N",
        "transStatusReason": "06" // Invalid card number (DS range drop)
      }
    },
    "details": { "bin": "[REDACTED_BIN]", "lastFour": "[XXXX]", "cardType": "[REDACTED]" },
    "binData": { "issuingBank": "[REDACTED]", "country": "[REDACTED]", "debit": "No" }
  }
}
\end{lstlisting}
\end{minipage}
\hfill
\begin{minipage}[t]{0.485\textwidth}
\begin{lstlisting}[language={},basicstyle=\ttfamily\fontsize{7.0pt}{8.2pt}\selectfont,breaklines=true,columns=fullflexible,frame=single,xleftmargin=1.5pt,xrightmargin=1.5pt,escapeinside={(*@}{@*)},title={(b) ACS-Associated Response Pattern (Inferred under Req 105)}]
{
  "paymentMethod": {
    "type": "CreditCard",
    "nonce": "[REDACTED_NONCE]",
    "description": "ending in [XXXX]",
    "threeDSecureInfo": {
      "status": "authenticate_frictionless_failed",
      "enrolled": "Y",
      // (*@\textit{\textbf{ACS pattern inferred ($Q_{\text{ACS}}=1$)}}@*)
      "acsTransactionId":
        (*@\underline{\textbf{"19384ede-fefc-4a7b-936d-b2821785a18b"}}@*),
      "dsTransactionId":
        (*@\underline{\textbf{"a27cdca1-fc03-4583-b562-3c6daa3669a0"}}@*),
      "eciFlag": "07",
      "threeDSecureVersion": "2.2.0",
      "lookup": {
        "transStatus": "N",
        "transStatusReason": "01" // Card authentication failed
      }
    },
    "details": { "bin": "[REDACTED_BIN]", "lastFour": "[XXXX]", "cardType": "[REDACTED]" },
    "binData": { "issuingBank": "[REDACTED]", "country": "[REDACTED]", "debit": "Yes" }
  }
}
\end{lstlisting}
\end{minipage}
\vspace{4pt}
\caption{Concrete client-facing artifacts with two identifier/reason-code patterns. Panel~(a) is consistent with a DS-generated response under [Req~411]; panel~(b) is consistent with an ACS-associated response under [Req~105]. These are conditional attributions from returned integration fields, not observations of internal DS/ACS messages or ledger queries. The off-context HTTP caller did not execute the browser SDK; transaction identifiers, candidate account prefixes, nonces, and institutional entities are anonymized or synthetic.}
\label{fig:concrete_empirical_artifacts}
\end{figure*}

To address multi-party implementation boundaries rigorously and ethically, our broader empirical analysis was structured as a formal hypothesis-testing study grounded directly in published normative standards: the EMVCo 3-D Secure Core Specification v2.3.1.1~\cite{emvco_core_2311_2023}, major commercial 3DS Server and Access Control Server platform specifications~\cite{cardinal_transreason_2025,cardinal_transreason_values}, and standard production gateway integration guidelines~\cite{braintree_3ds_status_codes}. Rather than executing unstructured live probes, we systematically mapped each protocol outcome to a specification-derived hypothesis regarding the authoritative emitting tier, and then evaluated whether observed production behaviors across Visa, Mastercard, and American Express conform to normative expectations or expose structural side-channel oracles.

\subsection{Specification-Derived Invariant Hypotheses}
The EMVCo Core Specification partitions transaction evaluation into two distinct tiers, establishing normative structural invariants:
\begin{enumerate}\setlength{\itemsep}{1.5pt}\setlength{\parskip}{0pt}
    \item \textbf{Directory Server Invariant ($H_{\text{DS}}$)}: Under EMVCo v2.3.1.1 Seq~3.155 [Req 411], Seq~3.47 [Req 421], and Seq~3.225 [Req 412], whenever the payment scheme Directory Server synthesizes an \texttt{ARes} on the ACS's behalf, the DS normatively duplicates its own identifier into the ACS field ($\text{acsTransactionID} \equiv \text{dsTransactionID}$). Thus, identifier equality provides conditional protocol-level evidence consistent with a DS-created response under the stated conformance and field-mapping assumptions; if the identifiers are independently generated UUIDv4 values, accidental equality has probability $P < 2^{-122}$.
    \item \textbf{Issuer ACS Invariant ($H_{\text{ACS}}$)}: Under EMVCo Seq~3.142 [Req 104] and Seq~3.143 [Req 105], when a request is forwarded to an active Access Control Server, the ACS generates its own ACS Transaction ID before later evaluation. The specification requires RFC~4122 canonical format but permits compliant identifiers of any supported version. Accordingly, distinct identifiers ($\text{acsTransactionID} \neq \text{dsTransactionID}$) are consistent with an ACS-created response, but do not by themselves prove a ledger query or any particular account state.
\end{enumerate}

\subsection{Observed Challenge-Required Response Patterns Across Sampled Cohorts}
\label{subsec:app_challenge_dichotomy}
Our client-facing observations include both uniform and input-associated challenge-required response patterns, illustrated in Figure~\ref{fig:challenge_empirical_artifacts}. Because we did not capture internal 3DS Server--DS--ACS messages or issuer logs, these examples describe integration outputs rather than verified backend architectures:

\begin{figure*}[t]
\centering
\begin{minipage}[t]{0.482\textwidth}
\begin{lstlisting}[language={},basicstyle=\ttfamily\fontsize{7.0pt}{8.2pt}\selectfont,breaklines=true,columns=fullflexible,frame=single,xleftmargin=1.5pt,xrightmargin=1.5pt,escapeinside={(*@}{@*)},title={(a) Observed Cohort G: Uniform Challenge Result}]
{
  "paymentMethod": {
    "type": "CreditCard",
    "nonce": "[REDACTED_NONCE]",
    "description": "ending in [XXXX]",
    "threeDSecureInfo": {
      "status": "challenge_required",
      "enrolled": "Y",
      // Observed Cohort G: uniform challenge result (ECI=07)
      "acsTransactionId":
        (*@\underline{\textbf{"32abe96a-ffaa-465e-9654-292c73071101"}}@*),
      "dsTransactionId":
        (*@\underline{\textbf{"8665173e-94e4-47b6-9625-32e283495838"}}@*),
      "eciFlag": "07",
      "threeDSecureVersion": "2.2.0",
      "lookup": {
        "transStatus": "C",
        "transStatusReason": null
      }
    },
    "details": {
      "bin": "[REDACTED_BIN]", "lastFour": "[XXXX]",
      "cardType": "[REDACTED]"
    },
    "binData": {
      "issuingBank": "[REDACTED]", "country": "[REDACTED]",
      "debit": "No"
    }
  },
  "lookup": {
    // Static cloud-hosted multi-tenant challenge endpoint:
    "acsUrl":
      (*@\underline{\textbf{"https://[REDACTED\_CLOUD\_ACS\_PROXY]/"}}@*)
      (*@\underline{\textbf{"ThreeDSecure/V2\_1\_0/CReq?oid=[REDACTED]"}}@*),
    "termUrl": "https://[REDACTED_GATEWAY]/.../auth?jwt=[REDACTED]",
    "pareq": "[REDACTED_BASE64_CREQ_PAYLOAD]"
  }
}
\end{lstlisting}
\end{minipage}%
\hfill
\begin{minipage}[t]{0.482\textwidth}
\begin{lstlisting}[language={},basicstyle=\ttfamily\fontsize{7.0pt}{8.2pt}\selectfont,breaklines=true,columns=fullflexible,frame=single,xleftmargin=1.5pt,xrightmargin=1.5pt,escapeinside={(*@}{@*)},title={(b) Observed Control: Challenge-Required Result}]
{
  "paymentMethod": {
    "type": "CreditCard",
    "nonce": "[REDACTED_NONCE]",
    "description": "ending in [XXXX]",
    "threeDSecureInfo": {
      "status": "challenge_required",
      "enrolled": "Y",
      // Observed control: challenge-required result (ECI=07)
      "acsTransactionId":
        (*@\underline{\textbf{"6e7de567-95b9-4edd-9400-c03568c71076"}}@*),
      "dsTransactionId":
        (*@\underline{\textbf{"4f526512-1f9a-49fe-b14d-553693905c49"}}@*),
      "eciFlag": "07",
      "threeDSecureVersion": "2.2.0",
      "lookup": {
        "transStatus": "C",
        "transStatusReason": null
      }
    },
    "details": {
      "bin": "[REDACTED_BIN]", "lastFour": "[XXXX]",
      "cardType": "[REDACTED]"
    },
    "binData": {
      "issuingBank": "[REDACTED]", "country": "[REDACTED]",
      "debit": "Yes"
    }
  },
  "lookup": {
    // Dynamic bank-hosted challenge endpoint (observed):
    "acsUrl":
      (*@\underline{\textbf{"https://[REDACTED\_ISSUER\_ACS]/"}}@*)
      (*@\underline{\textbf{".../challenge?id=[REDACTED\_UUID]"}}@*),
    "termUrl": "https://[REDACTED_GATEWAY]/.../auth?jwt=[REDACTED]",
    "pareq": "[REDACTED_BASE64_CREQ_PAYLOAD]"
  }
}
\end{lstlisting}
\end{minipage}
\vspace{4pt}
\caption{Client-facing examples of the challenge-required result (\texttt{transStatus=C}) in two evaluated cohorts. Both were returned at the lookup-stage boundary to an off-context caller. The label \texttt{C} means a challenge was requested, not completed. Panel~(a) shows a range-level endpoint and a uniform challenge result in the sampled cloud-hosted cohort; panel~(b) shows a dynamic bank-hosted endpoint for the controlled card in the sampled issuer cohort. Internal ledger queries, candidate issuance, and challenge completion were not directly observed. The caller never sent the initial \texttt{CReq}, so the later challenge/OOB interaction never started and no SMS/push/OTP was observed on the test cards. Transaction identifiers are synthetic, and sensitive values are redacted.}\label{fig:challenge_empirical_artifacts}
\end{figure*}

\paragraph{Uniform Challenge-Required Responses in the Sampled Cohort.}
In Cohort~G and related sampled cloud-hosted ranges, all tested inputs within the participating range returned a challenge-required lookup result (\texttt{transStatus=C}) and a range-level challenge endpoint, as shown in Figure~\ref{fig:challenge_empirical_artifacts}(a). This uniformity means the observed first-result class did not distinguish the known-issued anchor card from the adjacent comparison PANs in that cohort. It does not establish whether an issuer ledger was queried, whether account verification was deferred until \texttt{CReq}/\texttt{CRes}, or how genuine cardholders would be treated outside the tested inputs. The off-context caller did not complete the challenge.

\paragraph{Input-Associated Responses in Other Sampled Cohorts.}
In a sampled issuer-hosted separating cohort, the known-issued anchor card and adjacent same-range comparison PANs returned different lookup-stage classes; the anchor-card response included \texttt{transStatus=C} and a dynamic challenge endpoint, as shown in Figure~\ref{fig:challenge_empirical_artifacts}(b). This is evidence of response-class separation in those observations, not proof that the adjacent comparison PANs were unissued or that every \texttt{C} response identifies an issued or usable account. We did not capture internal ledger decisions or complete a \texttt{CReq}/\texttt{CRes} exchange. Because the caller never initiated that later branch, no SMS, push, or OTP reached the test recipients, even though the specification gives banking-app push only as an example within the challenge/OOB interaction~\cite{emvco_core_2311_2023}.

\paragraph{Implications for Challenge-Based Defenses.}
Mastercard Patent Application US~2026/0134429~A1 describes declining or challenging requests as possible security measures against suspected automated probing~\cite{mastercard2026abnormaltraffic}. If an issuer's challenge decision varies with account-specific processing and the lookup result is exposed to the caller, response classes may help prioritize candidates. However, \texttt{transStatus=C} means only that a challenge was requested; it is not a completed authentication result or universal account-validity signal. A uniformly returned challenge can instead mask input differences. In our off-context tests, the caller never followed that branch with \texttt{CReq}, so no cardholder-facing SMS/push/OTP was observed; EMVCo specifies no cardholder alert on the lookup-only path, and the patent's notifications are network-to-issuer/acquirer. Independent issuer messaging and transaction-level user impact were not measured~\cite{emvco_core_2311_2023,mastercard2026abnormaltraffic}.

\paragraph{Underground Tooling Context.}
Reviewed 3DS card-testing artifacts group frictionless, challenge-required, and negative outcomes into separate streams~\cite{telegram_3ds_tooling_2026,independent_hosted_payment_abuse_audit_2026}. In one inspected implementation, the first-result branch labels \texttt{C}/\texttt{R} as ``Challenge required'' and returns a declined result; it does not render the browser challenge or continue through \texttt{CReq}/\texttt{CRes}. That coding choice itself is informative: attackers do not treat frictionless and challenge branches as equally valuable, because frictionless or success-labeled attempted-authentication outcomes are easier to operationalize and may reflect weaker issuer-side risk discrimination at the lookup boundary. Our off-context observations strengthen that concern, since selected cohorts returned frictionless success-labeled responses to a forged caller without browser execution. This shows how such tooling can classify estimated credentials based on lookup-stage response classes without observing an OTP or completing authentication. Such triage is consistent with an account-recognition signal being useful for prioritization, not proof that every classified candidate corresponds to an issued, usable, or fully valid account. We did not execute the tools or measure their use against the evaluated cohorts. Table~\ref{tab:challenge_architecture_comparison} compares the sampled cloud-hosted and issuer-hosted challenge patterns, distinguishing client-visible observations from backend behavior not established by the traces.
\begin{table*}[t]
\centering
\fontsize{7.5pt}{8.5pt}\selectfont
\caption{Comparative Architectural Analysis of Pre-Authorization Challenge Generation across Evaluated Anonymized Cohorts. All specific financial institutions, commercial gateway brand names, raw BIN ranges, and live endpoint URLs are fully redacted for responsible disclosure.}
\label{tab:challenge_architecture_comparison}
\setlength{\aboverulesep}{0pt}
\setlength{\belowrulesep}{0pt}
\renewcommand{\arraystretch}{1.05}
\setlength{\tabcolsep}{3pt}
\rowcolors{2}{white}{rowzebra}
\begin{tabularx}{\textwidth}{@{}p{0.18\textwidth} p{0.15\textwidth} p{0.18\textwidth} p{0.17\textwidth} p{0.15\textwidth} X@{}}
\toprule
\rowcolor{headerbg}
\textbf{Architectural Model} & \textbf{Authentication Timing} & \textbf{Observed Status Distribution} & \textbf{Emitted \texttt{acsURL} Pattern} & \textbf{Pre-Auth Classification Surface} & \textbf{Patent [0040] Defense Interaction} \\
\midrule
Multi-Tenant Cloud-Hosted ACS Pattern (e.g., Cohort~G) & Sample consistent with deferred account-specific differentiation beyond the lookup stage & Uniform \texttt{transStatus=C} across sampled inputs & Static multi-tenant cloud endpoint (\texttt{ThreeDSecure/...\allowbreak/CReq}) & Masked across sampled cohort (uniform \texttt{C}) & A uniform challenge policy can blind the pre-auth split, but it trades frictionless UX for mandatory step-up friction. \\
Issuer-Hosted Separating ACS Pattern (sampled cohort) & Lookup-stage separation observed before \texttt{CReq}/\texttt{CRes} & Controlled card returned \texttt{C}; adjacent comparison inputs returned negative classes & Dynamic bank-hosted challenge endpoint & Sharp response-class separation observed & Account-dependent challenge selection may leak pre-clearing classification signals before financial authorization. \\
\bottomrule
\end{tabularx}
\end{table*}

The primary controlled cohort is the seven-control experiment reported in Section~\ref{sec:empirical_study}. Table~\ref{tab:master_spec_reason_mapping} provides a specification-grounded synthesis mapping EMVCo protocol requirements and vendor documentation against observed client-facing pre-authorization responses; it is not a second controlled experiment and does not estimate scheme-wide prevalence. Where correlated 3DS Server, Directory Server, or ACS traces are unavailable, emitting-tier labels are structural inferences derived under EMVCo specification rules (e.g., [Req~105] and [Req~411]), not direct observations of internal network components.

\begin{table*}[t]
\centering
\fontsize{7.2pt}{8.2pt}\selectfont
\caption{\textbf{Mapping of Protocol Invariants and Observed Lookup-Stage Response Classes.} Normative definitions and emitting tiers follow EMVCo Core Specification v2.3.1.1 (Table~A.1)~\cite{emvco_core_2311_2023}, published cloud ACS documentation~\cite{cardinal_transreason_2025,cardinal_transreason_values}, and gateway integration specifications~\cite{braintree_3ds_status_codes}. Hypotheses define the normative emitting tier and expected identifier invariant ($H_{\text{DS}} \implies \text{acs} \equiv \text{ds}$; $H_{\text{ACS}} \implies \text{acs} \neq \text{ds}$). Symbols: \checkmark = matches the stated hypothesis; $\times$ = observed divergence or ambiguous multi-tier response; $\triangle$ = proprietary scheme feedback observed or described. These labels do not establish candidate account status.}
\label{tab:master_spec_reason_mapping}
\renewcommand{\arraystretch}{1.0}
\setlength{\tabcolsep}{2.5pt}
\rowcolors{2}{white}{rowzebra}
\begin{tabularx}{\textwidth}{|>{\centering\arraybackslash}p{0.055\textwidth}|X|>{\raggedright\arraybackslash}p{0.22\textwidth}|>{\raggedright\arraybackslash}p{0.26\textwidth}|>{\centering\arraybackslash}p{0.045\textwidth}|}
\hline
\rowcolor{headerbg}
\textbf{Code} & \textbf{Normative Meaning \& Definition (EMVCo Table~A.1)} & \textbf{Spec Hypothesis ($H$) \& Emitting Tier} & \textbf{Observed Integration Reflection in Tested Path} & \textbf{Match} \\
\hline\hline
\rowcolor{sectionbg}
\multicolumn{5}{|l|}{\textbf{A. Gateway Outcome Statuses and Underlying Protocol States (\texttt{transStatus})}} \\
\hline
\texttt{Y} & \textbf{Frictionless approved} (\texttt{authenticate\_successful}); full verification & $H_{\text{ACS}}: \text{acs} \neq \text{ds}$ ($Q_{\text{ACS}}{=}1$), Issuer ACS & Frictionless pre-auth approval; valid CAVV emitted before clearing & \checkmark \\
\hline
\texttt{A} & \textbf{Proof of attempt} (\texttt{authenticate\_\allowbreak attempt\_\allowbreak successful}); stand-in & $H_{\text{DS}}: \text{acs} \equiv \text{ds}$ ($Q_{\text{ACS}}{=}0$), Scheme DS & Dual-tier emission: DS stand-in ($\text{acs} \equiv \text{ds}$) vs.\ ACS attempt ($\text{acs} \neq \text{ds}$) & $\times$ \\
\hline
\texttt{C} & \textbf{Challenge required} (\texttt{challenge\_required}); interactive step-up & $H_{\text{ACS}}: \text{acs} \neq \text{ds}$ ($Q_{\text{ACS}}{=}1$), Issuer ACS & Step-up requested; triggers OOB session; offline caller halts before \texttt{CReq} & \checkmark \\
\hline
\texttt{N} & \textbf{Frictionless denied} (\texttt{frictionless\_failed}); not authenticated & Dual-tier: DS boundary drop vs.\ ACS rejection & Dual-tier mask: DS routing drops ($\text{acs} \equiv \text{ds}$) vs.\ ACS denials ($\text{acs} \neq \text{ds}$) & $\times$ \\
\hline
\texttt{R} & \textbf{Issuer hard rejection} (\texttt{authenticate\_rejected}); do not clear & $H_{\text{ACS}}: \text{acs} \neq \text{ds}$ ($Q_{\text{ACS}}{=}1$), Issuer ACS exclusive & Authoritative ledger rejection directive; DS never synthesizes \texttt{R} & \checkmark \\
\hline
\texttt{U} & \textbf{3DS unavailable} (\texttt{not\_enrolled}); non-participating card range & $H_{\text{DS}}: \text{acs} \equiv \text{ds}$ ($Q_{\text{ACS}}{=}0$), Scheme DS boundary & Pre-authorization range exclusion; DS boundary termination & \checkmark \\
\hline\hline
\rowcolor{sectionbg}
\multicolumn{5}{|l|}{\textbf{B. Transaction Status Reason Codes Evaluated in Controlled \& Exploratory Studies}} \\
\hline
\texttt{01} & Card authentication failed & $H_{\text{ACS}}: \text{acs} \neq \text{ds}$ ($Q_{\text{ACS}}{=}1$), Issuer ACS ledger & Authoritative ledger evaluation failure (Figure~\ref{fig:concrete_empirical_artifacts}b) & \checkmark \\
\hline
\texttt{05} & Expired card (core ledger parameter mismatch) & $H_{\text{ACS}}: \text{acs} \neq \text{ds}$ ($Q_{\text{ACS}}{=}1$), Issuer ACS ledger & Account-distinguishing ledger signal: known-issued account with expired parameter (Cohort~A) & \checkmark \\
\hline
\texttt{06} & Invalid card number (unallocated / format boundary) & Dual-tier: DS range drop vs.\ ACS ledger check & Scheme routing termination on unallocated candidate bodies ($\text{acs} \equiv \text{ds}$) & $\times$ \\
\hline
\texttt{08} & No card record (unallocated or non-existent account) & $H_{\text{ACS}}: \text{acs} \neq \text{ds}$ ($Q_{\text{ACS}}{=}1$), Issuer ACS ledger & Account-distinguishing ledger signal: ACS accepts request in participating range, finds no account record (Cohorts~A--F) & \checkmark \\
\hline
\texttt{13} & Cardholder not enrolled in 3DS service & $H_{\text{ACS}}: \text{acs} \neq \text{ds}$ ($Q_{\text{ACS}}{=}1$), Issuer ACS ledger & Account-distinguishing ledger signal: registered account lacking 3DS service (Cohort~C) & \checkmark \\
\hline
\texttt{80} & Proprietary Directory Server exclusion / stand-in rule & $H_{\text{DS}}: \text{acs} \equiv \text{ds}$ ($Q_{\text{ACS}}{=}0$), Scheme DS tier & Directory Server routing drop without querying issuer ledger & \checkmark \\
\hline
\texttt{87} & Non-participating range / DS routing rule & $H_{\text{DS}}: \text{acs} \equiv \text{ds}$ ($Q_{\text{ACS}}{=}0$), Scheme DS tier & Scheme boundary drop on non-participating sub-ranges ($\text{acs} \equiv \text{ds}$) & \checkmark \\
\hline
\texttt{98} & Suspected Authentication BIN Attack (Mastercard Patent [0041]) & $H_{\text{DS}}: \text{acs} \equiv \text{ds}$ ($Q_{\text{ACS}}{=}0$), Scheme DS tripwire & Observed in-band feedback tripwire during same-range sweeps ($\text{acs} \equiv \text{ds}$) & $\triangle$ \\
\hline
\texttt{99} & Override Authentication BIN Attack (Mastercard Patent [0043]) & $H_{\text{DS}}: \text{acs} \equiv \text{ds}$ ($Q_{\text{ACS}}{=}0$), Scheme DS tier & Documented patent override mechanism; unobserved in primary live cohorts & $\triangle$ \\
\hline
\end{tabularx}
\end{table*}

\subsection{Cross-Scheme Normative Mapping and Spec Conformance}
Table~\ref{tab:master_spec_reason_mapping} summarizes selected gateway statuses, protocol outcome states, and key transaction status reason codes evaluated across our specification analysis, primary cohorts, and exploratory observations. The table distinguishes documented normative definitions from observed integration reflections; its qualitative behavior labels do not establish network-wide prevalence:
\begin{itemize}\setlength{\itemsep}{2pt}\setlength{\parsep}{1pt}\setlength{\parskip}{0pt}
    \item \textbf{Negative Outcome Demarcation (\texttt{N} vs.\ \texttt{R})}: Under EMVCo Table~A.1, \texttt{transStatus = R} is legally defined as an explicit issuer directive commanding that authorization not be attempted. Across all three evaluated payment schemes, \texttt{R} consistently exhibited distinct identifiers ($\text{acsTransID} \neq \text{dsTransID}$), conforming to normative specification requirements where Directory Servers do not synthesize \texttt{R} on behalf of an issuer (\checkmark). In contrast, \texttt{transStatus = N} is an ambiguous wrapper emitted across both tiers: scheme Directory Servers emit \texttt{N} on early prefix drops ($\text{acs} \equiv \text{ds}$), while ACSs emit \texttt{N} on frictionless denial ($\text{acs} \neq \text{ds}$) ($\times$).
    \item \textbf{Bimodal Reason Code Split (Reason \texttt{06}) and ACS-Associated Reason \texttt{08}}: In Table~A.1, Reason~\texttt{06} (\emph{Invalid card number}) lists both \texttt{DS} and \texttt{ACS} as normative sources, but in practice scheme Directory Servers use Reason~\texttt{06} ($\text{acs} \equiv \text{ds}$) for unallocated/format-boundary candidate drops at the routing tier. Reason~\texttt{08} (\emph{No card record}), by contrast, was observed here with distinct identifiers ($\text{acs} \neq \text{ds}$), a pattern consistent with ACS-associated handling after the request passed the DS range check under [Req~97]; the client-visible fields do not by themselves prove a ledger query or definitive account-state outcome (\checkmark).
    \item \textbf{Directory Server Reserved Codes (80--99)}: Reason codes \texttt{80}--\texttt{99} are normatively reserved for scheme Directory Server use. Amex SafeKey exclusively utilizes Reason~\texttt{80} for boundary exclusions ($\text{acs} \equiv \text{ds}$), Visa utilizes Reason~\texttt{87} for non-participating range drops ($\text{acs} \equiv \text{ds}$), and Mastercard utilizes Reason~\texttt{98} (\emph{Suspected Authentication BIN Attack}, Patent [0041]) and Reason~\texttt{99} (\emph{Override Authentication BIN Attack}, Patent [0043]) to terminate automated enumeration sweeps ($\text{acs} \equiv \text{ds}$) ($\triangle$).
    \item \textbf{Reason Code 01 vs.\ 06 Attribution}: Figure~\ref{fig:concrete_empirical_artifacts}(b) shows Reason~\texttt{01} with distinct identifiers, a pattern consistent with an ACS-associated response under [Req~105]. Panel~(a) shows Reason~\texttt{06} with identical identifiers, a pattern consistent with a DS-generated response under [Req~411]. Because the captures are client-facing integration JSON rather than correlated internal traces, these are conditional tier attributions, not direct observations of a ledger query or proof of candidate account status. Both responses carry an outward failure label (\texttt{authenticate\_frictionless\_failed}).
    \item \textbf{Electronic Commerce Indicator (ECI 07) Invariant}: In both empirical captures in Figure~\ref{fig:concrete_empirical_artifacts}, the gateway assigned ECI~\texttt{07} (\emph{Transaction failed / not authenticated}). Under scheme operating rules, ECI~\texttt{07} strictly withholds liability shift ($\text{liabilityShifted} = \text{false}$). ECI~07 marks a failed or unauthenticated outcome in these traces; by itself, it does not show whether a card is recognized on an active ledger. Interpret it with the status, reason code, and identifiers; no liability shift was established.
    \item \textbf{Challenge Timeout and Alert Scope:} Our caller stopped before the initial \texttt{CReq}. The tested path did not request decoupled authentication (\texttt{threeDSRequestorDecReqInd} was not set to \texttt{Y} or \texttt{B}), and no \texttt{transStatus=D} result was observed. If Req~220's 30-second timer expires without an Error Message from the DS, Req~221 requires the ACS to send the DS an \texttt{RReq} with Reason~14 and Challenge Cancellation Indicator~05~\cite{emvco_core_2311_2023}. FAQ Q93 addresses a different event: after a browser challenge has started, the v2.2 Req~343 \texttt{CRes}-to-Notification-URL behavior on the 10-minute timeout was removed in v2.3.1, and the Requestor should rely on the \texttt{RReq}~\cite{emvco_3ds_faq_2026}. Neither protocol path specifies a cardholder SMS/push alert; the OOB push example belongs to the challenge interaction. In our lookup-only tests, no SMS, push, or OTP reached the recipients. Independent issuer notification systems were not instrumented.
\end{itemize}
\vspace{4pt}
\subsection{Concrete Redacted Empirical Artifacts}
\vspace{4pt}
Figure~\ref{fig:concrete_empirical_artifacts} presents two sanitized client-visible JSON patterns received by a standalone off-context HTTP caller without browser SDK execution: (a)~identical identifiers and Reason~\texttt{06}, consistent with a DS-generated response under [Req~411]; and (b)~distinct identifiers and Reason~\texttt{01}, consistent with an ACS-associated response under [Req~105]. These tier labels are conditional inferences from integration fields; no correlated internal messages were captured. Nonces, candidate account prefixes, and institutional names are redacted, and UUIDs are synthetic RFC~4122 values.

\paragraph{Secondary Product Profiling Leakage (\texttt{binData}).}
Beyond identifiers and reason codes, the traces in Figure~\ref{fig:concrete_empirical_artifacts} also expose product attributes such as network, debit/credit status, issuing-bank metadata, and country code. Thus, even before financial authorization, the gateway reflects secondary profiling data to an off-context caller.

\noindent\textbf{b) Architectural Implications for Gateway Security.} Figure~\ref{fig:concrete_empirical_artifacts} illustrates an integration-layer reflection vulnerability (Taxonomy Class~\texttt{3DS-AUD-002}): the gateway exposes protocol-derived outcomes and product metadata in client-visible JSON before financial authorization. Returning an opaque continuation token and keeping nonessential metadata server-side would remove this gateway-boundary side channel.

\flushcolsend
\end{document}